\documentclass[12pt]{article}
\usepackage{amsmath, amsthm, amssymb}
\usepackage{setspace}
\usepackage{multirow}
\usepackage{graphicx}
\usepackage{apacite}
\usepackage{natbib}
\usepackage{color}
\usepackage{subcaption}
\usepackage{hyperref}
\usepackage{xr}

\newcommand{\uA}       {\mbox{\boldmath$A$}}

\newcommand{\uc}       {\mbox{\boldmath$c$}}
\newcommand{\uD}       {\mbox{\boldmath$D$}}
\newcommand{\ud}       {\mbox{\boldmath$d$}}
\newcommand{\uE}       {\mbox{\boldmath$E$}}

\newcommand{\uI}       {\mbox{\boldmath$I$}}

\newcommand{\uL}       {\mbox{\boldmath$L$}}

\newcommand{\uM}       {\mbox{\boldmath$M$}}

\newcommand{\uR}       {\mbox{\boldmath$R$}}
\newcommand{\ur}       {\mbox{\boldmath$r$}}
\newcommand{\uS}       {\mbox{\boldmath$S$}}

\newcommand{\uT}       {\mbox{\boldmath$T$}}

\newcommand{\uu}       {\mbox{\boldmath$u$}}
\newcommand{\uV}       {\mbox{\boldmath$V$}}

\newcommand{\uW}       {\mbox{\boldmath$W$}}
\newcommand{\uw}       {\mbox{\boldmath$w$}}
\newcommand{\uX}       {\mbox{\boldmath$X$}}
\newcommand{\ux}       {\mbox{\boldmath$x$}}
\newcommand{\uY}       {\mbox{\boldmath$Y$}}

\newcommand{\uZ}       {\mbox{\boldmath$Z$}}
\newcommand{\uz}       {\mbox{\boldmath$z$}}

\newcommand{\ualpha}       {\mbox{\boldmath$\alpha$}}
\newcommand{\ubeta}       {\mbox{\boldmath$\beta$}}
\newcommand{\ugamma}       {\mbox{\boldmath$\gamma$}}
\newcommand{\uDelta}       {\mbox{\boldmath$\Delta$}}
\newcommand{\udelta}       {\mbox{\boldmath$\delta$}}
\newcommand{\uepsilon}       {\mbox{\boldmath$\epsilon$}}

\newcommand{\ueta}       {\mbox{\boldmath$\eta$}}
\newcommand{\utheta}       {\mbox{\boldmath$\theta$}}

\newcommand{\ulambda}       {\mbox{\boldmath$\lambda$}}
\newcommand{\umu}       {\mbox{\boldmath$\mu$}}
\newcommand{\unu}       {\mbox{\boldmath$\nu$}}
\newcommand{\uxi}       {\mbox{\boldmath$\xi$}}

\newcommand{\uSigma}       {\mbox{\boldmath$\Sigma$}}
\newcommand{\usigma}       {\mbox{\boldmath$\sigma$}}
\newcommand{\uPhi}       {\mbox{\boldmath$\Phi$}}

\newcommand{\uvarphi}       {\mbox{\boldmath$\varphi$}}
\newcommand{\upsi}       {\mbox{\boldmath$\psi$}}
\newcommand{\uPsi}       {\mbox{\boldmath$\Psi$}}
\newcommand{\uomega}       {\mbox{\boldmath$\omega$}}

\newcommand{\uzero}       {\mbox{\boldmath$0$}}

\newcommand{\ahat} {\hat{\boldsymbol{\hspace{-2pt}\alpha}}}

\newcommand{\ohat} {\hat{\boldsymbol{\hspace{-2pt}\omega}}}

\newcommand{\utX} {\tilde{\boldsymbol{\hspace{-2pt}X}}}
\newcommand{\utZ} {\tilde{\boldsymbol{\hspace{-2pt}Z}}}

\newcommand{\utTheta} {\tilde{\boldsymbol{\hspace{-2pt}\theta}}}
\newcommand{\uty} {\tilde{\boldsymbol{\hspace{-2pt}y}}}
\newcommand{\utY} {\tilde{\boldsymbol{\hspace{-2pt}Y}}}

\newcommand{\ucX} {\check{\boldsymbol{\hspace{-0pt}X}}}
\newcommand{\ucy} {\check{\boldsymbol{\hspace{-0pt}y}}}
\newcommand{\ucY} {\check{\boldsymbol{\hspace{-0pt}Y}}}
\newcommand{\ucepsilon} {\check{\boldsymbol{\hspace{-0pt}\epsilon}}}

\title{Bayesian semiparametric modelling of covariance matrices for multivariate longitudinal data}

\author{Georgios Papageorgiou\\
Department of Economics, Mathematics and Statistics
\\Birkbeck, University of London\\  g.papageorgiou@bbk.ac.uk
}

\begin{document}
\maketitle

\begin{center}
\emph{Abstract}
\end{center}

The article develops marginal models for multivariate longitudinal responses. Overall, the model 
consists of five regression submodels, one for the mean and four for the covariance matrix, with the latter
resulting by considering various matrix decompositions. 
The decompositions that we employ are intuitive, easy to understand, and they do not rely on any assumptions 
such as the presence of an ordering among the multivariate responses. The regression submodels
are semiparametric, with unknown functions represented by basis function expansions. We use spike-slap 
priors for the regression coefficients to achieve variable selection and function regularization,
and to obtain parameter estimates that account for model uncertainty.
An efficient Markov chain Monte Carlo algorithm for posterior sampling is developed.
The simulation studies presented investigate the effects of priors on posteriors, the gains that one may have when 
considering multivariate longitudinal analyses instead of univariate ones, and whether these gains 
can counteract the negative effects of missing data. 
We apply the methods on a highly unbalanced longitudinal dataset with four responses observed over of period of $20$ years.

\section{Introduction}

There is a variety of systems that evolve over time and are too intricate to be appropriately characterized 
by a single outcome variable. Consider for example `cognition', the topic of the data example presented later in this article. 
According to the American Psychological Association's Dictionary of Psychology, cognition is a term that refers 
to `all forms of knowing and awareness, such as perceiving, conceiving, remembering, reasoning, judging, 
imagining, and problem solving'. Cognition is a complex, not directly observable, system that evolves over time. 
Measuring cognitive function relies on administering multiple psychometric tests, and investigating how it evolves 
and declines in the elderly requires observations over multiple years. Analysing datasets generated by such investigations 
requires flexible regression models for multivariate longitudinal responses. The main goal of this article is to 
develop, within a Bayesian framework, marginal models for multivariate longitudinal continuous responses, 
non-parametrically linking the mean vectors and covariance matrices to the covariates. The current literature
on modelling covariance matrices for multivariate longitudinal data relies on decompositions of the covariance matrices that are either 
difficult to interpret or that make assumptions that are not tenable. Here, we avoid all such decompositions and 
we only utilize those that are easy to understand and justify.     

In order to set the notation, let $\uY_{j}$ denote a vector of $p$ responses observed at time point $t_{j}$, $j=1,\dots,n$, 
$\uY = (\uY_{1}^{\top},\dots,\uY_{n}^{\top})^{\top}$ the $np$-dimensional vector of all responses, and  
$\uSigma$ the covariance matrix of $\uY$.
Matrix $\uSigma$ must satisfy the positive definiteness (pd) condition, and that creates a  
major challenge when modelling its elements as functions of covariates.  
To ensure that this condition is satisfied, it is necessary to factorized $\uSigma$ into a product of matrices. 
The decomposition \citep{Hamilton} 
\begin{equation} \label{fd}
\uL \uSigma \uL^{\top} = \uD 
\end{equation}
has been utilized for modelling multivariate longitudinal data by \citet{xu}, \cite{kim}, \citet{KOHLI2016} and \citet{Lee2019}. 
Matrices $\uL$ and $\uD$ have the following structures
\begin{eqnarray}\label{LD}
\begin{array}{cc}
\uL 
= \left[ \begin{array}{cccc}
\uI & \uzero & \dots & \uzero \\ 
-\uPhi_{21} & \uI & \dots & \uzero \\ 
\vdots & \vdots & \ddots & \vdots  \\
-\uPhi_{n1} & -\uPhi_{n_2} & \dots & \uI \\ 
\end{array}
\right],
&
\uD 
= \left[ \begin{array}{cccc}
\uD_{1} & \uzero & \dots & \uzero \\ 
\uzero & \uD_{2} & \dots & \uzero \\ 
\vdots & \vdots & \ddots & \vdots  \\
\uzero & \uzero & \dots & \uD_{n} \\ 
\end{array}
\right],
\end{array}
\end{eqnarray}
where submatrices $\uPhi_{jk}$ are unconstrained, $j=2,\dots,n, k<j,$ while submatrices $\uD_{j}$ are pd, $j=1,\dots,n$,
and all submatrices in $\uL$ and $\uD$ are of dimension $p \times p$. 
The following interpretation that has  appeared  in, among others,  
\citet{Poura1999} and \citet{Daniels2002} for the univariate case, and \citet{xu} and \citet{kim} for the multivariate
case, provides justification for the lack of constraints on the elements of $\uPhi_{jk}$ and the pd constraint on the elements of 
$\uD_{j}$. For the sake of simplicity we assume that vectors $\uY_{j}$ have zero mean.  
Let $\hat \uY_{j}$ denote the linear least squares predictor of $\uY_{j}$
based on its predecessors $\uY_{j-1}, \dots, \uY_{1},$ and let $\uepsilon_{j} = \uY_{j} - \hat \uY_{j}$ 
denote the prediction error. It can be shown that the predictor $\hat \uY_{j}$ is expressed in terms of the negatives of 
the submatrices of $\uL$ as
\begin{equation}\label{pred}
\hat{\uY}_{j} = \sum_{k=1}^{j-1}\uPhi_{jk}\uY_{k}.
\end{equation}
Further, the prediction error variance is given by the corresponding block of $\uD$ i.e. $\text{var}(\uepsilon_{j}) = \uD_{j}$.
To see how factorization (\ref{fd}) is satisfied, 
let $\uepsilon=(\uepsilon_{1}^{\top},\dots,\uepsilon_{n}^{\top})^{\top}$.
We have that $\text{var}(\uepsilon)=\text{var}(\uL \uY)=\uL\uSigma\uL^{\top}$,
and since consecutive prediction errors are uncorrelated, it follows that 
$\uL\uSigma\uL^{\top}=\uD$.
The sets of matrices $\{\uPhi_{jk}: j=2,\dots,n, k < j\}$ and 
$\{\uD_{j}: j=1,\dots,n\}$ have been termed the generalized autoregressive matrices and 
innovation covariance matrices of $\uSigma$ \citep{xu}.

A simple modification of (\ref{fd}) can provide an alternative interpretation. 
Since $\uL^{-1}$ has the same block triangular structure as $\uL$, the decomposition 
\begin{equation} \label{sd}
\uL^{-1} \uSigma (\uL^{-1})^{\top} = \uD^{\ast}, 
\end{equation}
also holds, where $\uD^{\ast}$ has the same structure as $\uD$. 
This decomposition, for univariate responses, has been studied by \citet{ZhangLeng},
and for multivariate ones by \citet{Feng}. With this decomposition, the submatrices of $\uL$ that are below the main diagonal can 
be interpreted as moving average coefficient matrices in a time series context, hence, again, they are unconstrained. 

In this article, we adopt the decomposition in (\ref{fd}) that leads to the interpretation in (\ref{pred}). 
Since  matrices $\uPhi_{jk}$ are unconstrained, their  elements can be modelled utilizing semiparametric regression models. 
These are described in Section \ref{two}. 

The decompositions in (\ref{fd}) and (\ref{sd}) have been unanimous among researchers modelling multivariate longitudinal data.
By contrast, there have been several proposals for modelling the innovation covariance matrices $\{\uD_{j}: j=1,\dots,n\}$.
These are reviewed in the next paragraph.   
The difficulty in modelling these matrices arises from having to satisfy the pd condition and from the absence of 
any natural ordering in the elements of the errors $\uepsilon_{j}$. 

Based on the spectral decomposition and the matrix logarithmic transformation \citep{Chiu96}, \citet{xu} model matrices $\uD_{j}, j=1,\dots,n,$ 
in terms of the explanatory variable `time', $t$. Further, based on the modified Cholesky decomposition \citep{Poura1999, Poura2000}, i.e. 
the decomposition in (\ref{fd}) applied for $p=1$, \citet{kim} model the innovation covariance matrices as functions of $t$.    
Lastly, \citet{KOHLI2016} avoid decompositions and instead model each $\uD_j^{-1}$ using linear covariance models \citep{anderson1973}, that is, 
$\uD_j^{-1} = \alpha_1 \uM_1 + \dots + \alpha_j \uM_j$, where  
$\uM_k$ are known pd matrices and $\alpha_k$ are unknown weights, $k=1,\dots,j$. 
The spectral decomposition and the modified Cholesky decomposition outside the context of longitudinal studies, lack simple statistical 
interpretation. Further, the linear covariance model does not express the matrices in terms of covariates and its implementation in practice 
can be difficult. For instance, the best performing model in the study of \citet{KOHLI2016} does not guarantee that the estimated covariance matrix is pd. 

A decomposition, however, that is statistically simple and intuitive, was proposed by \citet{Barnard00}. 
It separates the matrices $\uD_j$ into diagonal matrices of innovation variances $\uS_j$ 
and innovation correlation matrices $\uR_j$, $\uD_j = \uS_j^{1/2} \uR_j \uS_j^{1/2}$. 
With this decomposition it is easy to model the innovation variances in terms of covariates 
as the only constrained they must satisfy is the positiveness. This was the approach taken by \citet{Lee2019} 
and the approach that we take in this article too. 

As was remarked by \citet{Poura07}, of the available decompositions, the separation of variances and correlations is the least effective in 
satisfying the pd condition. Clearly, this condition must now be satisfied by the innovation correlation matrices
$\{\uR_{j}: j=1,\dots,n\}$. To overcome this condition, and model the elements of a single correlation matrix $\uR$ in terms of covariates, 
\citet{Zhang15} utilize the Cholesky decomposition $\uR = \uT \uT^{\top}$, parametrizing the non-zero elements of $\uT$ 
using hyperspherical coordinates. In the applications presented by \citet{Zhang15}, the elements of $\uR$ are modelled in terms 
of covariate `lag'. \citet{Lee2019} follow the same approach as \citet{Zhang15}, but merely reparametrise the innovation correlation matrices 
without modelling their elements in terms of covariates. In the data analysis they presented, 
this resulted in a common, time invariant, innovation correlation matrix, even though they stressed the importance 
of correctly specifying the correlation structure. 

Here, we take a different, more intuitive approach, by specifying a normal prior on the Fisher's $z$ transformation of the 
nonredundant elements of the matrices $\uR_j = \{r_{jkl}\}$, 
\begin{equation}
\log\{(1+r_{jkl})/(1-r_{jkl})\}/2 \sim N(\mu_{cj},\sigma^2_{cj}) I[\uR_j \in \mathcal{C}],\label{model1}
\end{equation}
where $\mathcal{C}$ denotes the space of correlation matrices and $I[.]$ the indicator function that maintains positive definiteness. 
In addition, the indicator function truncates the range of the correlations and induces dependence among them \citep{DanielsKass99}. 
Due to this truncation, parameters $\mu_{cj}$ and $\sigma^2_{cj}$ are no longer interpreted as the mean and variance of the distribution.
Here, we model both parameters as unknown functions of time $t$,
$\mu_{cj} = \eta_0 + f_{\mu}(t_j)$ and $\log \sigma^2_{cj} = \omega_0 + f_{\sigma}(t_j)$.   

The model in (\ref{model1}) can be restrictive because it only allows for a single function $\mu_{cj}$ that is common to all correlations. 
However, it is conceivable that correlations evolve differently over time. Failing to specify a prior
that allows for the needed flexibility can have a negative impact on the estimated correlations, especially in small samples \citep{DanielsKass99}. 
Here, this flexibility is reached by considering mixtures of normal distributions 
$\log\{(1+r_{jkl})/(1-r_{jkl})\}/2 \sim \sum_{h} \pi_h N(\mu_{cjh},\sigma^2_{cj}) I[\uR_j \in \mathcal{C}]$.
We refer to this model as the `grouped correlations model' \citep{Liechty}. We also consider a 
`grouped variables model' that clusters the variables instead of the correlations and it is more structured 
than the grouped correlations model.

The overall model consists of $5$ regression submodels. These are the models for the mean of the response vector, the 
elements of the generalized autoregressive matrices, the innovation variances, and the $2$ parameters 
of distribution of the innovation correlation matrices. In the approach presented here, each regression 
model involves nonparametrically modelled effects, represented utilizing basis function expansions.
We enable flexible estimation by utilizing several basis functions  
and we implement variable selection and function regularization by utilizing spike-slab priors (see e.g. \citet{george97}).
The work presented in this article builds upon the work of \citet{Chan06} and \citet{pap18}
who describe methods for univariate response regression with nonparametric models for the mean and variance, 
and the work of \citet{multi} who presented methods for multivariate response regression 
with nonparametric models for the means, the variances and the correlation matrix.  

We develop an efficient stochastic search variable selection algorithm by using Zellner's g-prior \citep{AZ} that allows integrating out the regression coefficients in 
both the mean function of the responses $E(\uY)$ and the parameter $\mu_{cj}$ (or $\mu_{cjh}$) of the innovation correlation matrices. 
In addition, the Markov chain Monte Carlo (MCMC) algorithm generates the variable selection indicators in blocks \citep{Chan06, pap18} and selects 
the MCMC tuning parameters adaptively \citep{Roberts2001c}. 

The remainder of this article is arranged as follows. Section \ref{two} describes the model in detail and Section \ref{three} 
describes the main elements of the MCMC algorithm. All the details of the MCMC algorithm are available in the supplementary material.
In Section \ref{sim} we present results from $2$ simulation studies. The first one investigates how posteriors, 
based on different priors, concentrate around the true covariance and correlation matrices, while the 
second one investigates the gains that one may have, in terms of reduced posterior mean squared error (MSE), 
when fitting multivariate longitudinal models instead of univariate ones. Additionally, in the second study we examine 
the effects that missing data have on the posterior MSE and whether adding more than one response to the model can 
counteract these effects. Section \ref{application} applies the methods on a  highly unbalanced dataset on cognitive function 
and depressive symptomatology, with $4$ responses observed over of period of $20$ years. 
Section \ref{discus} concludes the paper with a brief discussion. 
All the methods described in this article are freely available in the R package BNSP \citep{bnsp}.

\section{Multivariate longitudinal response model}\label{two}

Let $\uY_{ij} = (Y_{ij1},\dots,Y_{ijp})^{\top}$ denote the vector of $p$ responses 
observed on individual $i, i=1,\dots,n,$ at time point $t_{ij}$, $j=1,\dots,n_i$. Here, we allow the observational time points $t_{ij}$ to be unequally spaced. 
We let $T$ denote the ordered set of all unique observational times, $T = \{t_{1},\dots,t_{M}\}$, and we denote its cardinality by $M$.
Further, let $\uu_{ij}$ denote the covariate vector that is observed along with $\uY_{ij}$ and that may include time, 
other time-dependent covariates and time-independent ones. 
In addition, let $\uY_{i}=(\uY_{i1}^{\top},\dots,\uY_{in_i}^{\top})^{\top}$ denote the $i$th response vector. 
With $\umu_i=E(\uY_{i})$ and $\uSigma_i = \text{cov}(\uY_i)$, the overall model takes the form
\begin{equation}
\uY_{i} \sim N(\umu_i, \uSigma_i), i=1,2,\dots,n. \nonumber \label{com1} 
\end{equation}    
In the following subsections we detail how the means $\umu_i$ and covariance matrices $\uSigma_i$ are modelled semiparametrically in terms of covariates. 

\subsection{Mean model}

The means $\mu_{ijk}=E(Y_{ijk}), k=1,\dots,p,$ are modelled utilizing semiparametric regression methods
\begin{equation}
\mu_{ijk} = \beta_{k0} + \sum_{l=1}^{K_1} u_{ijl} \beta_{kl} + \sum_{l=K_1+1}^{K} f_{\mu,k,l}(u_{ijl}),\label{mean2}
\end{equation}
where $u_{ijl}, l=1,\dots,K_1,$ are regressors with parametrically modelled effects, 
$u_{ijl}, l=K_1+1,\dots,K,$ are regressors with effects that are modelled as unknown functions, 
and $K$ denotes the total number of effects that enter the $p$ mean models.

Unknown functions are represented utilizing basis function expansions
\begin{equation}
f_{\mu,k,l}(u_{ijl}) = \sum_{s=1}^{q_{\mu l}} \beta_{kls} \kappa_{\mu ls}(u_{ijl}) = \ux_{ijl}^{\top} \ubeta_{kl},\label{star}
\end{equation}
where $q_{\mu l}$ is fixed and it represents the maximum number of basis functions that can be used in modelling $f_{\mu,k,l}(.)$. Further, 
$\ux_{ijl} = (\kappa_{\mu l1}(u_{ijl}),\dots,\kappa_{\mu lq_{\mu l}}(u_{ijl}))^{\top}$ 
is the vector of basis functions and 
$\ubeta_{kl} = (\beta_{kl1},\dots,\beta_{klq_{\mu l}})^{\top}$ 
is the vector of the corresponding coefficients. 
Here, the basis functions of choice are the radial basis functions, given by 
$\kappa_{1}(u)  = u, \kappa_2(u) = |u - \xi_{1}|^2 \log\left(|u-\xi_{1}|^2\right), \dots,
\kappa_{q-1}(u) = |u - \xi_{q-1}|^2 \log\left(|u-\xi_{q-1}|^2\right)$, where 
$\xi_{1},\dots,\xi_{q-1}$ are fixed knots. 

Now, model (\ref{mean2}) can be linearised as 
\begin{equation}
\mu_{ijk} = \beta_{k0} + \sum_{l=1}^{K_1} u_{ijl} \beta_{kl} + \sum_{l=K_1+1}^{K} \ux_{ijl}^{\top} \ubeta_{kl} = 
\beta_{k0} + \ux_{ij}^{\top} \ubeta_k, = (\ux_{ij}^{\ast})^{\top} \ubeta_k^{\ast}, \label{mean3}
\end{equation}
where $\ux_{ij} = (u_{ij1},\dots,u_{ijK_1},\ux^{\top}_{ij,K_1+1},\dots,\ux^{\top}_{ijK})^{\top}$
and $\ubeta_k=(\beta_{k1},\dots,\beta_{kK_1},\ubeta_{k,K_1+1}^{\top},\dots,\ubeta_{kK}^{\top})^{\top}$. 
Further, $\ux_{ij}^{\ast} = (1,\ux_{ij}^{\top})^{\top}$ and $\ubeta_{k}^{\ast} = (\beta_{k0}, \ubeta_k^{\top})^{\top}$.

The implied model for the mean of vector $\uY_{ij}$ is
\begin{equation}
\umu_{ij} = E(\uY_{ij}) = \ubeta_0 + \uX_{ij} \ubeta = \uX_{ij}^{\ast} \ubeta^{\ast}, \nonumber
\end{equation} 
where $\ubeta_0 = (\beta_{10},\dots,\beta_{p0})^{\top}$, $\uX_{ij} = \uI_{p} \otimes \ux_{ij}^{\top}$,
and $\ubeta = (\ubeta_{1}^{\top},\dots,\ubeta_{p}^{\top})^{\top}$.
In the final expression, $\uX_{ij}^{\ast}$ and $\ubeta^{\ast}$ are as $\uX_{ij}$ and $\ubeta$ but with 
$\ux_{ij}$ and $\ubeta_k$ replaced by $\ux_{ij}^{\ast}$ and $\ubeta_{k}^{\ast}, k=1,\dots,p$.
   
Recalling that $\uY_{i}$ denotes the $i$th response vector, we may write the mean model as 
$E(\uY_{i}) = \uX_{i}^{\ast} \ubeta^{\ast}$, where 
$ \uX_{i}^{\ast} = [(\uX_{i1}^{\ast})^{\top}, (\uX_{i2}^{\ast})^{\top}, \hdots, 
(\uX_{i n_i}^{\ast})^{\top}]^{\top}$.
Alternatively, the model can be written in the usual form $E(\uY) = \uX^{\ast} \ubeta^{\ast},$
where $\uY=(\uY_1^{\top},\dots,\uY_n^{\top})^{\top}$ and $\uX^{\ast} = [(\uX_1^{\ast})^{\top},\dots,(\uX_n^{\ast})^{\top}]^{\top}$.

We note that $\ux_{ij}$ is a maximal vector covariates and it is common to the $k$ responses. By 
introducing binary indicators for variable selection, we allow each response to have its own set of covariates. 
Vector $\ugamma_k=(\gamma_{k1},\dots,\gamma_{kK_1},\ugamma_{k, K_1+1}^{\top},\dots,\ugamma_{kK}^{\top})^{\top}$ 
has the same length and structure are vector $\ubeta_k$, introduced just below (\ref{mean3}), and its elements are binary indicators
that select which regressors enter the mean model of the $k$th response. 
Given $\ugamma_k$, model (\ref{mean3}) is written as 
\begin{eqnarray}\label{meanfinal}
\mu_{ijk} = \beta_{k0} + \ux_{ijk}^{\top} \ubeta_{\gamma_k} = (\ux_{ijk}^{\ast})^{\top} \ubeta_{\gamma_k}^{\ast}, \nonumber
\end{eqnarray}
where $\ubeta_{\gamma_k}$ consists of all non-zero elements of $\ubeta_k$
and $\ux_{ijk}$ of the corresponding elements of $\ux_{ij}$.
 
Letting $\ugamma = (\ugamma_1^{\top},\dots,\ugamma_p^{\top})^{\top}$ denote the vector of binary indicators and $\ubeta_{\gamma}^{\ast} = ((\ubeta_{\gamma_1}^{\ast})^{\top},\dots,(\ubeta_{\gamma_k}^{\ast})^{\top})^{\top}$ the vector of non-zero regression coefficients for the $p$ responses, 
we can write the regression model for the mean of $\uY_i$ as
$E(\uY_i) = \uX_{\gamma i}^{\ast} \ubeta_{\gamma}^{\ast}$. Similarly, we can write  
$E(\uY) = \uX_{\gamma}^{\ast} \ubeta_{\gamma}^{\ast}$. 

\subsection{Covariance model}

Let $\uSigma_i$ denote the covariance matrix of $\uY_i$. 
As was discussed in the introduction, to model the elements of $\uSigma_i$ in terms of covariates, we beging by considering the 
factorization $\uL_i \uSigma_i \uL_i^{\top} = \uD_i$, where  
matrices $\uL_i$ and $\uD_i$ have the form shown in (\ref{LD}).
The next two subsections describe semiparametric models for 
the generalized autoregressive matrices $\{\uPhi_{ijk}: j=2,\dots,n_i, k < j\}$ and the innovation covariance matrices 
$\{\uD_{ij}: j=1,\dots,n_i\}$.

\subsection{Generalized autoregressive matrices}

For $\phi_{ijklm}$, the $(l,m)$ element of $\uPhi_{ijk}$, $l,m=1,\dots,p$, we consider the following semiparametric model
\begin{equation}\label{phi}
\phi_{ijklm} = \psi_{lm0} + \sum_{b=1}^{B_1} v_{ijkb} \psi_{lmb} + \sum_{b=B_1+1}^{B} f_{\phi,l,m,b}(v_{ijkb}) = 
\psi_{lm0}  + \uz_{ijk}^{\top} \upsi_{lm} = (\uz^{\ast}_{ijk})^{\top} \upsi_{lm}^{\ast},
\end{equation} 
where the developments in (\ref{phi}) follow along the same lines as those for the mean model, detailed in (\ref{mean2}) - (\ref{mean3}). 
Functions $f_{\phi,l,m,b}(v_{ijkb}) = \uz_{ijkb}^{\top} \upsi_{lmb}$ are linearised utilizing a fixed number $q_{\phi b}$ of basis functions, $b=B_1+1,\dots,B$.
Vector $\uz^{\ast}_{ijk}$ denotes the maximal set covariates that is common to the $p^2$ autoregressive coefficients. 
Further, vector $\upsi_{lm}^{\ast} = (\psi_{lm0}, \psi_{lm1},\dots, \psi_{lmB_1}, \upsi_{lm,B_1+1}^{\top},\dots, \upsi_{lmB}^{\top})^{\top}$ consists of the regression coefficients, grouped by the effect they model.
Vector $\uxi_{lm}$ is the corresponding vector of binary indicators that selects which coefficients enter the model of the $(l,m)$ autoregressive coefficient.
We note that here the intercepts are subject to selection. 
Given $\uxi_{lk}$, model (\ref{phi}) is expressed as 
\begin{eqnarray}\label{phifinal}
\phi_{ijklm} = (\uz^{\ast}_{ijklm})^{\top} \upsi_{\xi_{lm}}^{\ast}, \label{phi2}
\end{eqnarray}
where $\upsi_{\xi_{lm}}^{\ast}$ consists of all non-zero elements of $\upsi_{lm}^{\ast}$
and $\uz_{ijklm}^{\ast}$ of the corresponding elements of $\uz_{ijk}^{\ast}$. 

Let $\upsi^{\ast} = ((\upsi_{11}^{\ast})^{\top},\dots,(\upsi_{1p}^{\ast})^{\top}, (\upsi_{21}^{\ast})^{\top}, 
\dots, (\upsi_{pp}^{\ast})^{\top})^{\top}$ be the vector of all regression coefficients, and 
$\uxi = (\uxi_{11}^{\top},\dots,\uxi_{1p}^{\top},\uxi_{21}^{\top}, \dots,\uxi_{pp}^{\top})^{\top}$
be the vector of all binary indicators. Further, let $\upsi_{\xi}^{\ast}$ be the vector of all non-zero coefficients. 
Now, from (\ref{pred}) and the definition of the prediction error, we have
\begin{equation}\label{pred2}
\uY_{ij} = \sum_{k=1}^{j-1}\uPhi_{ijk}\uY_{ik} + \uepsilon_{ij}
= \sum_{k=1}^{j-1} [\uI_p \otimes \uY_{ik}^{\top} \otimes (\uz^{\ast}_{ijk})^{\top} ]_{\xi} \upsi_{\xi}^{\ast} + \uepsilon_{ij} = 
\uV_{ij \xi}  \upsi_{\xi}^{\ast} + \uepsilon_{ij},
\end{equation}
that represents a dynamic linear model, since the design matrix $\uV_{ij \xi} = 
\sum_{k=1}^{j-1} [\uI_p \otimes \uY_{ik}^{\top} \otimes (\uz^{\ast}_{ijk})^{\top} ]_{\xi}$ involves the predecessors of $\uY_{ij}$. 
Note that, by using $\xi$ as subscript in a matrix $[.]_{\xi}$, we mean the matrix with the columns that correspond to the zero elements of $\uxi$ removed. 
When the mean of $\uY_{ij}$ is not zero, we 
replace $\uY_{ij}$ in (\ref{pred2}) by its centred version  $\uY_{ij}-\uX_{ij}^{\ast} \ubeta^{\ast}$. This leads to 
\begin{equation}
\uY_{ij} - \uX_{ij}^{\ast} \ubeta^{\ast} = \sum_{k=1}^{j-1}\uPhi_{ijk}(\uY_{ik} -\uX_{ik}^{\ast} \ubeta^{\ast})+ \uepsilon_{ij},\label{pred3} 
\end{equation}
which can be written in the more familiar form of a multivariate mixed model 
\begin{equation}\label{pred4}
\uY_{ij} = \uX_{ij}^{\ast} \ubeta^{\ast} + \uV_{ij \xi \beta} \upsi_{\xi}^{\ast} + \uepsilon_{ij}, 
\end{equation}
where the design matrix $\uV_{ij \xi \beta}$ depends on $\ubeta^{\ast}$. 

\subsection{Innovation covariance matrices}

For modelling the innovation covariance matrices $\uD_{ij}, i=1,\dots,n, j=1,\dots,n_i$,  
we begin by employing the separation strategy of \citet{Barnard00}, by which $\uD_{ij}$ is decomposed into a diagonal matrix of variances $\uS_{ij} = \text{diag}(\sigma^2_{ij1},\dots,\sigma^2_{ijp})$ and a correlation matrix $\uR_{ij}$,  
\begin{equation}
\uD_{ij} = \uS_{ij}^{1/2} \uR_{ij} \uS_{ij}^{1/2}.\label{sep}
\end{equation}
The next subsections consider models for the diagonal elements of $\uS_{ij}$ and the correlation matrix $\uR_{ij}$. 

\subsubsection{Diagonal innovation variance matrices}

It is easy to model matrix $\uS_{ij}$ in terms of covariates as the only requirement on its diagonal elements is that they are nonnegative. It is clear that the following semiparametric model satisfies this requirement
\begin{equation}
\log \sigma^2_{ijk} = \alpha_{k0} + \sum_{l=1}^{L_1} w_{ijl} \alpha_{kl} + \sum_{l=L_1+1}^{L} f_{\sigma,k,l}(w_{ijl}) =  \alpha_{k0} + \uw_{ij}^{\top} \ualpha_{k}. \label{var1}
\end{equation}
Similar models for innovation variances have appeared in \citet{Leng} and \citet{LinPan}.

Consider now vectors of indicator variables for selecting the elements of $\uw_{ij}$ that enter the $k$th variance regression model.
In line with the indicator variables for the mean and autoregressive models, these are denoted by $\udelta_k=(\delta_{k1},\dots,\delta_{kL_1},\udelta_{k,L_1+1}^{\top},\dots,\udelta_{kL}^{\top})^{\top}$.
Given $\udelta_k$, model (\ref{var1}) can be expressed as 
\begin{eqnarray}
\log \sigma^2_{ijk} = \alpha_{k0} + \uw^{\top}_{ijk} \ualpha_{\delta_k}, \nonumber
\end{eqnarray}
or equivalently
\begin{eqnarray}
\sigma^2_{ijk} = \exp(\alpha_{k0}) \exp(\uw^{\top}_{ijk} \ualpha_{\delta_k}) =
\sigma^2_{k} \exp(\uw^{\top}_{ijk} \ualpha_{\delta_k}).\label{lastvar}
\end{eqnarray}

\subsubsection{Time-varying innovation correlation matrices}

The approach we take for modelling the correlation matrices $\uR_{ij}, i=1,\dots,n, j=1,\dots,n_i$, extends the work of \citet{Liechty} and \citet{multi} who proposed models for correlation matrices in the context of cross-sectional studies. In the context of longitudinal studies, we propose to model these matrices utilizing time as the only covariate. Hence, we use symbols $\uR_{t}=\{r_{tkl}\}, t \in T,$ to denote these matrices. Below we describe three priors, termed `common correlations', `grouped correlations' and `grouped variables' priors.    

\subsubsection{Common correlations}

The common correlations model is defined as follows 
\begin{equation}\label{priorR}
f(\uR_{t}|\mu_{ct},\sigma^2_{ct}) = \pi(\mu_{ct},\sigma^2_{ct}) 
\prod_{k<l} \left\{\exp\{-(g(r_{tkl})-\mu_{ct})^2/2\sigma^2_{ct}\} J[g(r_{tkl}) \rightarrow r_{tkl}]\right\} I[\uR_{t} \in \mathcal{C}],
\end{equation}
where $\mathcal{C}$ denotes the space of correlation matrices, $I[.]$ is the indicator function that ensures that the correlation matrix is pd 
and $\pi(.,.)$ is the normalizing constant
\begin{eqnarray}\nonumber
\pi^{-1}(\mu_{ct},\sigma_{ct}^2) = \int_{\uR_{t} \in \mathcal{C}} \prod_{k<l} 
\left\{\exp\{-(g(r_{tkl})-\mu_{ct})^2/2\sigma^2_{ct}\} J[g(r_{tkl}) \rightarrow r_{tkl}]\right\} dr_{tkl}.
\end{eqnarray}
A typical choice for $g(r)$ is the Fisher's $z$ transformation $g(r) = \log([1+r]/[1-r])/2$ that leads to $J[g(r) \rightarrow r] = (1-r)^{-1} (1+r)^{-1}$. 

We model the parameters of this distribution using nonparametric regression, with time $t$ as the only independent variable 
\begin{eqnarray}
&& \mu_{ct} = \eta_0 + f_{\mu}(t) = \eta_0 + \uz_{\mu t}^{\top} \ueta, \label{muct}\\
&& \log \sigma^2_{ct} = \omega_0 + f_{\sigma}(t) = \omega_0 + \uz_{\sigma t}^{\top} \uomega. \label{logSigma}
\end{eqnarray}
Equivalently, we can write (\ref{logSigma}) as
\begin{eqnarray}
\sigma^2_{ct} = \exp(\omega_0) \exp(\uz_{\sigma t}^{\top} \uomega) = \sigma^2 \exp(\uz_{\sigma t}^{\top} \uomega). \label{Sigma} \nonumber
\end{eqnarray}

If the indicator function were not present in (\ref{priorR}), then the model described in 
(\ref{priorR}) - (\ref{logSigma}) could be thought of as a semi-parametric regression
model for the mean and the variance of normally distributed responses (see e.g. \citet{Chan06} and \citet{pap18}).  
Let vector $\ur_t$ consist of the non-redundant elements of $\uR_t, t \in T$,
and let $\ur = (\ur_{t_1}^{\top},\dots,\ur_{t_{M}}^{\top})^{\top}$. Subject to the pd constraint, we can express the common correlations model as 
\begin{eqnarray}
g(\ur) = \uZ^{\ast} \ueta^{\ast} + \uepsilon, \uepsilon \sim N(\uzero, \sigma^2 \uD^2(\uomega)), \label{linear}
\end{eqnarray}
where $\ueta^{\ast} = (\eta_0,\ueta^{\top})^{\top}$, $\uZ^{\ast}$ is a design matrix with rows equal to $\uz_{\mu t}^{\ast} = (1,\uz_{\mu t}^{\top})^{\top}$
and $\uD(\uomega)$ is a diagonal matrix with elements equal to 
$\exp(\uz_{\sigma t}^{\top} \uomega/2), t \in T$, where $\uz_{\mu t}^{\ast}$ and $\exp(\uz_{\sigma t}^{\top} \uomega/2)$ 
appear in the corresponding matrices $d=p(p-1)/2$ times, the number of unique elements of $\uR_t$. 
 
Introducing now vectors of indicators $\unu$ and $\uvarphi$ for selecting the elements of $\uz_{\mu t}$ and $\uz_{\sigma t}$ that enter the models for $\mu_{ct}$ and $\sigma^2_{ct}$, respectively, model (\ref{linear}) can be written as 
\begin{eqnarray}\label{modcor}
g(\ur) = \uZ^{\ast}_{\nu} \ueta_{\nu}^{\ast} + \uepsilon,
\uepsilon \sim N(\uzero, \sigma^2 \uD^2(\uomega_{\varphi})),
\end{eqnarray}
where $\ueta_{\nu}^{\ast}$ and $\uomega_{\varphi}$ consist of the non-zero elements of $\ueta^{\ast}$ and $\uomega$, and 
$\uZ^{\ast}_{\nu}$ consists of the columns of $\uZ^{\ast}$ that correspond to the non-zero elements of $\ueta^{\ast}$. 

\subsubsection{Grouped correlations}

The common correlations model is restrictive as it implies that all correlations evolve in the same way over time. 
It is important, however, from a theoretical standpoint to estimate the correlation structure correctly. 
Hence, here we generalize the model to include multiple surfaces by 
considering the following prior
\begin{eqnarray}\label{priorR3}
&&f(\uR_{t}|\umu_{ct},\sigma^2_{ct},\ulambda) = \pi(\umu_{ct},\sigma^2_{ct},\ulambda) \nonumber\\ 
&&\times \prod_{k<l} \left\{ 
\sum_{h=1}^H I[\lambda_{kl}=h] \exp[-(g(r_{tkl})-\mu_{cth})^2/2\sigma^2_{ct}]
\right\} J[g(r_{tkl}) \rightarrow r_{tkl}] I[\uR_{t} \in \mathcal{C}],
\end{eqnarray}
where $H$ denotes the number of surfaces in the model, $\lambda_{kl}$ is the surface assignment indicator and
\begin{eqnarray}
\mu_{cth} = \eta_{h0} + f_{\mu h}(t) = \eta_{h0} + \uz_{\mu t h}^{\top} \ueta_{\nu_h} =
(\uz_{\mu t h}^{\ast})^{\top}\ueta_{\nu_h}^{\ast}.\nonumber
\end{eqnarray}
In the above, $\ueta_{h}$ are surface specific regression coefficients and $\unu_h$ are surface specific variable selection indicators, $h=1,\dots,H$. 

Let vector $\ur_{th}$ consist of the non-redundant elements of $\uR_t, t \in T$, assigned to surface $h$.
Further, let $\ur_h = (\ur_{t_1 h}^{\top},\dots,\ur_{t_{M} h}^{\top})^{\top}$. Subject to the pd constraint, we can express model (\ref{priorR3}) as
\begin{eqnarray}
g(\ur_h) = \uZ^{\ast}_{\nu_h} \ueta_{\nu_h}^{\ast} + \uepsilon_h, h=1,\dots,H, \label{linearCluster}
\end{eqnarray}
where $\uepsilon_h \sim N(\uzero, \sigma^2 \uD^2_h(\uomega_{\varphi}))$.

\subsubsection{Grouped variables}

Here we describe another clustering model that, instead of correlations, clusters the variables.
We have the following prior
\begin{eqnarray}\label{priorR4}
&& f(\uR_{t}|\umu_{ct},\sigma^2_{ct},\ulambda) = \pi(\umu_{ct},\sigma^2_{ct},\ulambda) \nonumber\\ 
&& \times \prod_{k<l} \left\{ \sum_{h_1, h_2=1}^G I[\lambda_{k}=h_1] I[\lambda_{l}=h_2] 
\exp[-(g(r_{tkl})-\mu_{cth_1h_2})^2/2\sigma^2_{ct}] \right\} J[g(r_{tkl}) \rightarrow r_{tkl}]
I[\uR_{t} \in \mathcal{C}],\nonumber
\end{eqnarray}
where
\begin{eqnarray}
\mu_{cth_1h_2} = \eta_{h_1h_20} + f_{\mu h_1 h_2}(t) = \eta_{h_1h_20} + \uz_{\mu t h_1 h_2}^{\top} \ueta_{\nu_{h_1h_2}} = (\uz_{\mu t h_1 h_2}^{\ast})^{\top} \ueta_{\nu_{h_1h_2}}^{\ast},\nonumber
\end{eqnarray}
and $G$ is the number of groups in which the variables are distributed, creating $H=G(G+1)/2$ clusters 
for the correlations.  

\subsection{Prior specification}

In this section we specify the priors for the model parameters. 
We begin by describing the priors of the parameters of the mean model. 
For $\ubeta_{\gamma}^{\ast}$ we specify a g-prior \citep{AZ},
\begin{eqnarray}
\ubeta_{\gamma}^{\ast} | c_{\beta}, \ugamma, \uSigma \sim N(\uzero,c_{\beta} (\utX_{\gamma}^{\top} \utX_{\gamma} )^{-1}), \label{gprior1} 
\end{eqnarray}
where $\utX_{\gamma} = \uSigma^{-\frac{1}{2}} \uX^{\ast}_{\gamma}$, and $\uSigma$ is the covariance matrix of the $\uY$, that is $\uSigma = \text{diag}(\uSigma_i, i=1,\dots,n)$. 
Further, the prior for $c_{\beta}$ is taken to be an inverse Gamma, $c_{\beta} \sim \text{IG}(a_{\beta},b_{\beta})$.

Consider now the vectors $\ugamma_k=(\gamma_{k1},\dots,\gamma_{kK_1},\ugamma_{k,K_1+1}^{\top},\dots,\ugamma_{kK}^{\top})^{\top}, k=1,\dots,p,$ of 
variable selection indicators for the mean functions. We specify independent binomial priors for each of their $K$ subvectors, 
\begin{eqnarray}
P(\ugamma_{kl}|\pi_{\mu k l}) = \pi_{\mu k l}^{N(\gamma_{kl})} (1-\pi_{\mu k l})^{q_{\mu l}-N(\gamma_{kl})},  l=1,\dots,K,\label{priorGamma}
\end{eqnarray}
where, for parametric effects, $N(\gamma_{kl}) = \gamma_{kl}$ and $q_{\mu l}=1$, $l=1,\dots,K_1$,
while for nonparametric effects, $N(\gamma_{kl}) = \sum_{s=1}^{q_{\mu l}} \gamma_{kls}$
and $q_{\mu l}$ was defined in (\ref{star}), $l=K_1+1,\dots,K.$ Independent Beta priors are specified for 
$\pi_{\mu k l} \sim \text{Beta}(c_{\mu k l},d_{\mu k l})$, $k=1,\dots,p, l=1,\dots,K$.

Moving on to the priors of the parameter of the autoregressive coefficients, a normal prior is specified for the non-zero coefficients $\upsi_{\xi lm}^{\ast}, l,m=1,\dots,p,$ of the model in (\ref{phi2})
\begin{eqnarray}
\upsi_{\xi lm}^{\ast} \sim N(\uzero,c_{\psi lm}^2\uI). \nonumber
\end{eqnarray}
For the scale parameter $c_{\psi lm}^2$,  we consider inverse Gamma and half-normal priors, $c_{\psi lm}^2 \sim \text{IG}(a_{\psi lm},b_{\psi lm})$ and
$c_{\psi lm} \sim N(0,\phi^2_{\psi lm}) I[c_{\psi lm}>0]$. 
Further, the specification of the priors for the vectors of variable selection indicators  $\uxi_{lm}=(\xi_{lm0},\xi_{lm1},\dots,\xi_{lmB_1},\uxi_{lm,B_1+1}^{\top},\dots,\uxi_{lmB}^{\top})^{\top}, l,m=1,\dots,p,$ 
follows the same pattern as for $\ugamma_k$ indicators. That is, independent binomial priors are specified for each of the $1+B$ subvectors 
of $\uxi_{lm}$, 
\begin{eqnarray}
P(\uxi_{lmb}|\pi_{\phi lmb}) = \pi_{\phi lmb}^{N(\xi_{lmb})} (1-\pi_{\phi lmb})^{q_{\phi b}-N(\xi_{lmb})},  b=0,\dots,B,\nonumber
\end{eqnarray}
where $N(\xi_{lmb}) = \xi_{lmb}$ and $q_{\phi b}=1$ for $b=0,\dots,B_1,$ 
and $N(\xi_{lmb}) = \sum_{s=1}^{q_{\phi b}} \xi_{lmbs}$
and $q_{\phi b}$ was defined after (\ref{phi}) for $b=B_1+1,\dots,B.$ 
Independent Beta priors are specified for 
$\pi_{\phi lmb} \sim \text{Beta}(c_{\phi lmb},d_{\phi lmb}), l,m=1,\dots,p.$

Next, we describe the priors for the innovation variance parameters. First, for $\ualpha_{\delta_k}$ we specify independent normal priors 
\begin{eqnarray}
\ualpha_{\delta_k} | c_{\alpha k}, \udelta_k \sim N(\uzero,c_{\alpha k}^2 \uI), k=1,\dots,p.\nonumber
\end{eqnarray}
Further, we consider inverse Gamma and half-normal priors for $c_{\alpha k}$, 
$c_{\alpha k}^2 \sim \text{IG}(a_{\alpha k},b_{\alpha k})$ and
$c_{\alpha k} \sim N(0,\phi^2_{\alpha k}) I[c_{\alpha k} > 0], k=1,\dots,p.$
For the scale parameter $\sigma^2_{k}$ in (\ref{lastvar}), $k=1,\dots,p,$ we also consider inverse Gamma and half-normal priors, 
$\sigma^2_{k} \sim \text{IG}(a_{\sigma k},b_{\sigma k})$ and
$\sigma_{k} \sim N(0,\phi^2_{\sigma k}) I[\sigma_{k} > 0]$.
Continuing with the priors for the vectors of indicators 
$\udelta_k=(\delta_{k1},\dots,\delta_{kL_1},\udelta_{k,L_1+1}^{\top},\dots,\udelta_{kL}^{\top})^{\top}$, $k=1,\dots,p,$ we specify 
independent binomial priors for each of their $L$ subvectors, 
\begin{eqnarray}
P(\udelta_{kl}|\pi_{\sigma kl}) = \pi_{\sigma kl}^{N(\delta_{kl})} (1-\pi_{\sigma kl})^{q_{\sigma l}-N(\delta_{kl})},l=1,\dots,L,\nonumber
\end{eqnarray}
where $N(\delta_{kl})$ and $q_{\sigma l}$ are analogous to $N(\gamma_{kl})$ and $q_{\mu l}$ (and $N(\xi_{lmb})$ and $q_{\phi b}$).
We specify independent Beta priors for 
$\pi_{\sigma kl} \sim \text{Beta}(c_{\sigma kl},d_{\sigma kl}), k=1,\dots,p, l=1,\dots,L.$

Lastly, we describe the priors placed on the parameters of the models of the innovation correlation matrices. Starting with the `common correlations model' in (\ref{modcor}), let $\utZ = \uD(\uomega)^{-1} \uZ^{\ast}$. 
The prior for the non-zero part of $\ueta^{\ast}$ is the following g-prior
\begin{eqnarray}
\ueta_{\nu}^{\ast} | c_{\eta}, \sigma^2, \unu, \uomega, \uvarphi \sim 
N(\uzero,c_{\eta} \sigma^2 (\utZ_{\nu}^{\top} \utZ_{\nu} )^{-1}).\label{etanu}
\end{eqnarray}
Further, the prior for $c_{\eta}$ is specified as  $c_{\eta} \sim \text{IG}(a_{\eta},b_{\eta}).$
Furthermore, we specify a binomial prior for vector $\unu$ 
\begin{eqnarray}
P(\unu|\pi_{\nu}) = \pi_{\nu}^{N(\nu)} (1-\pi_{\nu})^{q_{R\mu}-N(\nu)},\nonumber
\end{eqnarray}
where $q_{R \mu}$ denotes the number of basis functions used in linearising $f_{\mu}(t)$ in (\ref{muct}), and $N(\nu) = \sum_{j=1}^{q_{R\mu}} \nu_j$.
A Beta prior is specified for $\pi_{\nu} \sim \text{Beta}(c_{\nu},d_{\nu})$.

Further, the prior for $\uomega_{\varphi}$ is specified as
\begin{eqnarray}
\uomega_{\varphi} | c_{\omega}, \uvarphi \sim N(\uzero,c_{\omega}^2 \uI).\nonumber
\end{eqnarray}
For $c_{\omega}$ we consider inverse Gamma and half-normal priors, 
$c_{\omega}^2 \sim \text{IG}(a_{\omega},b_{\omega})$ and $c_{\omega} \sim N(0,\phi^2_{c_{\omega}}) I[c_{\omega} > 0].$
Continuing with the prior for vector $\uvarphi$, this is taken to be 
\begin{eqnarray}
P(\uvarphi|\pi_{\varphi}) = \pi_{\varphi}^{N(\varphi)} (1-\pi_{\varphi})^{q_{R\sigma}-N(\varphi)},\nonumber
\end{eqnarray}
where $q_{R \sigma}$ denotes the number of basis functions used in linearising $f_{\sigma}(t)$ in (\ref{logSigma}) and $N(\varphi) = \sum_{j=1}^{q_{R\sigma}} \varphi_j$. 
We specify a Beta prior for
$\pi_{\varphi} \sim \text{Beta}(c_{\varphi},d_{\varphi})$.
Further, we specify inverse Gamma and half-normal priors for $\sigma^2$, 
$\sigma^2 \sim \text{IG}(a_{\sigma},b_{\sigma})$ and $\sigma \sim N(\sigma;0,\phi^2_{\sigma}) I[\sigma>0].$

Next we describe the priors for the cluster specific parameters $(\unu_h,\ueta_{h})$
and prior weights $w_h, h=1,\dots,H,$ of the grouped correlations model. 
Firstly, conditionally on $\unu_h$, the non-zero elements of $\ueta_{h}$
are specified to have the following prior
\begin{eqnarray}
\ueta_{\nu_h} | c_{\eta}, \sigma^2, \unu, \uomega, \uvarphi \sim 
N(\uzero,c_{\eta} \sigma^2 (\utZ_{\nu_h}^{\top} \utZ_{\nu_h} )^{-1}),&& \text{if the cluster is non-empty,}\label{m2etaA}\nonumber\\
\ueta_{\nu_h} | c_{\eta}, \sigma^2, \unu, \uomega, \uvarphi \sim 
N(\uzero,c_{\eta} \sigma^2 \uI),&& \text{if the cluster is empty}.\label{m2etaB}
\end{eqnarray}
Secondly, for vectors $\unu_h$ we specify binomial priors
\begin{eqnarray}
P(\unu_h|\pi_{\nu_h}) = \pi_{\nu_h}^{N(\nu_h)} (1-\pi_{\nu_h})^{q_{R\mu}-N(\nu_h)},\nonumber\nonumber
\end{eqnarray}
with the prior on $\pi_{\nu_h}$ being $\pi_{\nu_h} \sim \text{Beta}(c_{\nu},d_{\nu}), h=1,\dots,H.$

Lastly, the prior weights $w_h$ are constructed utilizing the so called stick-breaking process \citep{Fer73,sethuraman}.
Let $v_h, h=1,\dots,H-1,$ be independent draws from a $\text{Beta}(1,\alpha^{*})$ distribution. We have: 
$w_1 = v_1$, for $ 2 \leq l < H$, $w_l = v_l \prod_{h=1}^{l-1} (1-v_h)$, and 
$w_H = \prod_{h=1}^{H-1} (1-v_h)$. We take the concentration parameter $\alpha^{*}$ to be unknown
and we assign to it a gamma prior $\alpha^{*} \sim \text{Gamma}(a_{\alpha*},b_{\alpha*})$ with mean $a_{\alpha*}/b_{\alpha*}$.

In the numerical illustrations that we present in this article, we use the following priors. For $c_{\beta}$ we specify 
$\text{IG}(1/2, np/2),$ as a $p$-variate analogue to the prior of \citet{liang_mixtures_2008}. For $\pi_{\mu k l}, \pi_{\phi l m b}, \pi_{\sigma k l}, \pi_{\nu},$ and $\pi_{\varphi}$  we specify Beta$(1,1)$ priors. 
The priors on $c_{\alpha k}, k=1,\dots,p,$ and $c_{\omega}$ are specified to be IG$(1.1,1.1)$. Further, for $c_{\psi lm}, l,m=1,\dots,p$, $\sigma_k, k=1,\dots,p$, and $\sigma$ we define HN$(2)$ priors.
Furthermore, for $c_{\eta},$ we specify $\text{IG}(1/2, Md/2)$, again, as an analogue to the prior of \citet{liang_mixtures_2008}.  
Lastly, the prior on the DP concentration parameter is specified as $\alpha^{\ast} \sim \text{Gamma}(5,2)$.

\section{Posterior Sampling}\label{three}

Posterior sampling is carried out by expressing the likelihood function in different ways and using the expression that is more computationally convenient for each step of the MCMC algorithm. Further, where possible, we integrate out parameters to improve mixing, and where beneficial, we augment the model with additional parameters to make sampling easier. Here we only present the main concepts and a detailed description of all MCMC steps is available in the supplementary material. 

We first consider the likelihood that is obtained from the normality assumption, $\uY_i \sim N(\uX_{\gamma i}^{\ast} \ubeta^{\ast}_{\gamma},\uSigma_i), i=1,\dots,n$. 
Due to the factorization in (\ref{fd}), the quadratic form $Q = \sum_{i=1}^n (\uY_i - \uX_{\gamma i}^{\ast} \ubeta^{\ast}_{\gamma})^{\top} \uSigma_i^{-1} (\uY_i - \uX_{\gamma i}^{\ast} \ubeta^{\ast}_{\gamma})$ of the likelihood may be written as 
\begin{eqnarray}
Q  = \sum_{i=1}^n (\uL_i\ur_i)^{\top} \uD_{i}^{-1} (\uL_i \ur_i) 
= \sum_{i=1}^n \sum_{j=1}^{n_i} (\ur_{ij} - \hat{\ur}_{ij})^{\top} \uD_{ij}^{-1} (\ur_{ij} - \hat{\ur}_{ij}), \label{Q1}
\end{eqnarray}
where  $\ur_i = \uY_i - \uX_i^{\ast}  \ubeta^{\ast}$.

Further, recalling (\ref{pred3}) and the definition of prediction error, and the separation of variances and correlations in (\ref{sep}), $Q$ may be written as 
\begin{eqnarray}
Q = \sum_{i=1}^n \sum_{j=1}^{n_i} \uepsilon_{ij}^{\top} [\uS_{ij}^{1/2} \uR_{t_j} \uS_{ij}^{1/2}]^{-1} \uepsilon_{ij}
=\sum_{t \in T} \sum_{i \in O_t} \ucepsilon_{it}^{\top} \uR_{t}^{-1} \ucepsilon_{it},\label{Qcom}
\end{eqnarray} 
where $\ucepsilon_{ij} =  \uS_{ij}^{-1/2}\uepsilon_{ij}$. In the first equality, matrix $\uR$ is subscripted by $t_j$ because here we model its elements in terms of covariate `time' $t$ only. In the second equality, $T$ denotes the set of unique observational time points and $O_t$ the set of sampling units observed at time point $t$. We denote the cardinality of $O_t$ by $n_t$.

In addition, recalling (\ref{pred2}) and its centred versions in (\ref{pred3}) and (\ref{pred4}), we can write (\ref{Q1}) as
\begin{eqnarray} 
Q = \sum_{i=1}^n \sum_{j=1}^{n_i} (\ur_{ij} - \uV_{ij \xi \beta} \upsi_{\xi}^{\ast})^{\top} \uD_{ij}^{-1} (\ur_{ij} - \uV_{ij \xi \beta} \upsi_{\xi}^{\ast}),\label{Q4}
\end{eqnarray}
where $\uV_{ij \xi \beta} = 
\sum_{k=1}^{j-1} [\uI_p \otimes \ur_{ik}^{\top} \otimes (\uz^{\ast}_{ijk})^{\top} ]_{\xi}$.
We take $\uV_{ij \xi \beta}$ to be a matrix of zeros when $j=1$. 

To improve mixing of the MCMC algorithm, we can integrate out vector $\ubeta^{\ast}_{\gamma}$ from the likelihood of $\uY$
\begin{eqnarray}\label{marginalY}
f(\uY) = (2\pi)^{-\frac{Np}{2}} 
\big(\prod_{i=1}^n|\uSigma_i|^{-\frac{1}{2}}\big) 
(c_{\beta}+1)^{-\frac{N(\gamma)+p}{2}} \exp(-S/2),
\end{eqnarray}
where 
\begin{equation}
S = \utY^{\top} \big\{ I - \frac{c_{\beta}}{1+c_{\beta}} \utX_{\gamma} (\utX_{\gamma}^{\top}\utX_{\gamma})^{-1} \utX_{\gamma}^{\top} \big\} \utY,\nonumber
\end{equation}
$\utY =  \uSigma^{-\frac{1}{2}} \uY$, 
$N=\sum_{i=1}^n n_i$, 
$\utX_{\gamma} = \uSigma^{-\frac{1}{2}} \uX^{\ast}_{\gamma}$ and 
$N(\gamma)+p$ is the total number of columns in $\uX_{\gamma}^{\ast}$. 

To write function $S$ in a computationally convenient way, first note that $\utY^{\top} \utY =  \sum_{i=1}^n \uY_{i}^{\top} \uSigma_{i}^{-1} \uY_{i}$, 
which, due to (\ref{fd}), may be written as $\sum_{i=1}^n (\uL_i\uY_{i})^{\top} \uD_{i}^{-1} \uL_i \uY_{i}$. 
Further, due to the separation of variances and correlations in (\ref{sep}), we may further express the latter as
$\sum_{i=1}^n (\uS_{i}^{-1/2} \uL_i\uY_{i})^{\top} \uR_{i}^{-1} \uS_{i}^{-1/2} \uL_i \uY_{i}$ = 
$\sum_{i=1}^n \ucY_{i}^{\top} \uR_{i}^{-1} \ucY_{i}$, where $\ucY_{i} = \uS_{i}^{-1/2} \uL_i\uY_{i}$.
Furthermore, we write the last expression as 
$\sum_{i=1}^n \sum_{j=1}^{n_i} \ucy_{ij}^{\top} \uR_{t_{j}}^{-1} \ucy_{ij}$.
Likewise, we may write that
$\utX_{\gamma}^{\top}\utX_{\gamma} = 
\sum_{i=1}^n (\uL_i \uX_{\gamma i}^{\ast})^{\top} \uD_{i}^{-1} \uL_i \uX_{\gamma i}^{\ast} = 
\sum_{i=1}^n (\uS_{i}^{-1/2} \uL_i \uX_{\gamma i}^{\ast})^{\top} \uR_{i}^{-1} \uS_{i}^{-1/2} \uL_i \uX_{\gamma i}^{\ast}=$
$\sum_{i=1}^n \sum_{j=1}^{n_i} (\ucX_{\gamma i j}^{\ast})^{\top} \uR_{t_{j}}^{-1} \ucX_{\gamma i j}^{\ast}$.
In addition, 
$\utX_{\gamma}^{\top}\utY = \sum_{i=1}^n (\uL_i \uX_{\gamma i}^{\ast})^{\top} \uD_{i}^{-1} \uL_i \uY_{i} = 
\sum_{i=1}^n \sum_{j=1}^{n_i} (\ucX_{\gamma i j}^{\ast})^{\top} \uR_{t_{j}}^{-1} \ucy_{ij}$. 
It follows that
\begin{eqnarray}
&&S = \sum_{i=1}^n \sum_{j=1}^{n_i} \ucy_{ij}^{\top} \uR_{t_{j}}^{-1} \ucy_{ij} - \nonumber \\
&&c_{\beta} (1+c_{\beta})^{-1} 
[\sum_{i=1}^n \sum_{j=1}^{n_i} (\ucy_{ij})^{\top} \uR_{t_{j}}^{-1} \ucX_{\gamma i j}^{\ast}] 
[\sum_{i=1}^n \sum_{j=1}^{n_i} (\ucX_{\gamma i j}^{\ast})^{\top} \uR_{t_{j}}^{-1} \ucX_{\gamma i j}^{\ast}]^{-1} 
[\sum_{i=1}^n \sum_{j=1}^{n_i} (\ucX_{\gamma i j}^{\ast})^{\top} \uR_{t_{j}}^{-1} \ucy_{ij}].\nonumber
\end{eqnarray}
 
The most computationally expensive step of the MCMC algorithm is the one that samples from the posterior of the parameters of the correlation matrices. This is because acceptance probabilities of the Metropolis-Hastings step involve the ratio of the normalizing constants of the density in 
(\ref{priorR}) (or the ones in (\ref{priorR3}) and (\ref{priorR4})). Calculating this ratio is very computationally demanding, but it 
can be avoided by introducing the `shadow prior' \citep{Liechty}. 
The basic idea is to introduce latent variables $\theta_{tkl}$ between the correlations $r_{tkl}$ and the means $\mu_{ct}$. This modifies prior (\ref{priorR}) as follows
\begin{equation}\label{priorR2}
f(\uR_{t}|\utheta_{t},\tau^2) = \pi(\utheta_{t},\tau^2) 
\prod_{k<l} \exp[-(g(r_{tkl})-\theta_{tkl})^2/2\tau^2] J[g(r_{tkl}) \rightarrow r_{tkl}] I[\uR_{t} \in \mathcal{C}],
\end{equation}
where 
\begin{eqnarray}\nonumber
\pi^{-1}(\utheta_{t},\tau^2) = \int_{\uR_{t} \in \mathcal{C}} \prod_{k<l} 
\exp[-(g(r_{tkl})-\theta_{tkl})^2/2\tau^2] J[g(r_{tkl}) \rightarrow r_{tkl}] dr_{tkl},
\end{eqnarray}
and $\tau$ is a fixed, small constant. Further, variables $\theta_{tkl}$ are independently distributed with time specific means and variances,
$\theta_{tkl} \sim N(\mu_{ct},\sigma^2_{ct})$. Hence, the semiparametric model (\ref{modcor}) now applies on the unconstrained $\utheta$,
\begin{eqnarray}\label{modtheta}
\utheta = \uZ^{\ast}_{\nu} \ueta_{\nu}^{\ast} + \uepsilon, 
\uepsilon \sim N(\uzero, \sigma^2 \uD^2(\uomega_{\varphi})).
\end{eqnarray}
Sampling from the posterior of $\utheta_t=\{\theta_{tkl}\}$ still involves the ratio of the normalising constants of the density in (\ref{priorR2}), but that, as was argued by \citet{Liechty}, for small $\tau$, can reasonably be approximated by one. 

Considering now the `grouped correlations' model, with the introduction of the shadow prior, model (\ref{priorR3}) becomes the same as the one in (\ref{priorR2}). The difference is that here variables $\theta_{tkl}$ are independent with conditional distribution
$\theta_{tkl}|\lambda_{kl}=h \sim N(\mu_{cth},\sigma^2_{ct})$. Further, letting $\utheta_h$ be analogous to $\ur_h$, defined above (\ref{linearCluster}),
it is easy to see that semiparametric model (\ref{linearCluster}) now applies on the unconstrained $\utheta_h$
\begin{eqnarray}\label{modtheta2}
\utheta_h = \uZ^{\ast}_{\nu_h} \ueta_{\nu_h}^{*} + \uepsilon_h, h=1,\dots,H.
\end{eqnarray}
Additionally, letting $\utheta=(\utheta_1,\dots,\utheta_H)^{\top}$, model (\ref{modtheta2}) may be written in vectorized form in precisely the same way as (\ref{modtheta}).

To improve mixing of the MCMC algorithm over the parameters of the models of the correlation matrices, we can integrate out vector $\ueta$ from the likelihood of $\utheta$. The marginal of $\utheta$, computed from (\ref{modtheta}) and (\ref{etanu}), is
\begin{eqnarray}\label{marginalTh}
f(\utheta|\uomega,c_{\eta},\unu,\uvarphi,\sigma^2) \propto 
|\sigma^2 D^2(\uomega_{\varphi})|^{-\frac{1}{2}} (c_{\eta}+1)^{-\frac{N(\nu)+1}{2}} \exp\{-S^{\ast}/2\sigma^2\},
\end{eqnarray}
where $S^{\ast} = S^{\ast}(\utheta,\uomega,c_{\eta},\unu,\uvarphi) = \utTheta^{\top} \utTheta - \frac{c_{\eta}}{1+c_{\eta}}\utTheta^{\top}\utZ_{\nu}
(\utZ_{\nu}^{\top}\utZ_{\nu})^{-1}\utZ_{\nu}^{\top}\utTheta$,
and $\utTheta =  \uD^{-1}(\uomega_{\varphi}) \utheta$.

Lastly, under the grouped correlations model, the marginal likelihood of $\utheta$ is given by
\begin{eqnarray}\label{marginal2}
f(\utheta|\uomega,c_{\eta},\unu,\uvarphi,\sigma^2) \propto 
|\sigma^2 D^2(\uomega_{\varphi})|^{-\frac{1}{2}} (c_{\eta}+1)^{-\frac{N(\nu)+H}{2}} \exp\{-S^{\ast}/2\sigma^2\},\nonumber
\end{eqnarray}
where $N(\nu) = \sum_{h} N(\nu_h)$, and 
$S^{\ast} = S^{\ast}(\utheta,\uomega,\uc_{\eta},\unu,\uvarphi) = \sum_{h} S^{\ast}_h = \sum_{h} \utTheta^{\top}_h \left\{ \uI - 
\frac{c_{\eta}}{1+c_{\eta}} \utZ_{\nu_h}
(\utZ_{\nu_h}^{\top}\utZ_{\nu_h})^{-1}\utZ_{\nu_h}^{\top} \right\} \utTheta_h$.

\section{Simulation Studies}\label{sim}

We present results from two simulation studies. The first one examines how posteriors, based on different priors, concentrate around the true covariance and 
correlation matrices, while the second one investigates the gains that one may have, in terms of reduced posterior mean squared error (MSE), when fitting multivariate longitudinal models instead of univariate ones. Additionally, in the second study we examine the effects that missing data have on the posterior MSE and whether adding more than one response to the model can counteract these effects. 

\subsection{Effects of priors on the posteriors of matrices}    

We consider a trivariate system of responses, observed over $M=6$ equally spaced time 
points, $t=0,0.2,\dots,1$. 
The single covariate here is taken to be time, $t$. For the sake of simplicity we take the means of the three responses to be 
constant (zero), but we consider complex function for all other model parameters. For the autoregressive coefficients
we consider 
\begin{eqnarray}\label{simphi}
&&\phi_{ijk11} = 0.4 + 0.2 (t_{ij} - t_{ik}) -0.4(t_{ij} - t_{ik})^2-0.2(t_{ij} - t_{ik})^3, \nonumber\\
&&\phi_{ijk12} = -0.2 + 0.2 (t_{ij} - t_{ik}), \nonumber\\
&&\phi_{ijk13} = -N(t_{ij} - t_{ik},\mu=0.2,\sigma^2=0.025)/7, \nonumber\\
&&\phi_{ijk21} = -0.2 I[t_{ij} - t_{ik} < 0.21], \nonumber\\
&&\phi_{ijk22} = N(t_{ij} - t_{ik},\mu=0.2,\sigma^2=0.025)/7, \nonumber\\
&&\phi_{ijk23} = 0.1 + 0.1 I[t_{ij} - t_{ik} < 0.61], \nonumber\\
&&\phi_{ijk31} = 0.15 \sin(2 \pi (t_{ij} - t_{ik})), \nonumber\\
&&\phi_{ijk32} = (N(t_{ij} - t_{ik},\mu=0.2,\sigma^2=0.025) - N(t_{ij} - t_{ik},\mu=0.7,\sigma^2=0.1))/10, \nonumber\\
&&\phi_{ijk33} = (N(t_{ij} - t_{ik},\mu=0.2,\sigma^2=0.003)+N(t_{ij} - t_{ik},\mu=0.6,\sigma^2=0.05))/10.
\end{eqnarray}
Further, for the diagonal innovation variance matrices we consider 
\begin{eqnarray}\label{simsig}
&&\sigma^2_{ij1} = (N(t_{ij},\mu=0,\sigma^2=0.04)+N(t_{ij},\mu=0.6,\sigma^2=0.1))/2, \nonumber\\
&&\sigma^2_{ij2} = 0.8 + 0.5 \sin(2 \pi t_{ij}), \nonumber\\
&&\sigma^2_{ij3} = 0.6 - 0.5 t_{ij}.
\end{eqnarray}
Lastly, the innovation correlation matrices are taken to be 
\begin{eqnarray}\label{simR}
\uR_t =
\begin{bmatrix}
1 & \mu_{ct} & \mu_{ct}\\
\mu_{ct} & 1 & 0\\
\mu_{ct} & 0 & 1
\end{bmatrix},
\end{eqnarray}
where $\mu_{ct}$ are set equal to $0.50, -0.50, 0.0, 0.50, 0.65, 0.55$ for the $M=6$ time points, respectively. 
The innovation correlation matrices $\uR_t$ consist of two clusters of correlations, one 
including elements $(2,1)$ and $(3,1)$ and the other including element $(3,2)$. 
The cluster that includes two elements has parameter $\mu_{ct}$ that changes rapidly during the first $4$ time points and 
slowly during the last $2$, while the other cluster has constant zero parameter.
These matrices can also be thought of as consisting of two clusters of variables,
one includes variables $2$ and $3$ and the other includes variable $1$. Generally, with $2$ clusters of variables, at most $3$ 
parameters are needed: $2$ to describe the correlations within the clusters and $1$ to describe the correlations between the
clusters. However, because here there is a singleton cluster, we need $2$ parameters: the parameter
that describes the correlation between variables $2$ and $3$ (set to $0$), and the parameter that
describes the correlation between the $2$ clusters (set to $\mu_{ct}$). 

\begin{figure}
	\begin{center}
		\begin{tabular}{ccc}
			\includegraphics[width=0.25\textwidth]{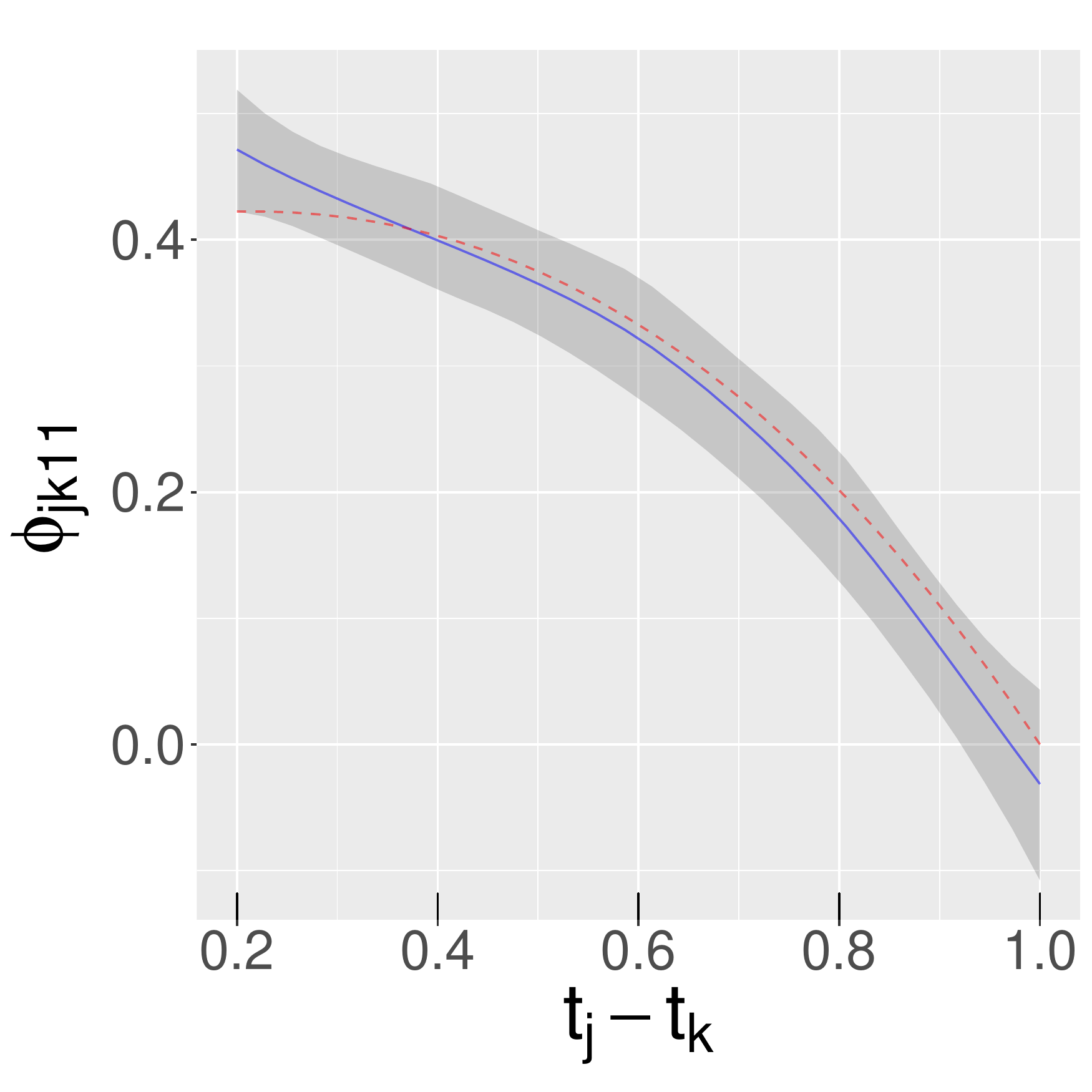} &  
			\includegraphics[width=0.25\textwidth]{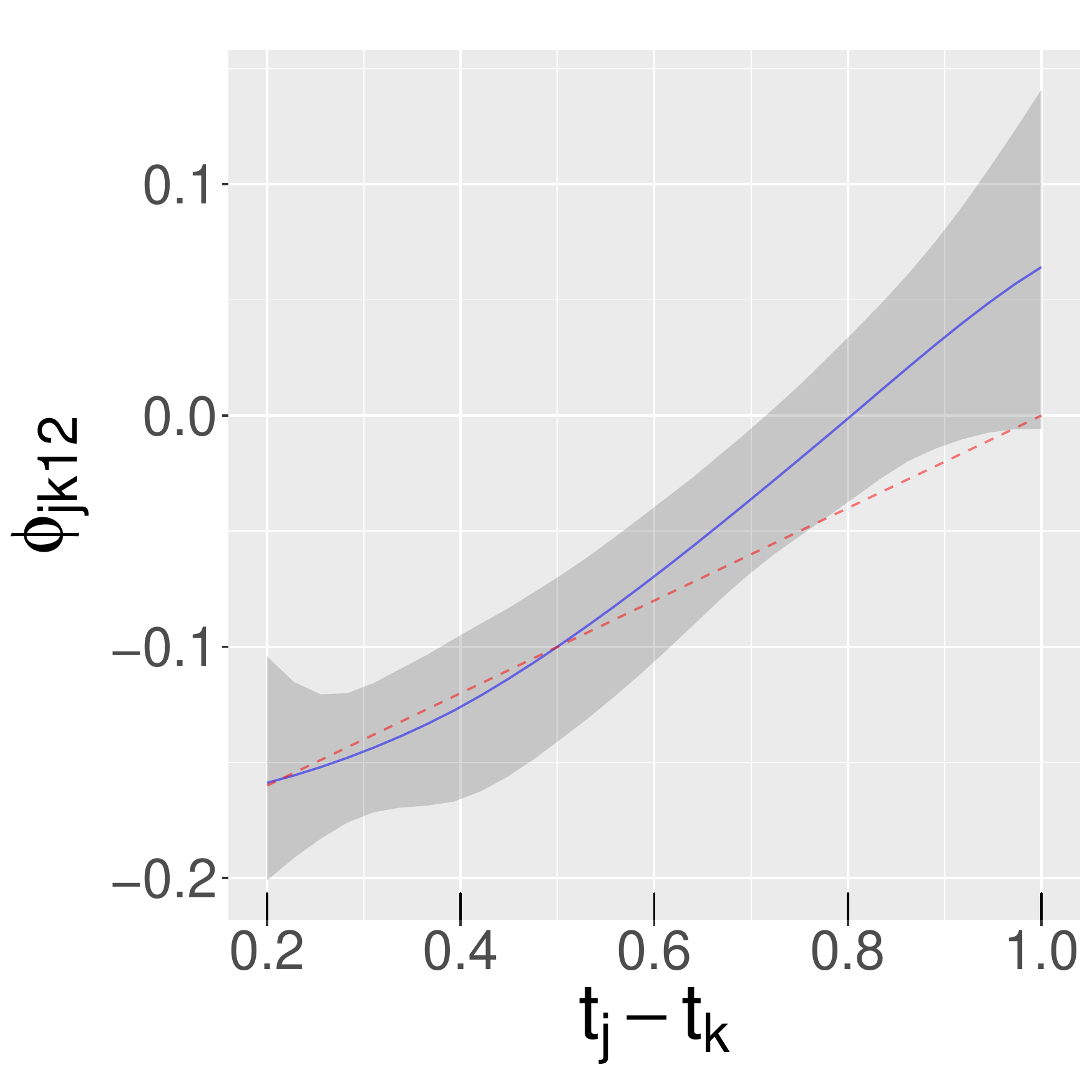} &  
			\includegraphics[width=0.25\textwidth]{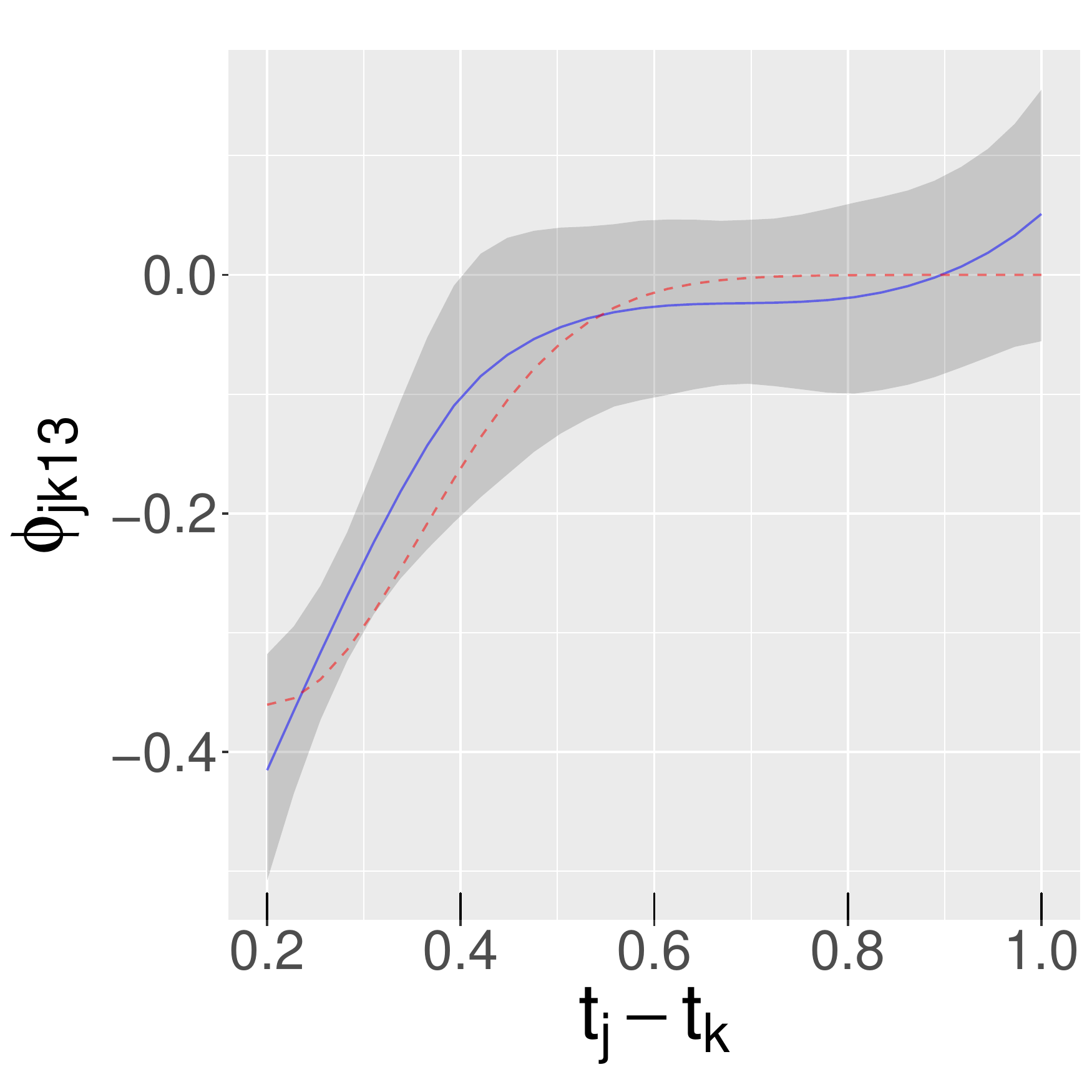} \\  
			\includegraphics[width=0.25\textwidth]{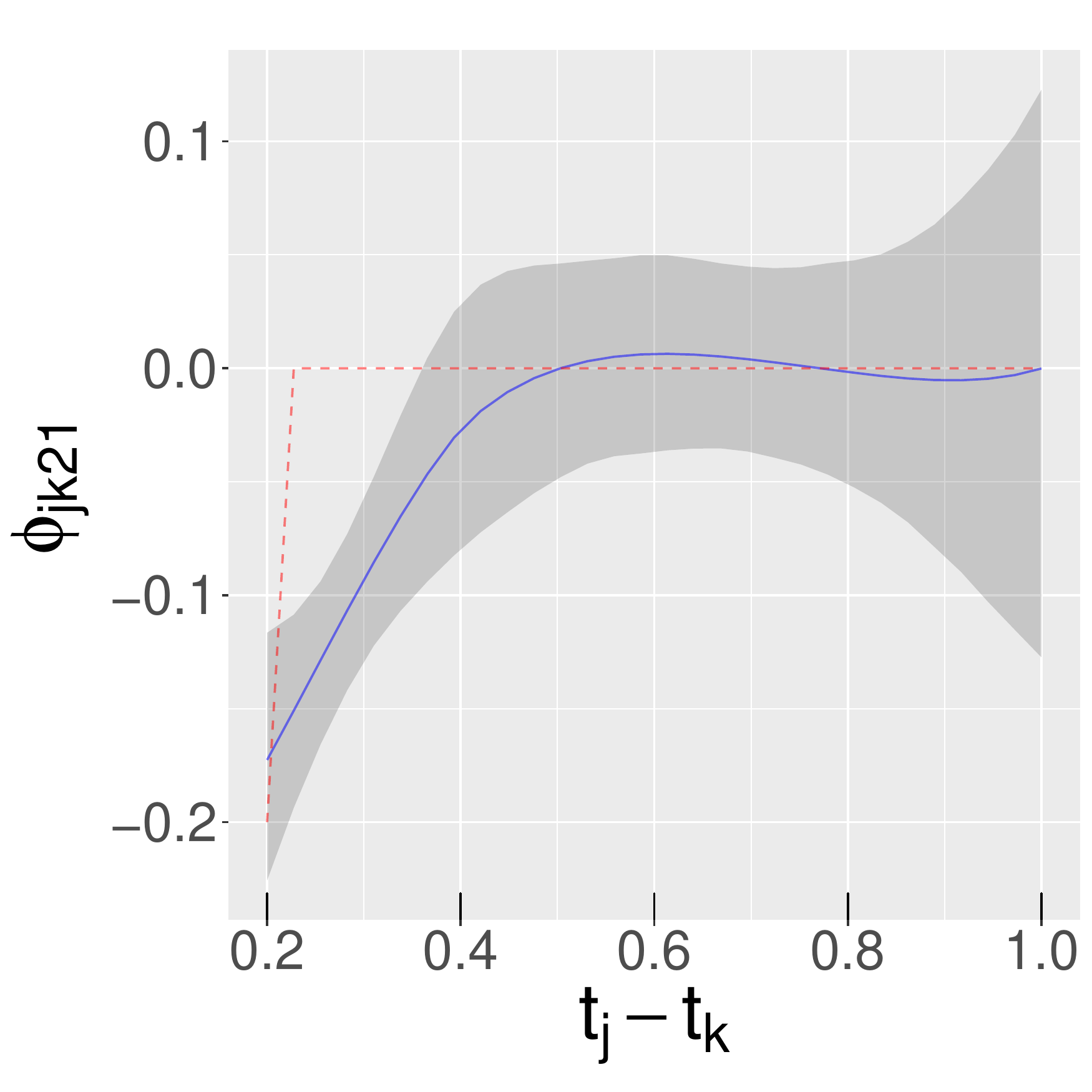} &  
			\includegraphics[width=0.25\textwidth]{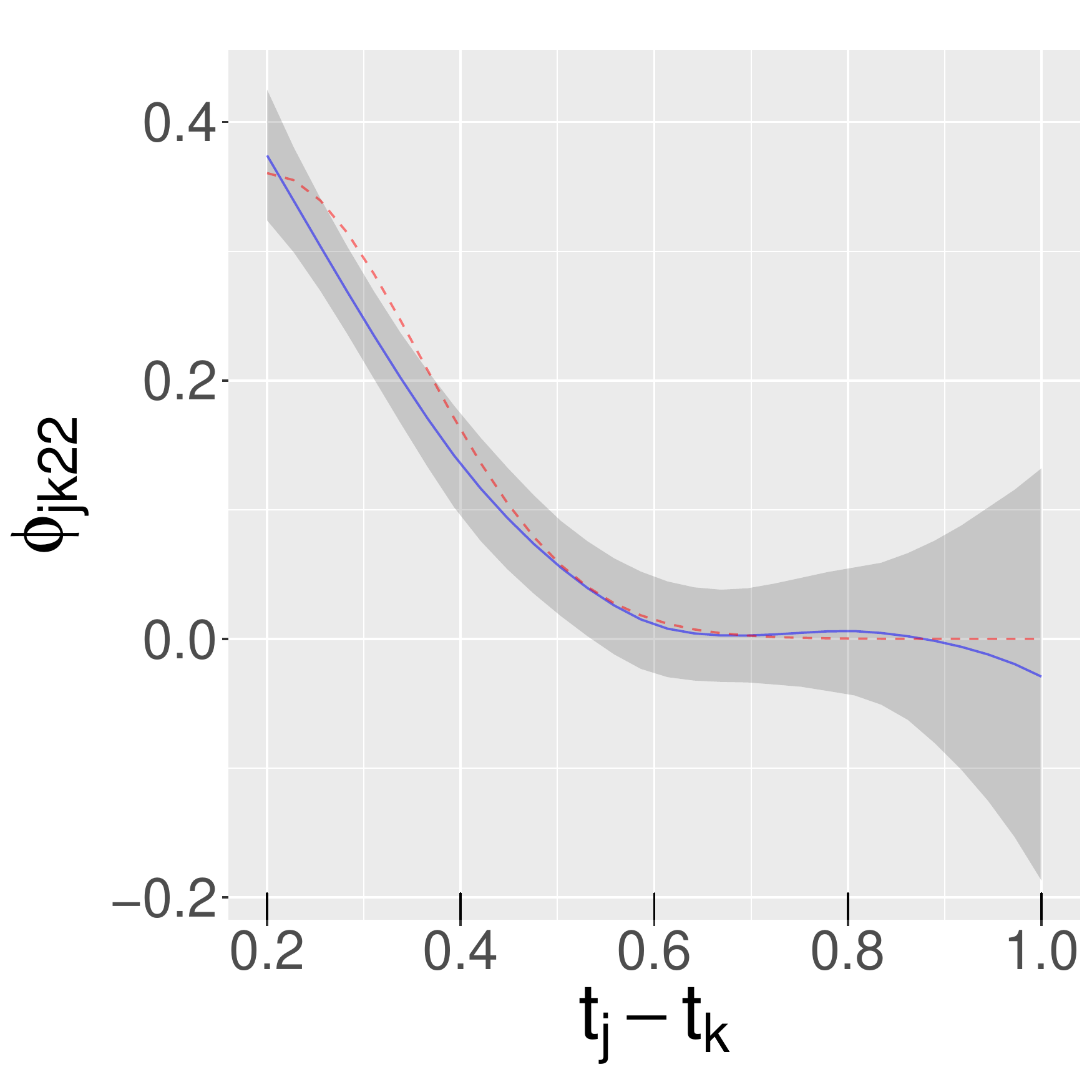} &  
			\includegraphics[width=0.25\textwidth]{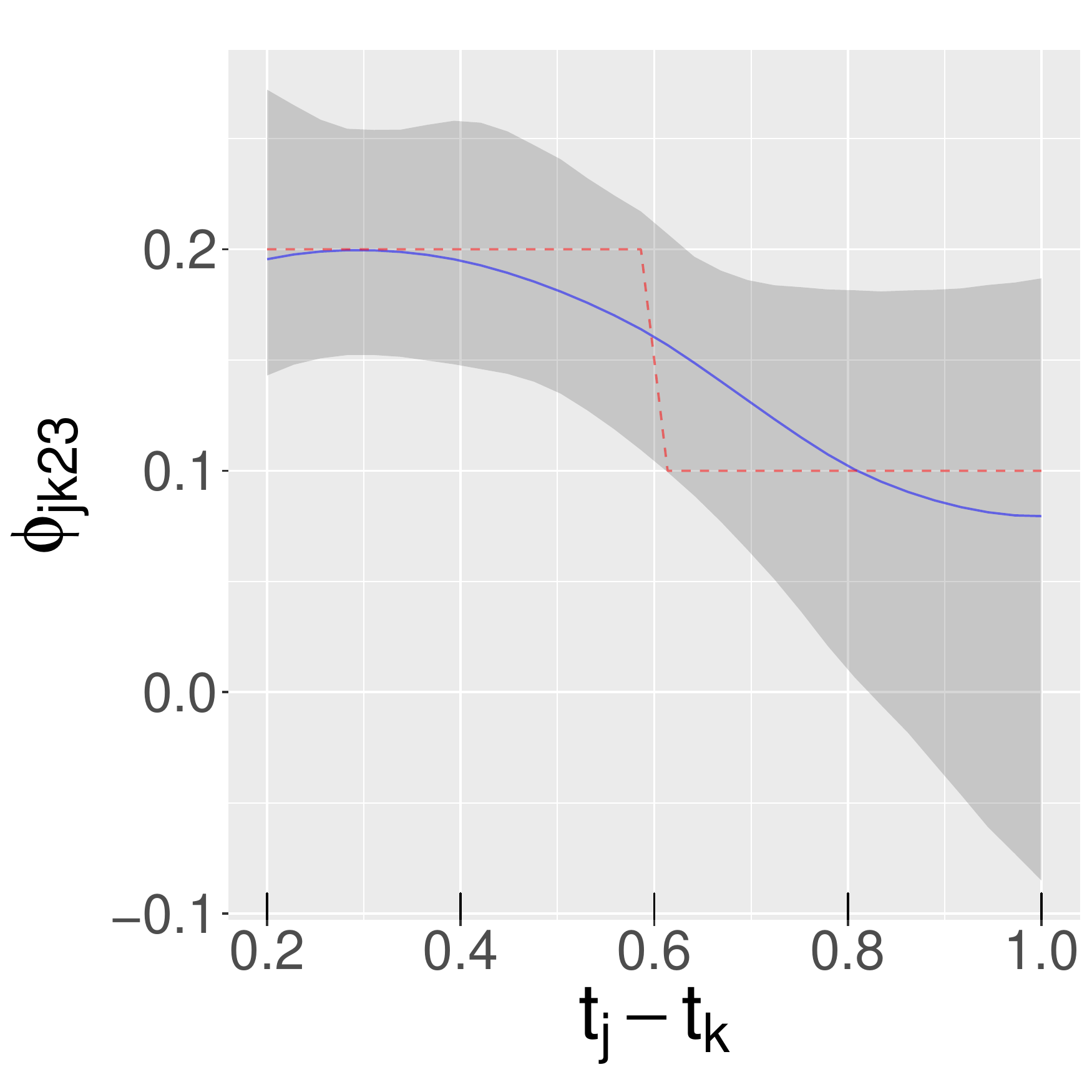} \\  
			\includegraphics[width=0.25\textwidth]{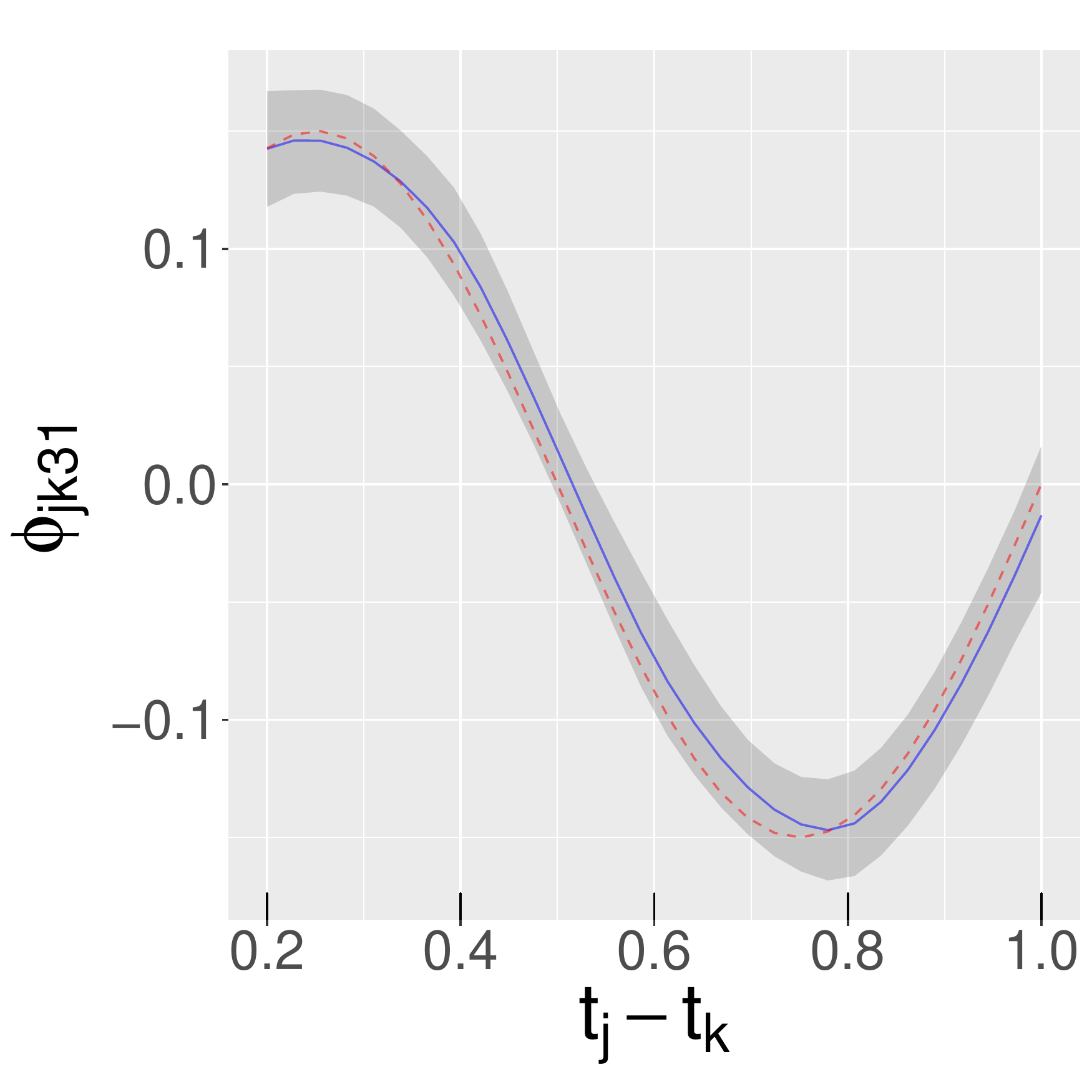} &  
			\includegraphics[width=0.25\textwidth]{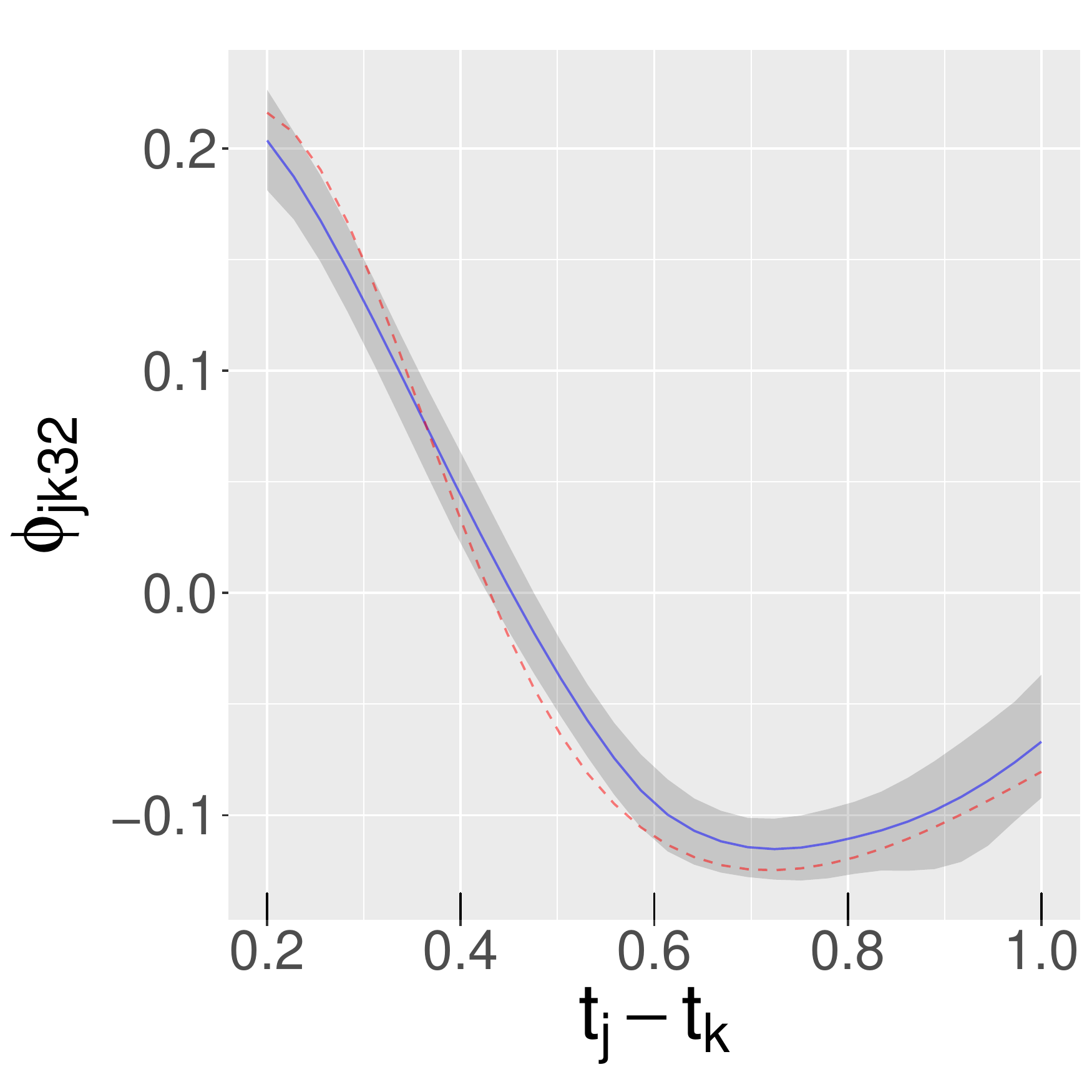} &  
			\includegraphics[width=0.25\textwidth]{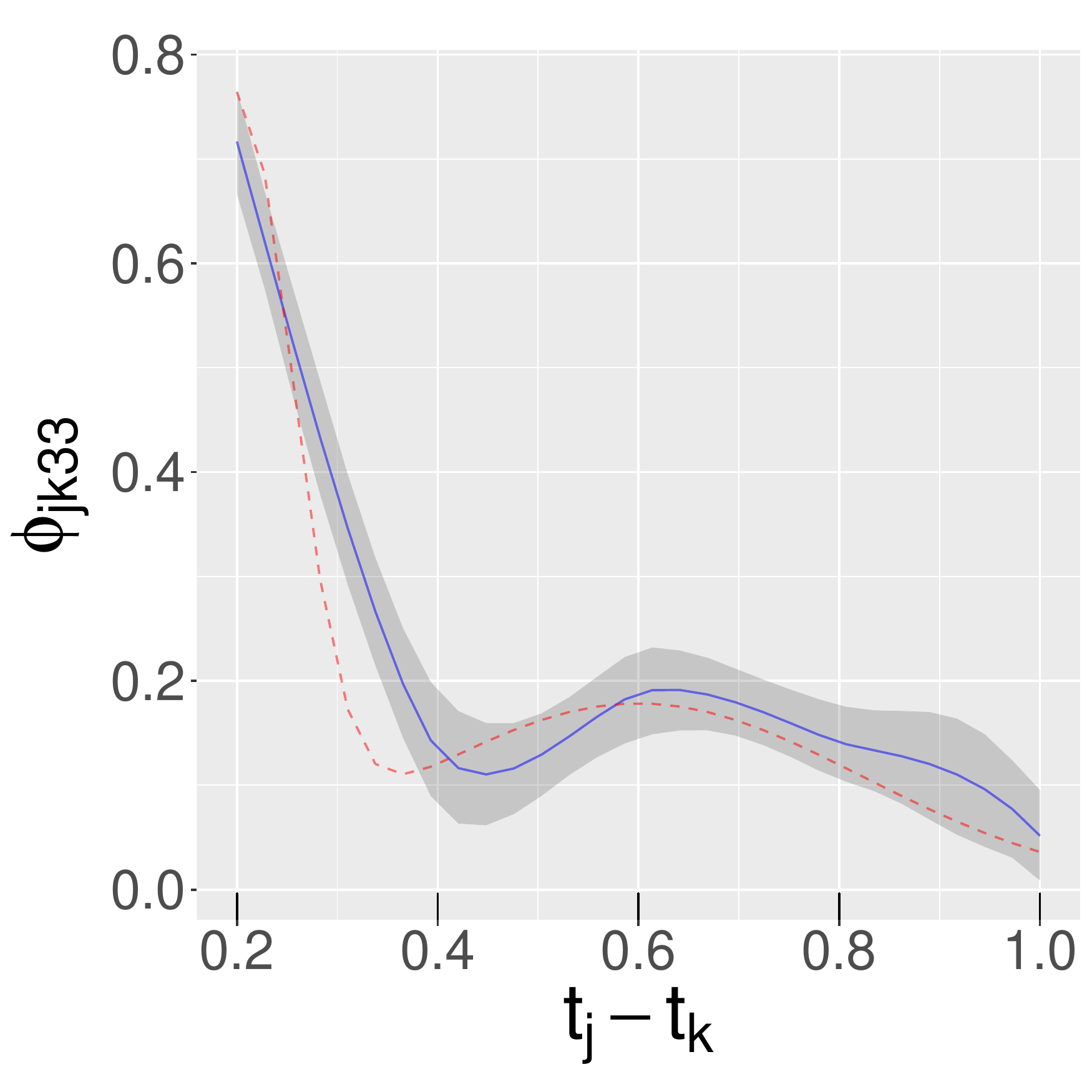} \\  
		\end{tabular}
	\end{center}
	\caption{First simulation study results: autoregressive coefficient regression models. The dashed (red) curves denote the true functions and the solid (blue) ones the posterior means that are displayed along with $90\%$ credible intervals. Plots are based on a single simulated dataset of size $n=100$.}\label{Fauto}
\end{figure}

\begin{figure}
	\begin{center}
		\begin{tabular}{ccc}
			\includegraphics[width=0.25\textwidth]{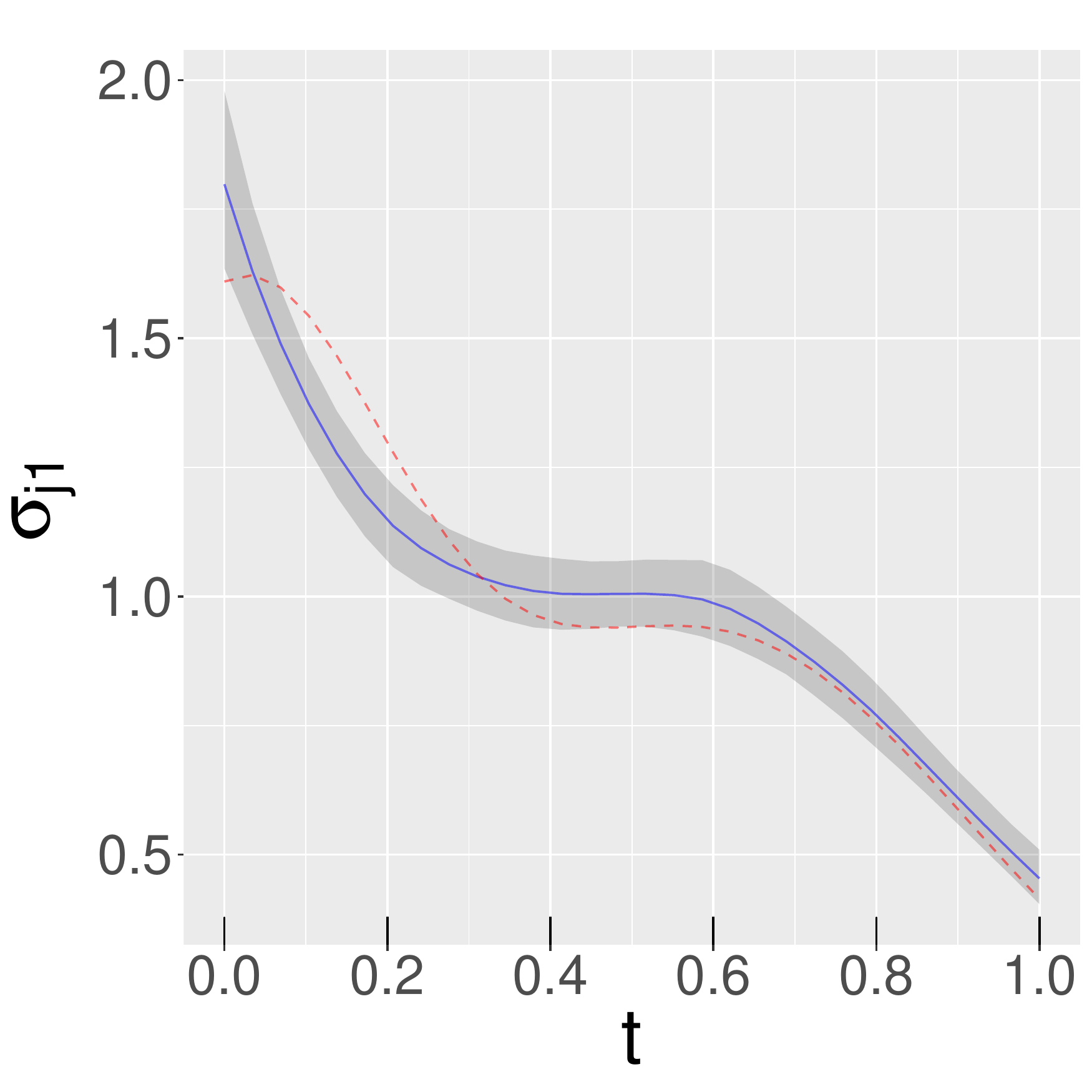} &  
			\includegraphics[width=0.25\textwidth]{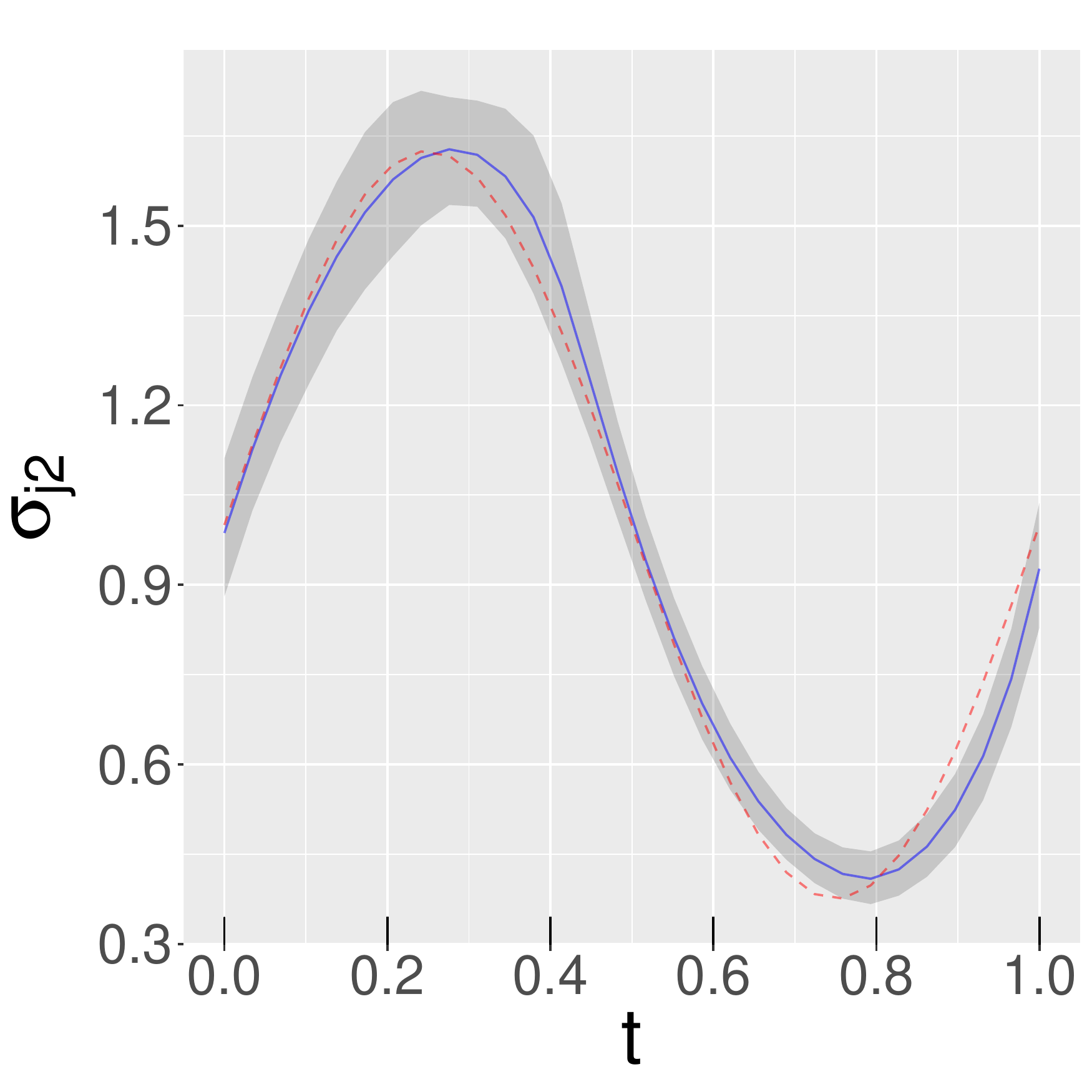} &
			\includegraphics[width=0.25\textwidth]{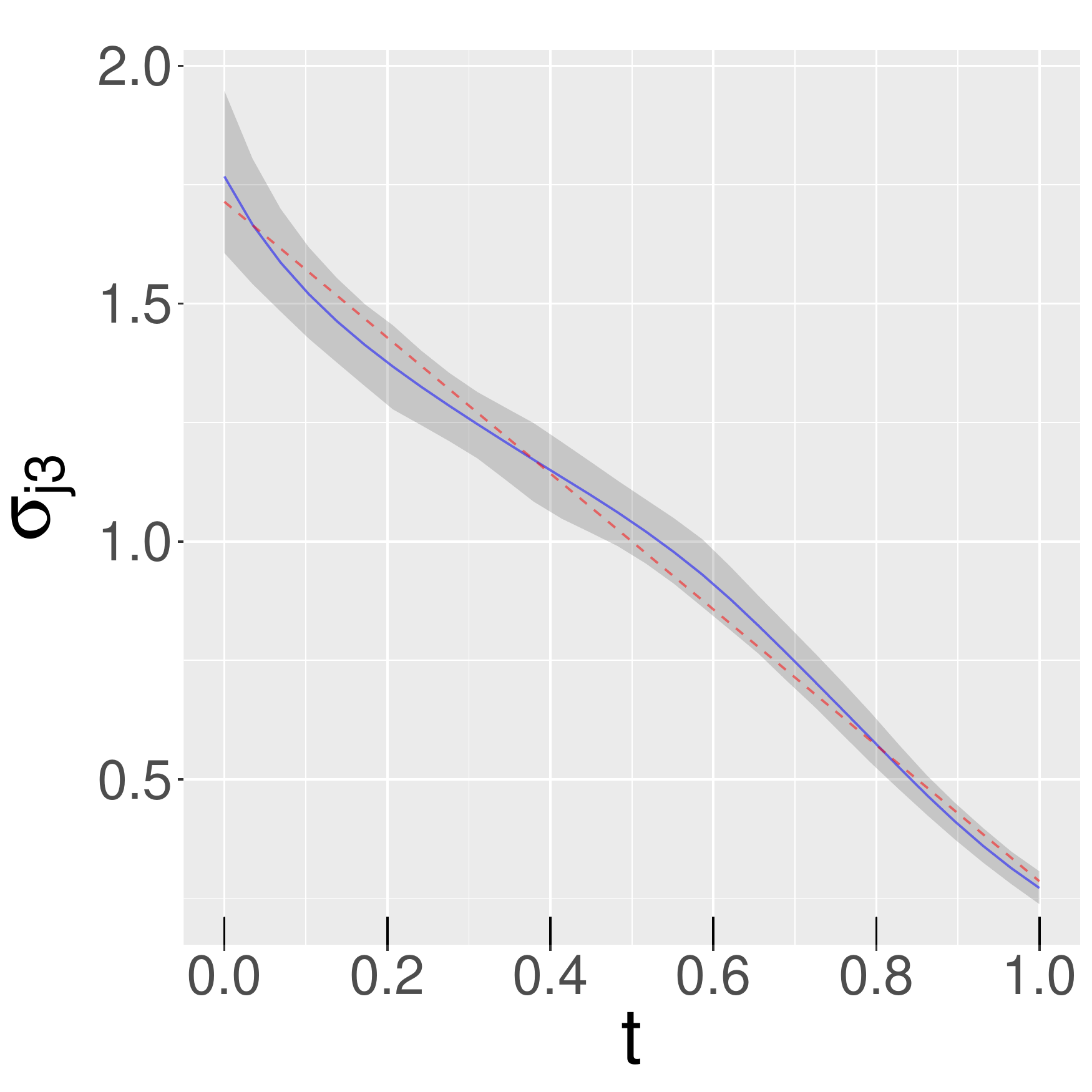}\\
		\end{tabular}
	\end{center}
	\caption{First simulation study results: innovation standard deviation regression models. The dashed (red) curves denote the true functions and the solid (blue) ones the posterior means that are displayed along with $90\%$ credible intervals. Plots are based on a single simulated dataset of size $n=100$.}\label{Fsd}
\end{figure}

\begin{figure}
	\begin{center}
		\begin{tabular}{c}
			\includegraphics[width=0.25\textwidth]{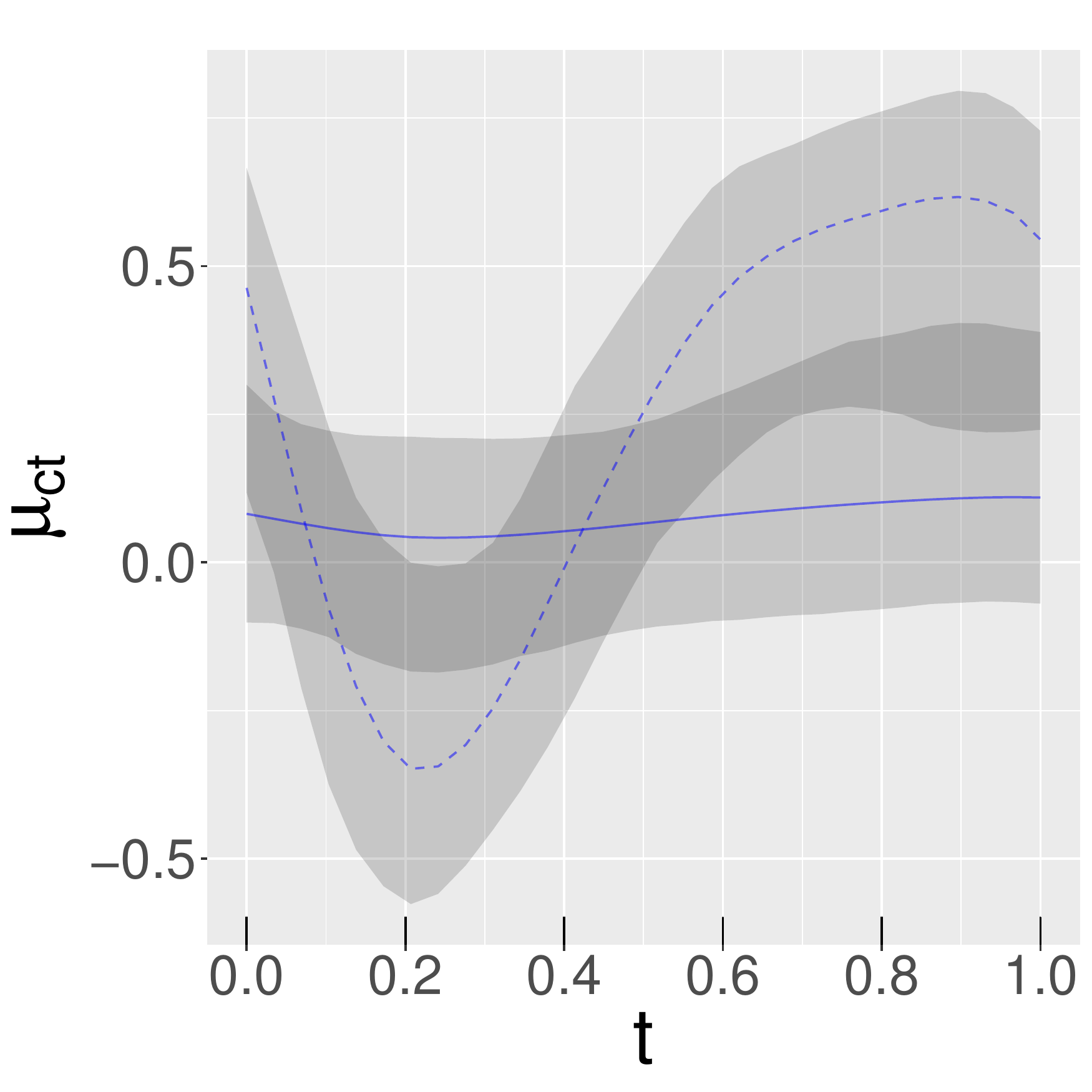}\\
		\end{tabular}
	\end{center}
	\caption{First simulation study results: centre parameter of the correlation regression model. The two curves denote the posterior means of the two clusters and they are displayed along with $90\%$ credible intervals. The plot is based on a single simulated dataset of size $n=100$.}\label{Fcor}
\end{figure}

Figures \ref{Fauto}, \ref{Fsd} and \ref{Fcor} show the true and fitted curves for the  parameters described above, 
displayed along with $90\%$ credible intervals. We note that fitted curves and credible intervals were obtained based on a simulated dataset of size
$n=100$.

Given the specifications of the autoregressive coefficients in (\ref{simphi}), we construct the matrices $\uPhi_{ijk}$ and from them, the lower triangular matrix $\uL_i$
in (\ref{LD}). Further, from the specifications in (\ref{simsig}) and (\ref{simR}), we construct the innovation covariance matrices in (\ref{sep})
and from them, the block diagonal matrix $\uD_i$ in (\ref{LD}). The $18 \times 18$  covariance matrix is obtained as $\uSigma_i = \uL_i^{-1} \uD_i (\uL_i^{\top})^{-1}$. We note that subscript $i$ is redundant here as the covariance matrix is common to all subjects and hence it is omitted in the sequel.  

Datasets are generated from a multivariate Gaussian with mean zero and covariance $\uSigma$. We consider $4$ sample sizes, $n=10,20,50,100$. To obtain results that are representative and independent of the generated dataset, for each sample size $n$ we generate $30$ replicate datasets and present results that are averaged over these.  

To each simulated dataset we fit $3$ models with common specifications for the mean, the autoregressive coefficients, the innovation variances, and the dispersion parameter of the correlations. The models only differ in their specification of the centre parameter of the correlations. Specifically, the means of the responses are modelled using only an intercept term, $\mu_{ijk} = \beta_{k0}$, the autoregressive coefficients are modelled using a smooth function of lag with $6$ basis functions, $\phi_{jklm} = \psi_{lm0} + f_{\phi,l,m}(t_j-t_k)$, where $f_{\phi,l,m}(t_j-t_k)=\sum_{s=1}^6 \psi_{lms} \kappa_{\phi s}(t_j-t_k)$, and the basis functions $\kappa(.)$ where defined above (\ref{mean3}).  
Further, the innovation variances are modelled using a smooth function of time, 
$\log \sigma^2_{jk} = \alpha_{k0} + f_{\sigma,k}(t_j)$, where $f_{\sigma,k}(t_j)=\sum_{s=1}^6 \alpha_{ks} \kappa_{\sigma s}(t_j)$. Lastly, the dispersion parameter of the correlations is modelled using only an intercept term, $\log \sigma^2_{ct} = \omega_{0}$. Concerning parameter $\mu_{ct}$, the first model $(M_1)$ specifies $\mu_{ct}$ to be constant $\mu_{ct} = \eta_0$, the second one $(M_2)$ specifies $\mu_{ct}$ to be a smooth function of time with $6$ basis functions, $\mu_{ct} = \eta_0 + \sum_{s=1}^6 \eta_{s} \kappa_{\mu c s}(t_j)$, while the third one $(M_3)$ specifies a grouped variables model, where the number of groups is $G=2$, and each group's mean is modelled as a smooth function of time using $6$ basis functions. We note that M$_3$ is the correctly specified model while the other $2$ are misspecified.        

For all models we run the MCMC algorithm for $3 \times 10^4$ iterations, discarding the first $10^4$ as burn in, and of the remaining $2 \times 10^4$ keeping one in two. This results in $10^4$ samples of innovation correlation matrices over the $6$ time points, $\uR_t^{(s)}, t=0,0.2,\dots,1,$ and covariance matrices $\uSigma^{(s)}$, $s=1,2,\dots,10^4$. 
To evaluate the effects of priors,
we compute the average distance between  $\uR_t^{(s)}, s=1,2,\dots,10^4,$ and $\uR_t$ using the loss function 
$L(\uR_t^{(s)},\uR_t) = \text{tr}(\uR_t^{(s)} \uR_t^{-1}-\uI)^2$, $t=0,0.2,\dots,1$, where the sampled matrices are based 
priors $M_1$, $M_2$, and $M_3$. We denote the average loss associated with model $M_d, d=1,2,3,$ by 
$D_{R_t}(M_d) = 10^{-4} \sum_{s} L(\uR_t^{(s)},\uR_t)$.
As there are $M=6$ correlation matrices $\uR_t$, results are presented in terms of the average risk over these matrices
$D_R(M_d) = M^{-1}\sum_{t} D_{R_t}(M_d)$. Using the same loss function, we compute the average distance between the sampled 
$\uSigma^{(s)}$ and $\uSigma$, $D_{\Sigma}(M_d) = 10^{-4} \sum_{s} L(\uSigma^{(s)},\uSigma), d=1,2,3$. 
Recall that $M_1$ is the simplest possible model with only an intercept term, while $M_2$ and $M_3$ allow for single and 
multiple curves to be fitted, respectively. We present results in terms of the percentage improvement   
of the two more complex models relative to the simplest model, $I_R(d) = D_R(M_d)/D_R(M_1)$ and 
$I_{\Sigma}(d) = D_{\Sigma}(M_d)/D_{\Sigma}(M_1), d=2,3$, that we refer to as the relative risks.     

Results are presented in Table \ref{table.sim1}. 
It is suggested that, for small samples, it is important to correctly model the innovation correlation matrices when either the innovation correlations themselves or the covariances/correlations among responses at each time point are of interest. From the Table we can see that  both $M_2$ and $M_3$ improve over model $M_1$, for the estimation of both $\uR_t$ and $\uSigma$. Further, the relative risk of $M_3$ is, with one exception, less than that of $M_2$.  
Lastly, the substantial reduction in risk observed for $n=10$, decreases as the sample sizes increases, and the relative risks get close to $100\%$. The exception occurs for the estimation of the $\uR_t$ with model $M_3$, for which  
an important gain of $13.12\%$ can be observed even for $n=100$.

\begin{table}
	\begin{center}
		\caption{First simulation study results: the entries of the table are the relative risks 
			$I_R(d) = D_R(M_d)/D_R(M_1)$ and $I_{\Sigma}(d) = D_{\Sigma}(M_d)/D_{\Sigma}(M_1), d=2,3,$
			expressed as percentages. 
			Rows refer to the sample size and columns to models $M_2$ and $M_3$. Results are based on $30$ replicate datasets.} \label{table.sim1}
		\begin{tabular}{l|cc|cc}
			\hline			
			n & $I_R(2)$ & $I_{\Sigma}(2)$ & $I_R(3)$ & $I_{\Sigma}(3)$\\
			\hline
			10  &  36.31 & 59.94 & 40.70 & 55.07\\
			20  &  90.52 & 87.20 & 80.18 & 83.71\\
			50  &  92.34 & 96.67 & 82.94 & 94.46\\
			100  &  98.73 & 99.16 & 86.88 & 96.38\\
			\hline 
		\end{tabular}
	\end{center}
\end{table}

\subsection{Gains}

To investigate the potential gains associated with fitting multivariate longitudinal models and the effects of missing data on posterior distributions, we consider a data-generating mechanism that consists of $3$ responses, $Y_{1t}, Y_{2t}, Y_{3t},$ observed over $M=6$ time points, $t=-0.5,-0.3,\dots,0.5$. The $3$ responses are generated from a multivariate normal distribution with 
mean $\umu_t = (\beta_{10} + \beta_{11} t,0,0)^{\top}$, that is, the mean of the first response is a linear function of $t$, 
while the other $2$ responses have constant zero mean. Further, the $18 \times 18$ covariance matrix is taken to be  
$\uSigma=\uSigma(\rho_1,\rho_2) = \uSigma_1(\rho_1) \otimes \uSigma_{23}(\rho_2)$,
where $\uSigma_1(\rho_1) = \{\rho_1^{|k-l|}\}$ is the marginal covariance matrix of $\uY_1$ and 
$\uSigma_{23}(\rho_2)$ has diagonal elements equal to $1$ and all off diagonal elements equal to $\rho_2$.
The choice of autoregressive covariance structure $\uSigma_1$ is a standard one. It implies that correlations between observations on $Y_1$ at different time points decrease exponentially fast as the lag increases. The value of $\rho_1$ is not of particular interest here, and hence it is fixed to $\rho_1=0.5$, implying that correlations get halved every time the lag increases by $0.2$. The structure of $\uSigma$ indicates that $Y_1$ is equally correlated with $Y_2$ and $Y_3$ at all time points. This correlation is equal to $\rho_2$ at lag $0$ and it gets halved when the lag increases by $0.2$. 

The chosen mean and covariance structures are fairly simple, and this suffices for the purposes of the current study. 
The main interest here is on the quality of the estimate of the mean function of the first response. 
We examine how this quality, as measured by the posterior MSE, or it's components bias and variance, depends on the dimension of the response in the fitted model, the value of the correlation coefficient $\rho_2$, the sample size $n$, and the percentage of missing data. The effect of the dimension of the response is evaluated by fitting $1$-, $2$-, and $3$-dimensional response models. The effect of the correlation coefficient is examined by letting $\rho_2$ take values in the set $\{0.2, 0.4, 0.6, 0.8\}$. 
The effect of the sample size is assessed by letting $n$ take values in the set $\{100,300\}$. Lastly, the effect of missing data is investigated by allowing $0\%$, $25\%$ and $50\%$ chance of missing observations, utilizing a missing completely at random mechanism.  

Considering a $3$-dimensional response model as an example, the mean functions are modelled using $\mu_{tk} = \beta_{k0} + \beta_{k1} t$,
$k=1,2,3$. This specification is correct for the first response and wrong for the other two responses. 
Further, the autoregressive coefficients and innovation variances are modelled using
$\phi_{jklm} = \phi_{lm0} + f_{\phi,l,m}(t_j-t_k)$ and $\log \sigma^2_{tk} = \alpha_{k0} + f_{\sigma,k}(t)$,
where $f_{\phi,l,m}(t_j-t_k)$ and $f_{\sigma,k}(t)$ include $5$ and $6$ knots, respectively, allowing for the possibility of non-linear effects. 
Lastly, we fit the common correlations model (\ref{priorR}) with constant mean, $\mu_{ct}=\eta_0,$ and constant variance, $\log \sigma^2_{ct} = \omega_0$.  
These choices for $\mu_{ct}$ and $\log \sigma^2_{ct}$ were arrived at after experimentation with more complex models that showed 
the more complex models to not be necessary. 

The regression coefficients are taken to be $\beta_{10}=0$ and $\beta_{11}=2.95$, where the value of $\beta_{11}$ was so chosen to achieve a signal-to-noise ratio (SNR) equal to $1$. The SNR is defined as $\text{SNR} = (\text{SST} - \text{SSE}) / \text{SSE}$, where SST the total sum of squares SST=$\sum_{i,j} (y_{ij1}-\bar{y}_1)^2$ and SSE the error sum of squares SSE=$\sum_{i,j} (y_{ij1}-\hat{y}_{ij1})^2$. Predictions $\hat{y}_{ij1}$ were obtained by fitting the true data-generating model.  

For all models we obtain $3 \times 10^4$ posterior samples of which we discard the first $10^4$ as burn in, and of the remaining $2 \times 10^4$ we keep one in two. Based on the $10^4$ retained samples, we obtain samples for the parameter of main interest, 
$\mu_{t1} = E(Y_{t1}|t) = \beta_{10} + \beta_{11} t$, by replacing the regression coefficients $\beta_{10}$ and $\beta_{11}$ by the corresponding sampled values,  $\mu_{t1}^{(s)} = \beta_{10}^{(s)} + \beta_{11}^{(s)} t, s=1,\dots,10^4$. 
We compare fitted models in terms of their posterior bias and variance in estimating $\mu_{t1}$.
Using subscript $p$ to denote quantities obtained from models with fitted response dimension $p=1,2,3$, 
the posterior bias and variance are computed as 
$B_p(t) = \sum_s (\mu_{t1,p}^{(s)} - \bar \mu_{t1,p})^2$ and $V_p(t) = (\bar \mu_{t1,p} - \mu_{t1})^2$,
where $\bar \mu_{t1,p}$ is the mean of $\mu_{t1,p}^{(s)}, s=1,\dots,10^4$.
Further, because $B_p(t)$ and $V_p(t)$ are computed for $t=-0.5,-0.3,\dots,0.5$, results are summarised by computing sums:
$B_p=\sum_t B_p(t)$ and $V_p=\sum_t V_p(t)$. Results presented below are based on $30$ replicate datasets. 

Table \ref{bias.sim} compares model performance by reporting the ratio $100 B_p/B_1$, 
that we refer to as relative bias, while Table \ref{var.sim} compares model performance
by reporting the ratio $100 V_p/V_1$, that we refer to as relative variance, $p=2,3$.  
We can see from Table \ref{bias.sim} that relative bias decreases as the correlation $\rho_2$ between 
the responses increases, for all $n$, $p$, and missingness probabilities. 
Comparing the left half of the Table, that refers to $2$-dimensional response models, to the right half, that refers to 
$3$-dimensional models, we see that the comparison depends on the missingness probability. 
When it is $0\%$, $2$- and $3$-dimensional models fair almost equally, but as the missingness probability 
increases, the bias of the $3$-dimensional model becomes lower than that of the $2$-dimensional model. 
Table \ref{var.sim} displays the results on relative variances. We can observe that  
as correlations between the responses increase, the relative variance decreases, for all $n$, $p$, and missingness probabilities.
The relative variance for $3$-dimensional response models is lower than that of $2$-dimensional models, for
all missingness probabilities. These results are very much in line with what one would expect given the results of \citet{Zellner62}. 

Lastly, we study the effects of missing data on the posterior bias and variance. To do so, we introduce a second subscript:
we let $B_{p,m}$ and $V_{p,m}$ denote the bias and variance in estimating $\mu_{t1}$ based on a $p$-dimensional response model
where the observations in the dataset are missing with probability $m$ . Table \ref{bmiss.sim} presents results on the 
ratios $100 B_{p,m}/B_{1,0}, p=1,2,3, m=0.25,0.50$, that is, it compares the bias of univariate response models
that are based on the full dataset with the performance of univariate and multivariate response models that are based on  
random subsets of the full dataset. First, looking at univariate response models, $p=1$, when the probability of missingness is $m=0.25$, 
bias increases by $8.4\%$ and $8.7\%$ for $n=100$ and $n=300$,
respectively. These two percentages are derived by averaging the $4$ relevant numbers in the Table.
Adding a second response to the model is enough to counteract the increase in bias due to missing $25\%$ of the data. We can see that almost all the $8$ relevant entries in the Table are less than $100\%$, and as low as $75.19\%$. Adding a third response slightly improves the model performance relative to the model with $2$ responses. Further, when the probability of missingness increases to $m=0.50$, the bias increases, on average, by $29\%$ and $37.9\%$ for the $2$ sample sizes(again computed by averaging the $4$ relevant numbers in the Table).  Adding a second response improves performances, but it cancels out the increase in the bias only when the correlation $\rho_2$ is as high as $0.80$. Adding a third response does not further improve model performance.

Table \ref{vmiss.sim} presents results on the ratios $100 V_{p,m}/V_{1,0}, p=1,2,3, m=0.25,0.50$.    
Concerning univariate response models, $p=1$, when the probability of missingness is $m=0.25$, 
the variance increases by a modest $4.6\%$ and $1.1\%$ for $n=100$ and $n=300$, respectively. 
Adding a second response to the model counterbalances the effect of missing $25\%$ of the data. 
It can be seen in the Table that, with only $1$ exception, all entries are below $100\%$, and as low as $79.71\%$. 
The addition of a third response further improves model performance. Further, when the probability of missingness is $m=0.50$, 
the variance increases, on average, by $13.8\%$ and $12.2\%$ for the $2$ sample sizes. Adding a second response 
completely cancels out the increase in the variance when $\rho_2=0.80$ and it almost cancels it out 
when $\rho_2=0.6$, for both sample sizes. A third response further improves performances and 
it neutralizes the increase in the variances when $\rho_2$ is $0.60$ or higher.

\begin{table}
	\begin{center}
		\caption{Second simulation study results: the entries of the table are the relative biases $100 B_p/B_1, p=2,3$. 
			Rows refer to the sample size $n = 100, 300,$ columns to the correlation coefficient $\rho_2=
			0.2,0.4,0.6,0.8$, and the $3$ parts of the table to the probability of missingness $0\%, 25\%, 50\%$. 
			Results are based on $30$ replicate datasets per sample size by correlation combination.} \label{bias.sim}
		\begin{tabular}{l|rrrr|l|rrrr}
			\multicolumn{10}{l}{Probability of missingness: $0\%$}\\
			\hline
			$p=2$  &  0.2   &    0.4  &   0.6  &   0.8  &   $p=3$  &  0.2   &    0.4  &   0.6  &   0.8  \\
			\hline
			100  &  93.79 & 85.81 & 76.42 & 67.07 & 100 & 96.37 & 87.48 & 76.00 & 66.49 \\
			300 &  95.79 & 89.59 & 81.43 & 71.70 &  300 & 96.79 & 90.03 &  82.50 & 73.78 \\
			\hline
			\multicolumn{10}{l}{Probability of missingness: $25\%$}\\
			\hline
			$p=2$  &  0.2   &    0.4  &   0.6  &   0.8  &   $p=3$  &  0.2   &    0.4  &   0.6  &   0.8  \\
			\hline
			100  &  95.69  & 88.12 & 77.99 & 67.12 & 100 & 95.04 & 84.01 & 74.29 & 64.48 \\
			300 &  93.27  & 86.67 & 79.45 & 70.32 & 300 & 92.43 & 85.44 & 78.51 & 70.58 \\
			\hline
			\multicolumn{10}{l}{Probability of missingness: $50\%$}\\
			\hline
			$p=2$  &  0.2   &    0.4  &   0.6  &   0.8  &   $p=3$  &  0.2   &    0.4  &   0.6  &   0.8  \\
			\hline
			100  &  96.39 & 91.55 & 84.72 & 75.56 & 100 & 94.53 & 87.70 & 78.89 & 68.02 \\
			300 &  100.80 & 94.8 & 87.24 & 79.93 & 300 & 94.71 & 86.6 & 78.04 & 66.83\\
			\hline			
		\end{tabular}
	\end{center}
\end{table}

\begin{table}
	\begin{center}
		\caption{Second simulation study results: the entries of the table are the relative variances $100 V_p/V_1, p=2,3$.
			Rows refer to the sample size $n = 100, 300,$ columns to the correlation coefficient $\rho_2=
			0.2,0.4,0.6,0.8$, and the $3$ parts of the table to the probability of missingness $0\%, 25\%, 50\%$. 
			Results are based on $30$ replicate datasets per sample size by correlation combination.}\label{var.sim}
		\begin{tabular}{l|rrrr|l|rrrr}
			\multicolumn{10}{l}{Probability of missingness: $0\%$}\\
			\hline
			$p=2$  &  0.2   &    0.4  &   0.6  &   0.8  &   $p=3$  &  0.2   &    0.4  &   0.6  &   0.8  \\
			\hline
			100  & 99.06 & 95.74 & 90.36 & 81.64 & 100 & 97.90 & 93.10 & 85.67 & 75.57 \\
			300 &  98.77 & 95.59 & 89.47 & 79.73 & 300 & 97.42 & 92.19 & 84.91 & 74.13 \\
			\hline
			\multicolumn{10}{l}{Probability of missingness: $25\%$}\\
			\hline
			$p=2$  &  0.2   &    0.4  &   0.6  &   0.8  &   $p=3$  &  0.2   &    0.4  &   0.6  &   0.8  \\
			\hline
			100  & 98.58 & 95.13 & 90.04 & 81.38 & 100 & 97.58 & 92.97 & 85.48 & 75.09\\
			300 &  98.87 & 94.96 & 88.76 & 78.84 & 300 & 98.27 & 92.07 & 83.95 & 72.72\\
			\hline
			\multicolumn{10}{l}{Probability of missingness: $50\%$}\\
			\hline
			$p=2$  &  0.2   &    0.4  &   0.6  &   0.8  &   $p=3$  &  0.2   &    0.4  &   0.6  &   0.8  \\
			\hline
			100  & 98.35 & 95.13 & 90.22 & 80.74 & 100 & 97.40 & 92.56 & 85.41 & 75.22\\
			300 &  98.94 & 95.25 & 89.37 & 80.18 & 300 & 98.13 & 93.21 & 85.07 & 73.65\\
			\hline			
		\end{tabular}
	\end{center}
\end{table}

\begin{table}
	\begin{center}
		\caption{Second simulation study results: the entries of the table are the relative biases 
			$100 B_{p,m}/B_{1,0}, p=1,2,3, m=0.25,0.50$.
			Rows refer to the sample size $n = 100, 300,$ columns to the correlation coefficient $\rho_2=
			0.2,0.4,0.6,0.8$, and the $2$ parts of the table to the probability of missingness $m=0.25, 0.50$.
			Results are based on $30$ replicate datasets per sample size by correlation combination.} \label{bmiss.sim}
		\begin{tabular}{l|rrrr|l|rrrr}
			\multicolumn{5}{l|}{Probability of missingness: $0.25$} & \multicolumn{5}{l}{Probability of missingness: $0.50$}\\
			\hline
			$p=1$  &  0.2   &    0.4  &   0.6  &   0.8  &   $p=1$  &  0.2   &    0.4  &   0.6  &   0.8  \\
			\hline
			100 & 105.33 & 107.67 & 108.65 & 112.03 & 100 & 131.80 & 130.82 & 127.64 & 125.61 \\
			300 & 110.74 & 109.26 & 107.88 & 107.11 & 300 & 141.21 & 139.64 & 137.11 & 133.82 \\
			\hline
			$p=2$  &  0.2   &    0.4  &   0.6  &   0.8  &   $p=2$  &  0.2   &    0.4  &   0.6  &   0.8  \\
			\hline
			100 & 100.79 & 94.87 & 84.74 & 75.19 & 100 & 127.04 & 119.77 & 108.13 & 94.91\\
			300 & 103.28 & 94.70 & 85.72 & 75.32 & 300 & 133.49 & 122.47 & 108.17 & 91.02\\
			\hline
			$p=3$  &  0.2   &    0.4  &   0.6  &   0.8  &   $p=3$  &  0.2   &    0.4  &   0.6  &   0.8  \\
			\hline
			100 & 100.11 & 90.45 & 80.72 & 72.24 & 100 & 132.86 & 124.01 & 111.35 & 100.40\\
			300 & 102.36 & 93.35 & 84.70 & 75.59 & 300 & 133.73 & 120.93 & 107.00 & 89.42\\
			\hline			
		\end{tabular}
	\end{center}
\end{table}

\begin{table}
	\begin{center}
		\caption{Second simulation study results: the entries of the table are the relative variances 
			$100 V_{p,m}/V_{1,0}, p=1,2,3, m=0.25,0.50$.
			Rows refer to the sample size $n = 100, 300,$ columns to the correlation coefficient $\rho_2=
			0.2,0.4,0.6,0.8$, and the $2$ parts of the table to the probability of missingness $m=0.25, 0.50$. 
			Results are based on $30$ replicate datasets per sample size by correlation combination.} \label{vmiss.sim}
		\begin{tabular}{l|rrrr|l|rrrr}
			\multicolumn{5}{l|}{Probability of missingness: $0.25$} & \multicolumn{5}{l}{Probability of missingness: $0.50$}\\
			\hline
			$p=1$  &  0.2   &    0.4  &   0.6  &   0.8  &   $p=1$  &  0.2   &    0.4  &   0.6  &   0.8  \\
			\hline
			100 & 104.35 & 104.54 & 104.41 & 104.98 & 100 & 113.97 & 114.17 & 113.45 & 113.60  \\
			300 & 100.79 & 101.19 & 101.36 & 101.11 & 300 & 111.57 & 112.18 & 112.77 & 112.45 \\
			\hline
			$p=2$  &  0.2   &    0.4  &   0.6  &   0.8  &   $p=2$  &  0.2   &    0.4  &   0.6  &   0.8  \\
			\hline
			100 & 102.87 & 99.45 & 94.00 & 85.43 & 100 & 112.09 & 108.60 & 102.36 & 91.71\\
			300 &  99.65 & 96.09 & 89.97 & 79.71 & 300 & 110.39 & 106.85 & 100.79 & 90.16\\
			\hline
			$p=3$  &  0.2   &    0.4  &   0.6  &   0.8  &   $p=3$  &  0.2   &    0.4  &   0.6  &   0.8  \\
			\hline
			100 & 101.82 & 97.19 & 89.25 & 78.83 & 100 & 111.01 & 105.68 & 96.90 & 85.44\\
			300 & 99.05 & 93.17 & 85.10 & 73.53 & 300 & 109.48 & 104.57 & 95.93 & 82.81\\
			\hline			
		\end{tabular}
	\end{center}
\end{table}

\section{Application}\label{application}

The dataset that we analyse here is a random subsample of $n=500$ subjects from the Paquid prospective cohort study \citep{LCDB94}
and it is available in the R package \texttt{lcmm} \citep{lcmm}. The Paquid study aimed at investigating cerebral and functional aging in individuals aged 65 years and older, and it involved individuals living in south-western France. Longitudinal observations over a maximum period of $20$ years were collected on $3$ cognitive tests and depressive symptomatology. The cognitive tests evaluated the global mental status using the Mini Mental State Examination ($y_1$), verbal fluency using the Isaacs Set Test ($y_2$), and visual memory using the Benton's Visual Retention Test ($y_3$). Depressive symptomatology was measured by the Center for Epidemiological Study Depression scale ($y_4$). 
We note that in the first $3$ responses higher scores indicate better cognitive function performance, while in the fourth response, higher scores indicate more symptoms.  
The available covariates are a binary gender indicator, $x_1=1$ for males and $x_1=0$ for females, a binary indicator of educational level, $x_2=1$ for individuals who graduated from primary school and $x_2=0$ otherwise, age at entry in the cohort $x_3$, and time of observation $t$.  

Because the distribution of $y_1$ is very asymmetric, we analyse a normalized version of it \citep{normpsy}. 
Furthermore, to avoid computational problems, we scale the time of observation, $t=0,1,2,\dots,20$, by dividing by $20$,
hence, $t=0,0.05,0.1,\dots,1$.
Further, we centre and scale the age at entry $x_3$ by subtracting the minimum age and dividing by $15$.
Lastly, we note that the dataset is highly unbalanced, 
with individuals observed at no more than $9$ time points of the possible $21$. 
The number of individuals observed at each time point decreases quickly: 
there are about $300$ observations at $t=0.1$, about $200$ at $t=0.2$, 
about $150$ between $t=0.3$ and $t=0.4$, and thereafter the number of observations
stays well below $100$, with only $18$ individuals observed at the last time point.

We model the mean using
\begin{equation}\label{appmean}
\mu_{ijk} = \beta_{k0} + \beta_{k1} x_{i1} + \beta_{k2} x_{i2} +  f_{\mu,k,3}(x_{3i}) + f_{\mu,k,4}(t_{ij}), k=1,\dots,4,     
\end{equation}
where $f_{\mu,k,3}(x_{3i})$ and $f_{\mu,k,4}(t_{ij})$ are modelled using up to $10$ and $8$ knots,
equivalently $11$ and $9$ basis functions, respectively.
These knots are chosen as the unique quantiles of $x_3$ 
and $t$ that correspond to the $10$ equally spaced probabilities between $0$ and $1$. Although for $x_3$
there are $10$ unique quantiles, for $t$ there are only $8$, hence the difference in the number of knots for modelling these 2 functions. 
Thus, for modelling the means there are in total $92$ parameters, of which $88$ are subject to selection. 

Further, the $84 \times 84$ covariance matrix includes $3570$ unique elements. 
These are modelled using $4$ regression models that we describe next. First, 
the autoregressive coefficients are modelled using
\begin{equation}\label{appauto}
\phi_{ijklm} = \psi_{lm0} + f_{\phi,l,m}(t_{ij}-t_{ik}),  l,m=1,\dots,4,
\end{equation}
where $f_{\phi,l,m}(t_{ij}-t_{ik})$ is modelled using up to $10$ knots. These are selected as the unique quantiles of 
lag, $t_{ij}-t_{ik}$, that correspond to the $10$ equally spaced probabilities between $0$ and $1$.
Thus, there are up to $192$ parameters for modelling the autoregressive coefficients and all of these are subject to selection. 

The innovation variances are modelled as
\begin{equation}\label{appinv}
\log \sigma^2_{ijk} = \alpha_{k0} + f_{\sigma,k,1}(x_{3i}) + f_{\sigma,k,2}(t_{ij}), k=1,\dots,4,
\end{equation}
where $f_{\sigma,k,1}(x_{3i})$ and $f_{\sigma,k,2}(t_{ij})$ are modelled using up to $10$ and $8$ knots, just as was done for the mean functions. 
Hence, there are up to $84$ parameters for modelling the innovation variances, of which $4$ are always in the model, while
the other $80$ are subject to selection. 

Lastly, we describe the model for the correlation matrices. Since the $4$ responses can be naturally clustered into the group of $3$ variables 
that measure cognitive function and the group of $1$ variable that measures depressive symptomatology, a grouped variables model is a priori justifiable. Despite that, we prefer the less structured grouped correlations model, and let the data decide if a grouped variables model is preferable. 
The parameters of the correlation matrices are modelled using 
$\mu_{cth} = \eta_{h0} + f_{\mu h}(t_{ij})$ and 
$\log \sigma^2_{ct} = \omega_{0} + f_{\sigma}(t_{ij})$ 
As the $4 \times 4$ correlation matrices have $6$ unique elements, we allow up to $H=6$ different trajectories for the means $\mu_{cth}, h=1,\dots,6$.
Functions $f_{\mu h}(t_{ij})$ and $f_{\sigma}(t_{ij})$ are modelled using up to $10$ knots. 
These are selected as the unique quantiles of $t$ that correspond to the $10$ equally spaced probabilities between $0$ and $1$. Although for modelling innovation variances there are $8$ unique quantiles, here there are $10$, as there is a single correlation matrix observed at each time point.
Hence, there are up to $72$ parameters for modelling $\mu_{cth}$, of which $66$ are subject to selection. 
For modelling $\log \sigma^2_{ct}$ there are $12$ parameters, of which $11$ are subject to selection. 
This brings the number of parameters for modelling the covariance matrix to $360$, $10$ times smaller 
than the number of unique elements in the covariance matrix. 

Results presented below are based on $25 \times 10^3$ posterior samples that are obtained from $10^5$ iterations of the MCMC sampler, with the first $5 \times 10^4$ 
discarded as burn-in, and of the remaining keeping every second sample.

Concerning the mean regression models, results in the form of posterior means and $80\%$ credible intervals are presented in the Figure of the supplementary material.  
Covariate $x_1$ (gender) has little to no effect on the means of three cognitive function responses but it has an important effect
on the mean of depressive symptomatology, indicating that males have, on average, less symptoms than females. 
Further, covariate $x_2$ (education) has important effects on the means of the responses relating to cognitive function,
with those who graduated from primary school having higher mean cognitive function scores.
In addition, $x_2$ appears to have no effect on mean depressive symptomatology.
Covariates $x_3$ (age) and $t$ (time) have negative linear (or almost linear) associations with the means of the $3$ responses that relate to cognitive function.
Furthermore, $x_3$ has a more complex than linear relationship with mean depressive symptomatology, but, due to the high 
uncertainty, this relationship could also be seen as flat.
Lastly, $t$ has positive linear association with the mean of fourth response, indicating that the average depressive
symptomatology increases with time. 
Visually, the covariates have simple relationships with the mean responses. This can also be confirmed by the small number
of regression coefficients that are selected to fit the mean functions: of the $88$, on average, $16.1$ or $18.3\%$ were selected during MCMC sampling.

\begin{figure}
	\begin{center}
		\begin{tabular}{cccc}
			\includegraphics[width=0.2\textwidth]{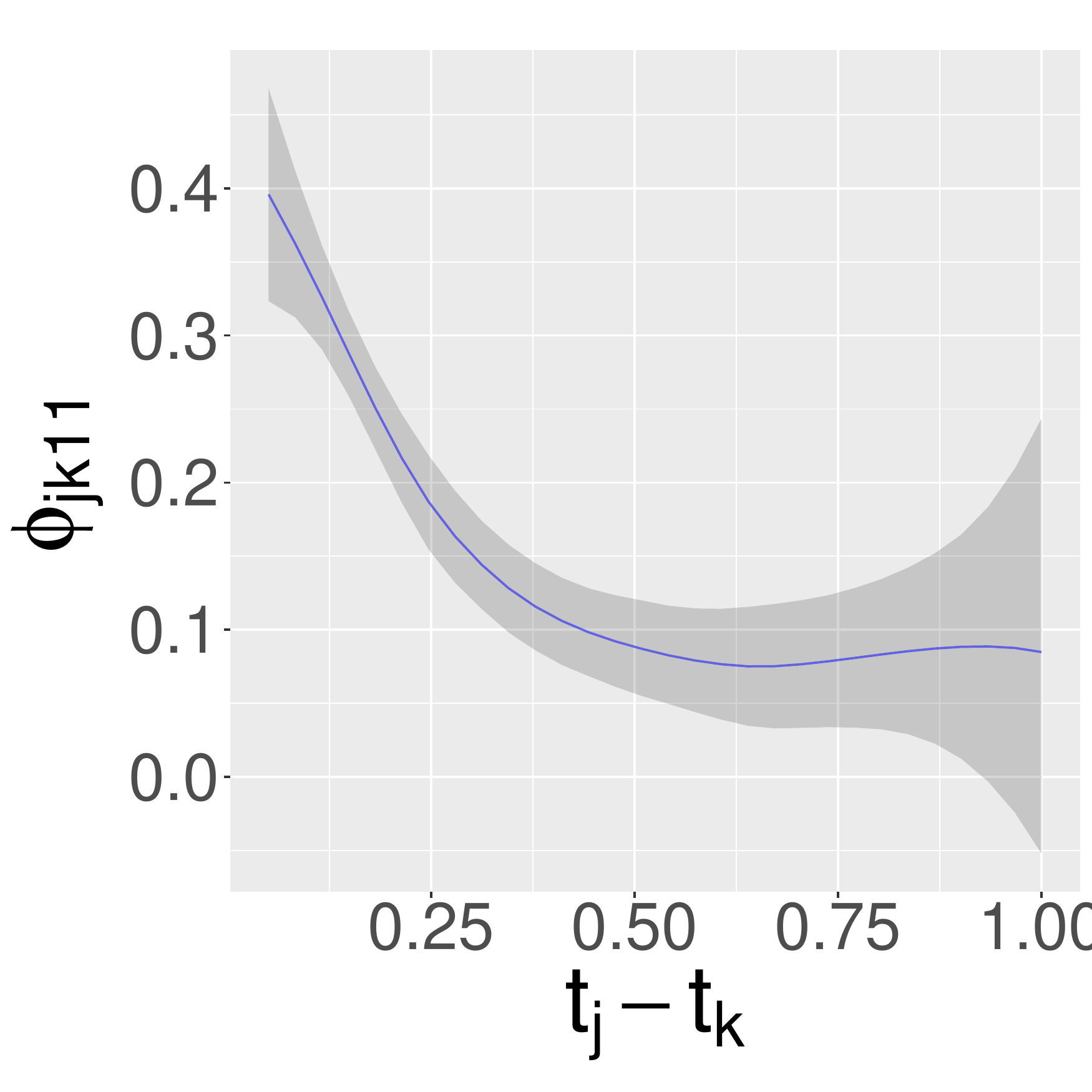} &  
			\includegraphics[width=0.2\textwidth]{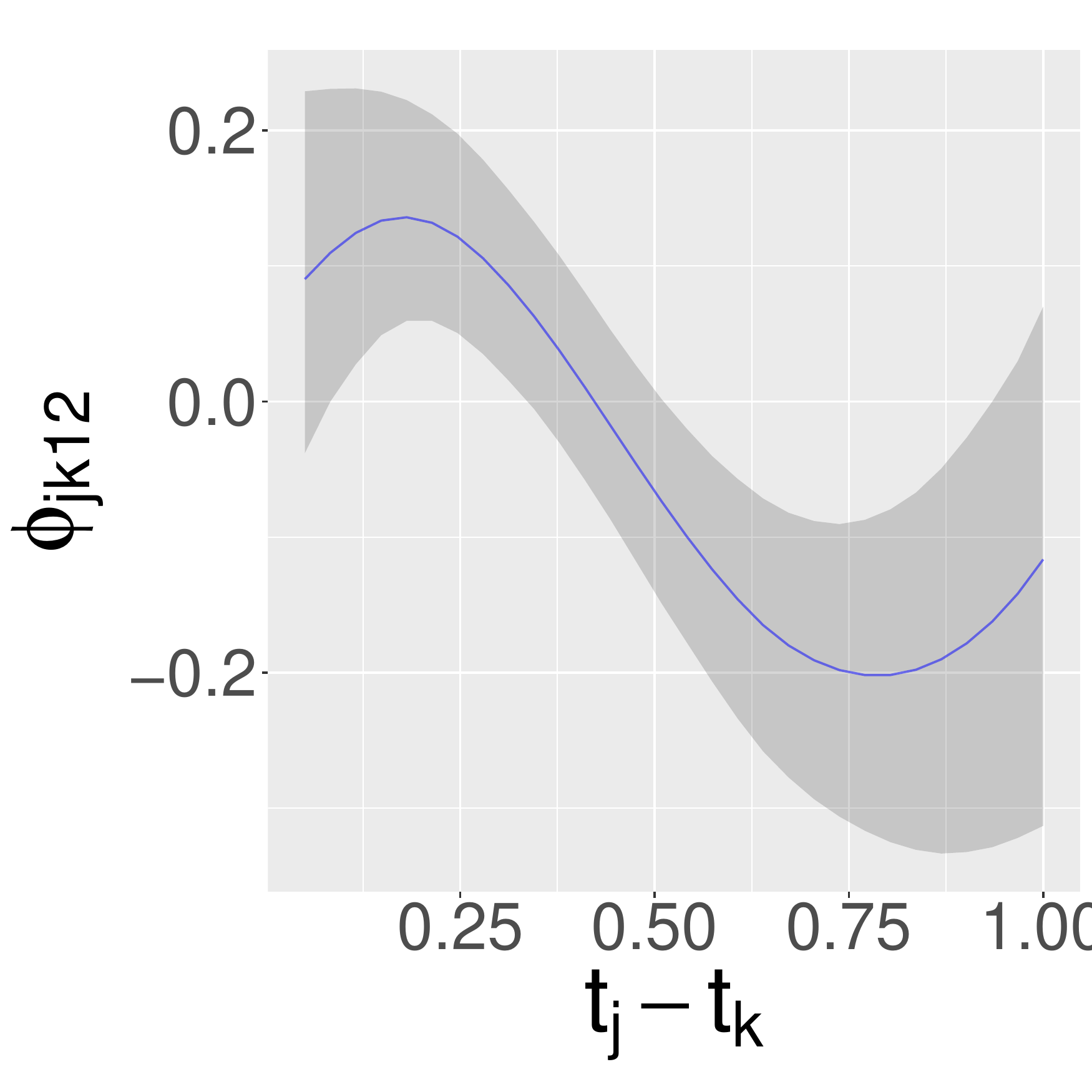} &  
			\includegraphics[width=0.2\textwidth]{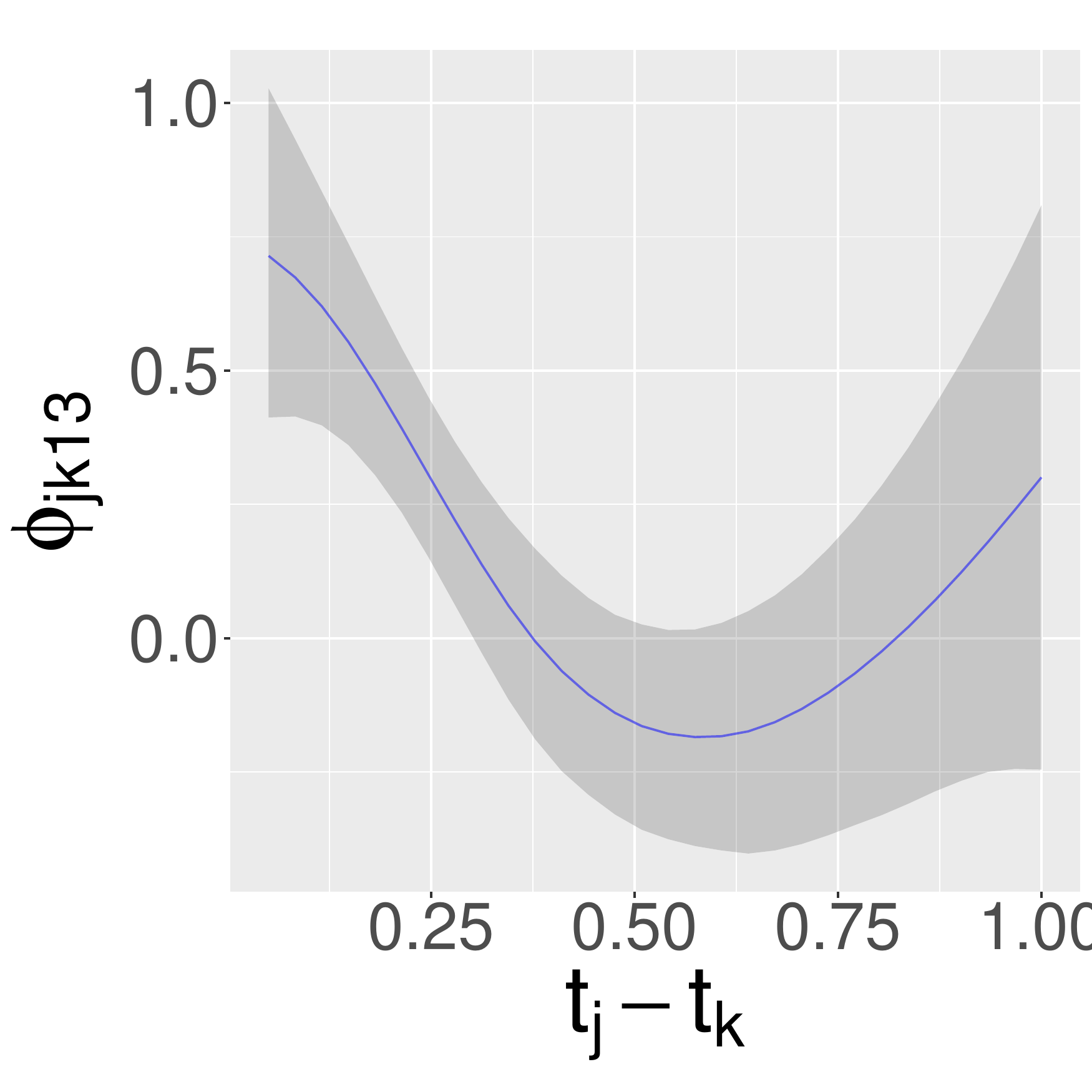} &
			\includegraphics[width=0.2\textwidth]{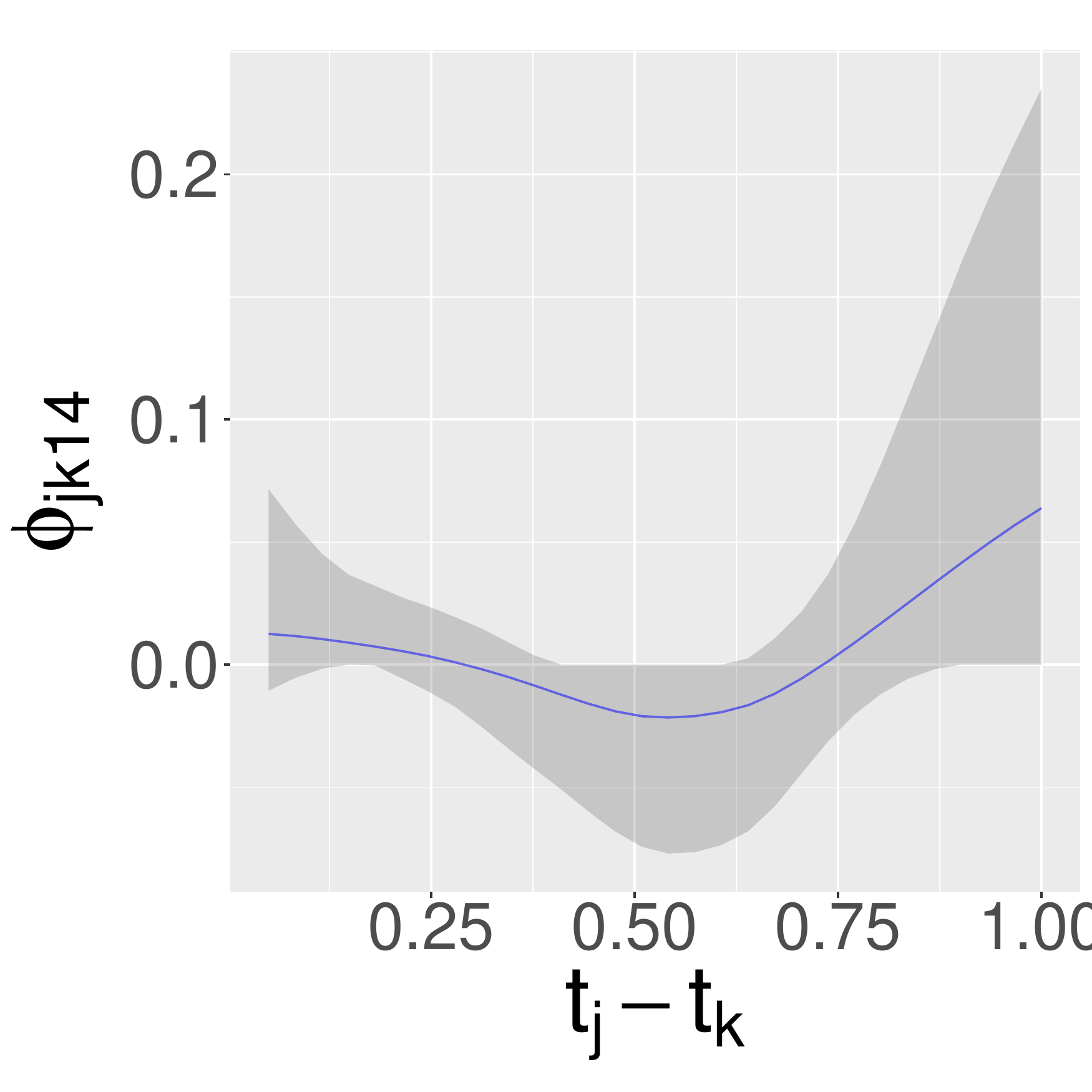} \\    
			\includegraphics[width=0.2\textwidth]{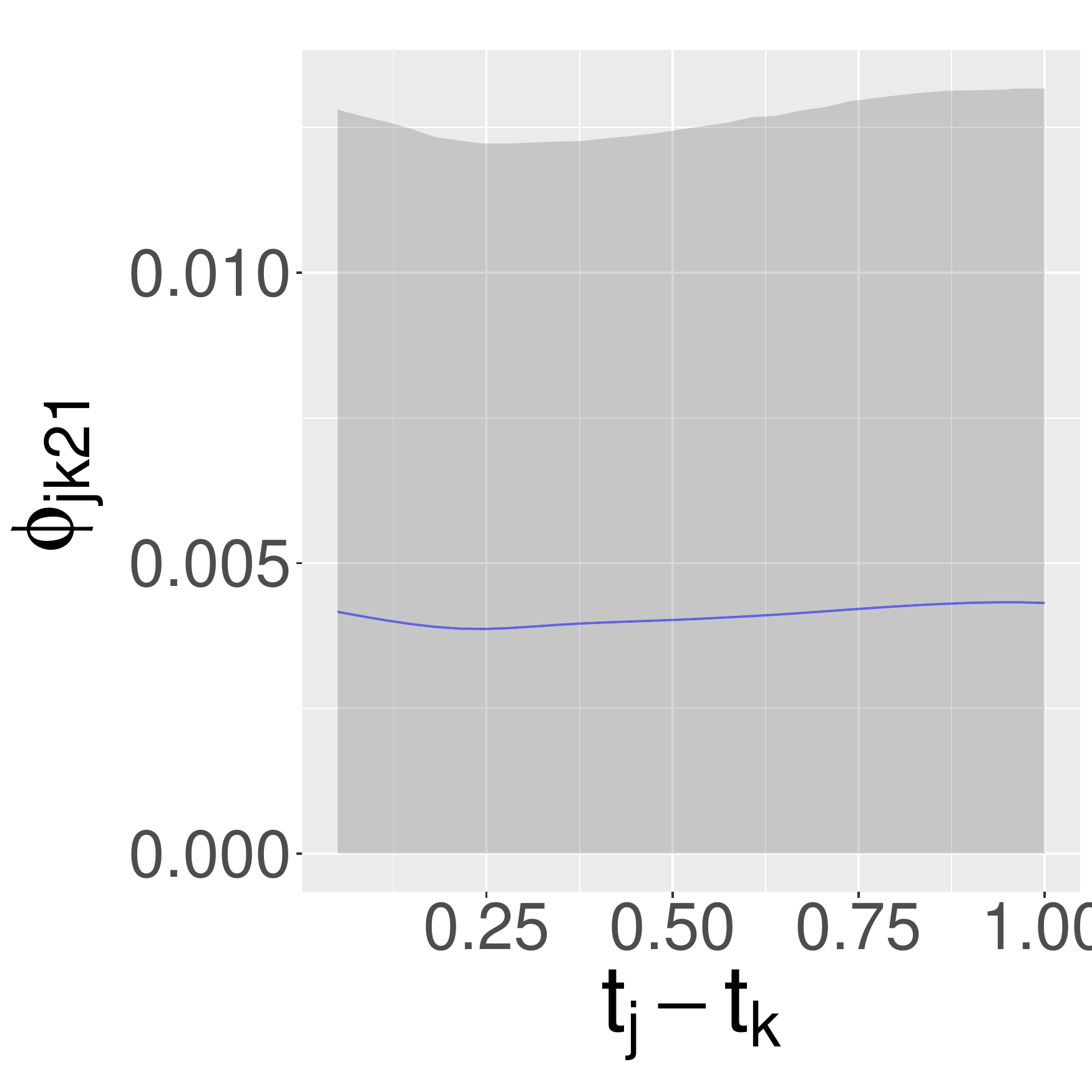} &  
			\includegraphics[width=0.2\textwidth]{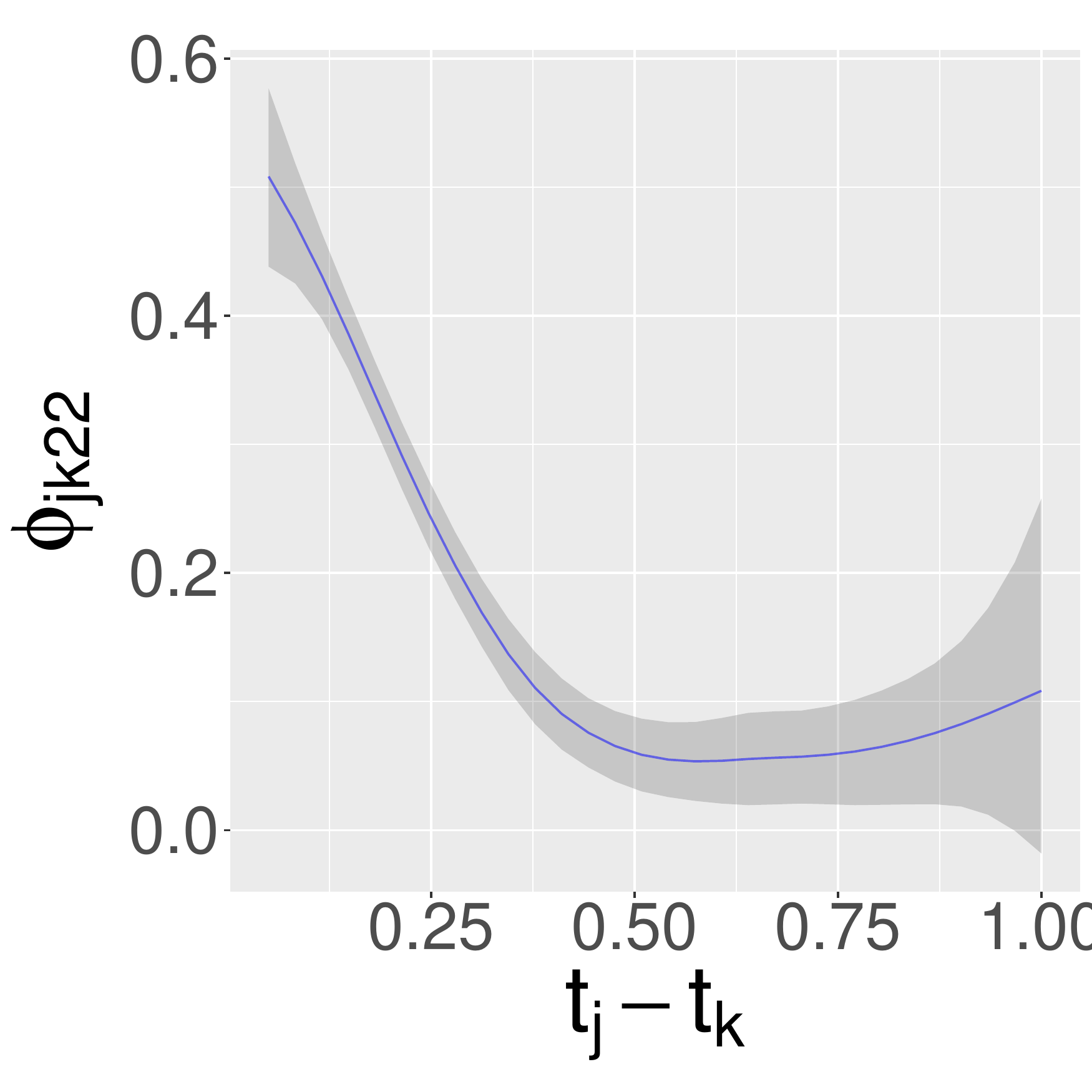} &  
			\includegraphics[width=0.2\textwidth]{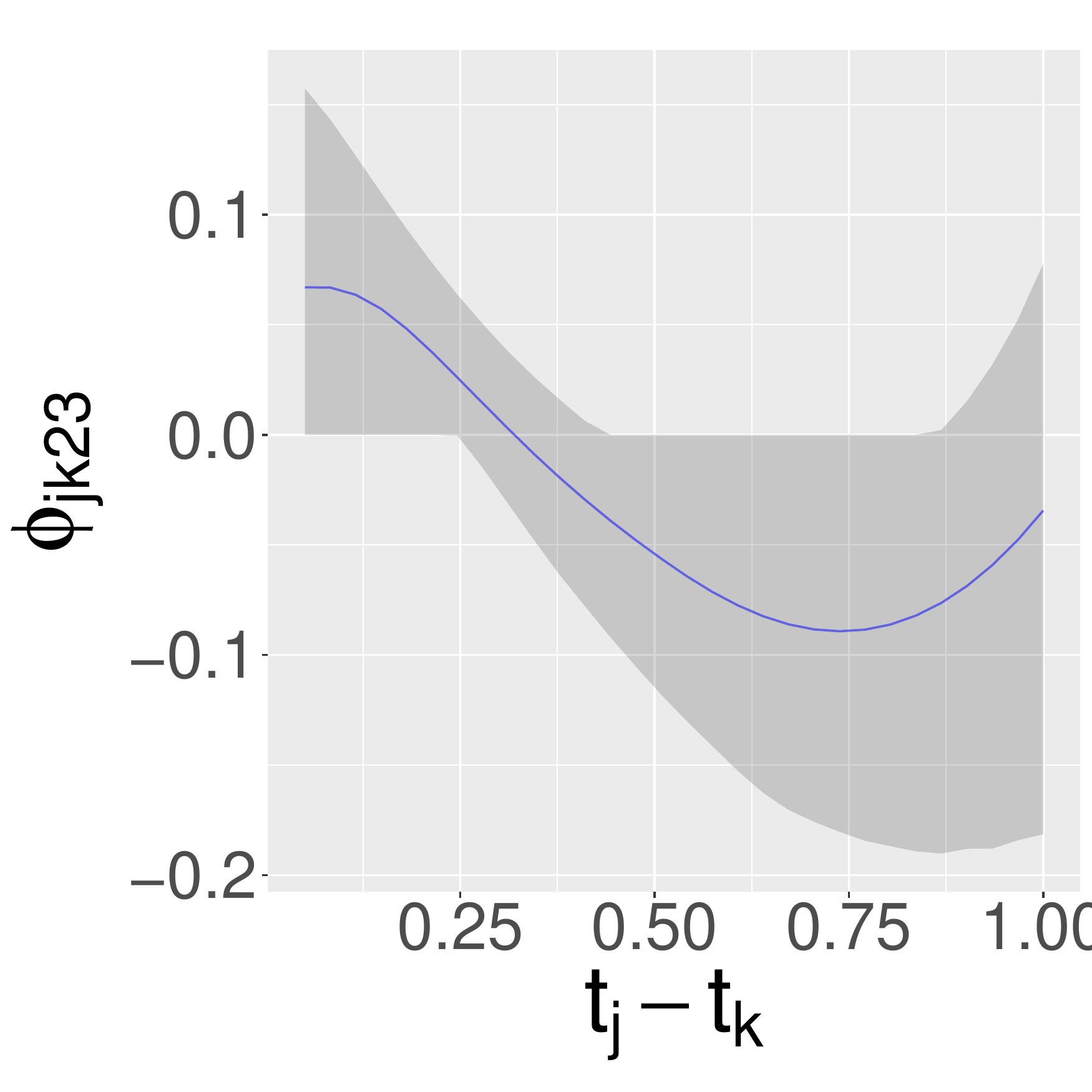} &
			\includegraphics[width=0.2\textwidth]{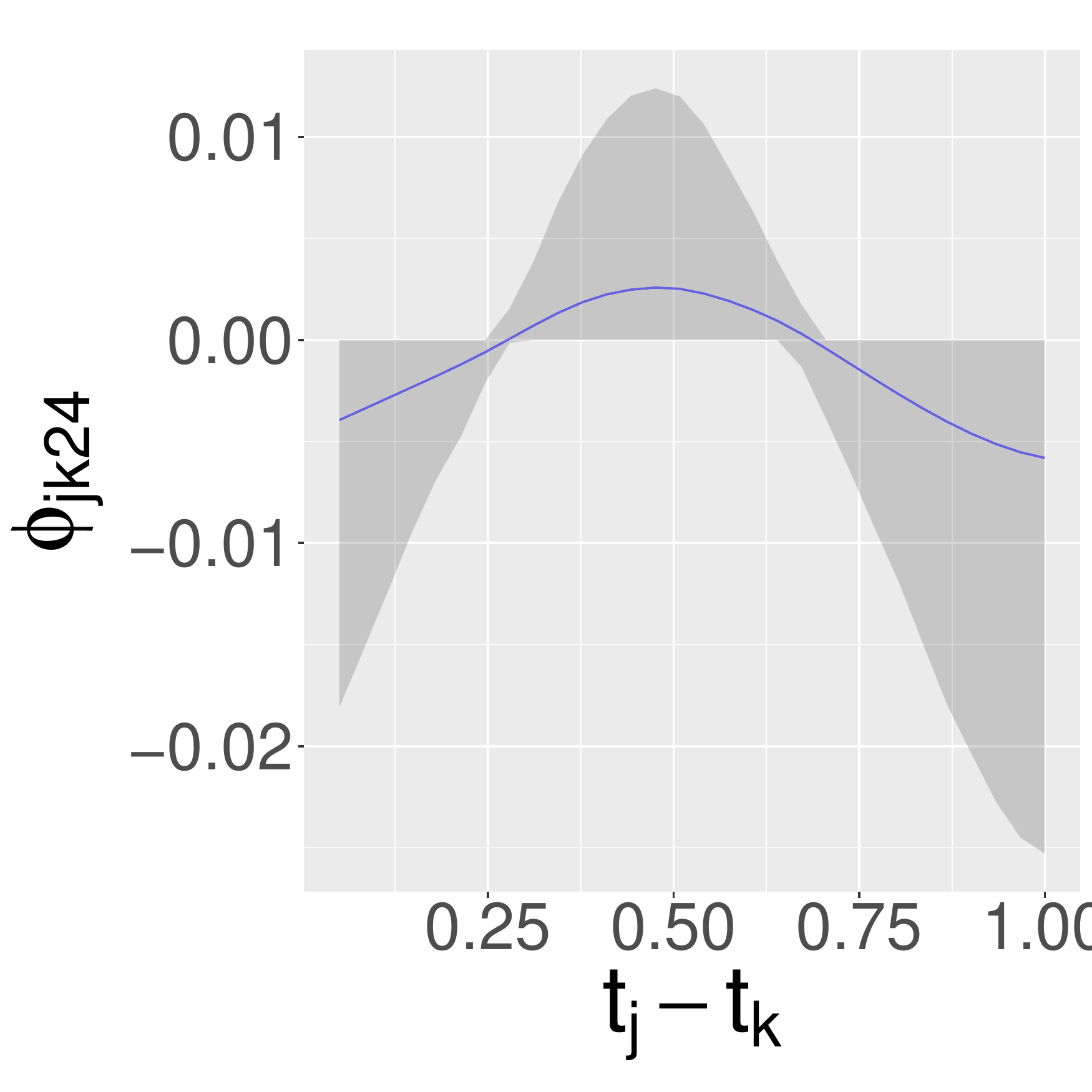} \\    
			\includegraphics[width=0.2\textwidth]{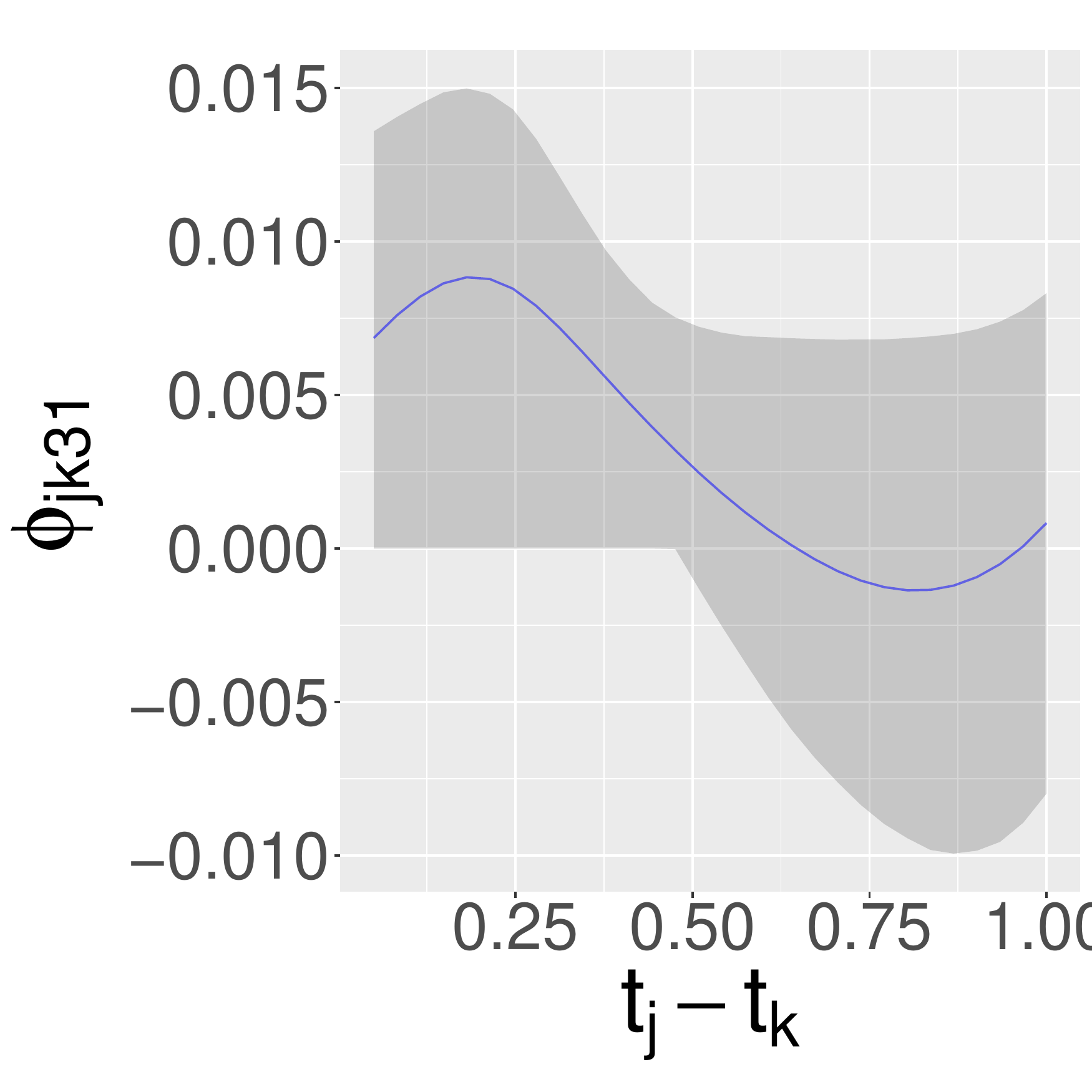} &  
			\includegraphics[width=0.2\textwidth]{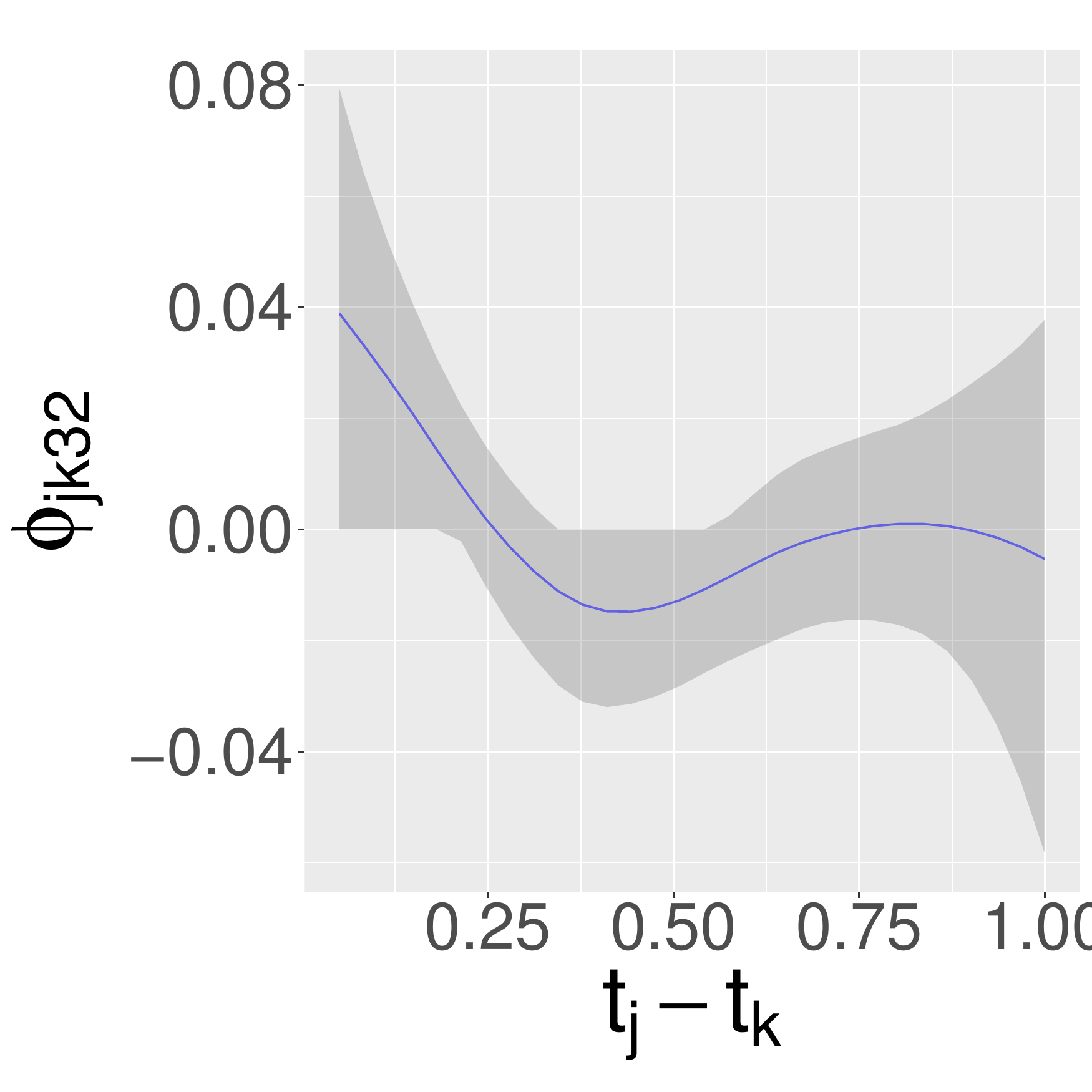} &  
			\includegraphics[width=0.2\textwidth]{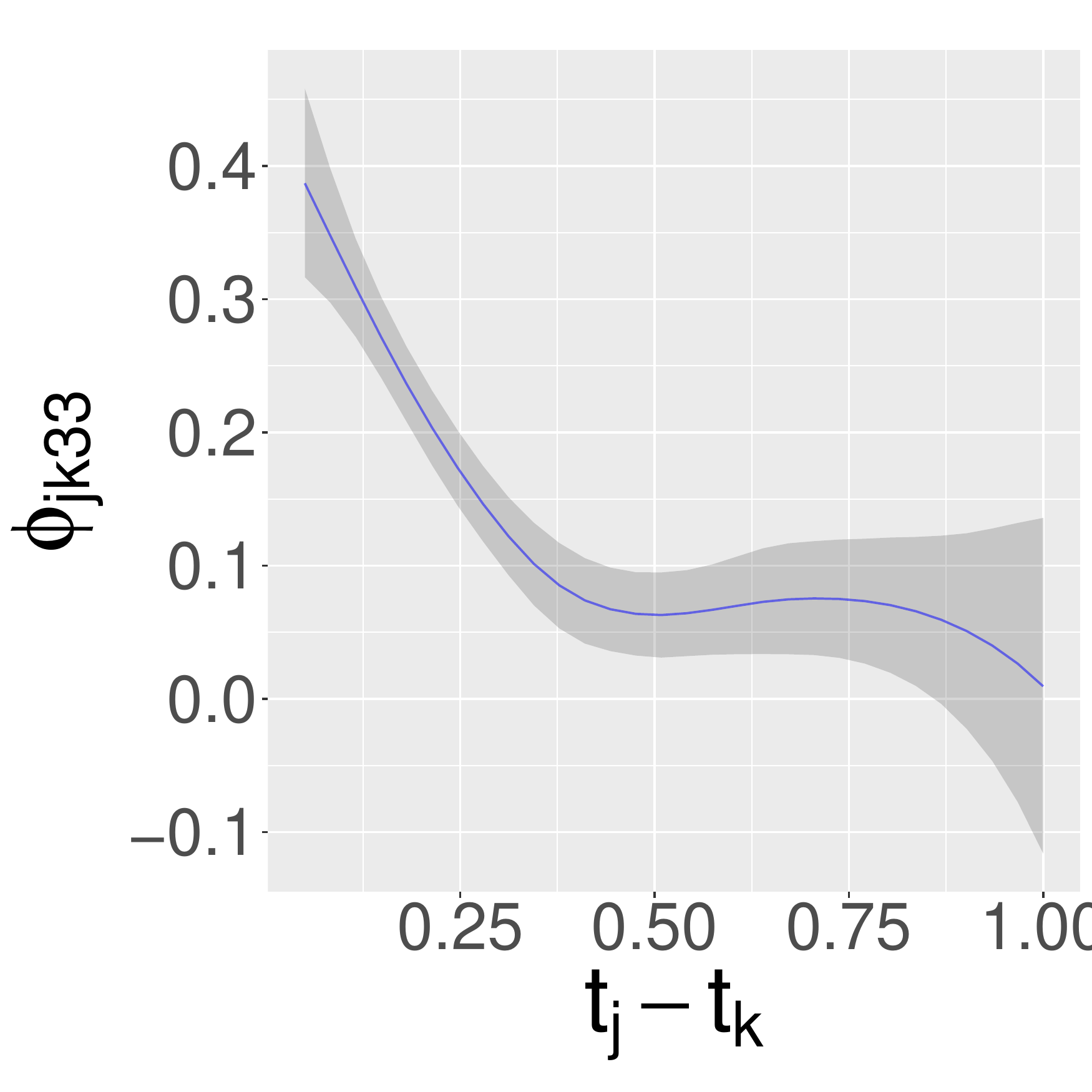} &
			\includegraphics[width=0.2\textwidth]{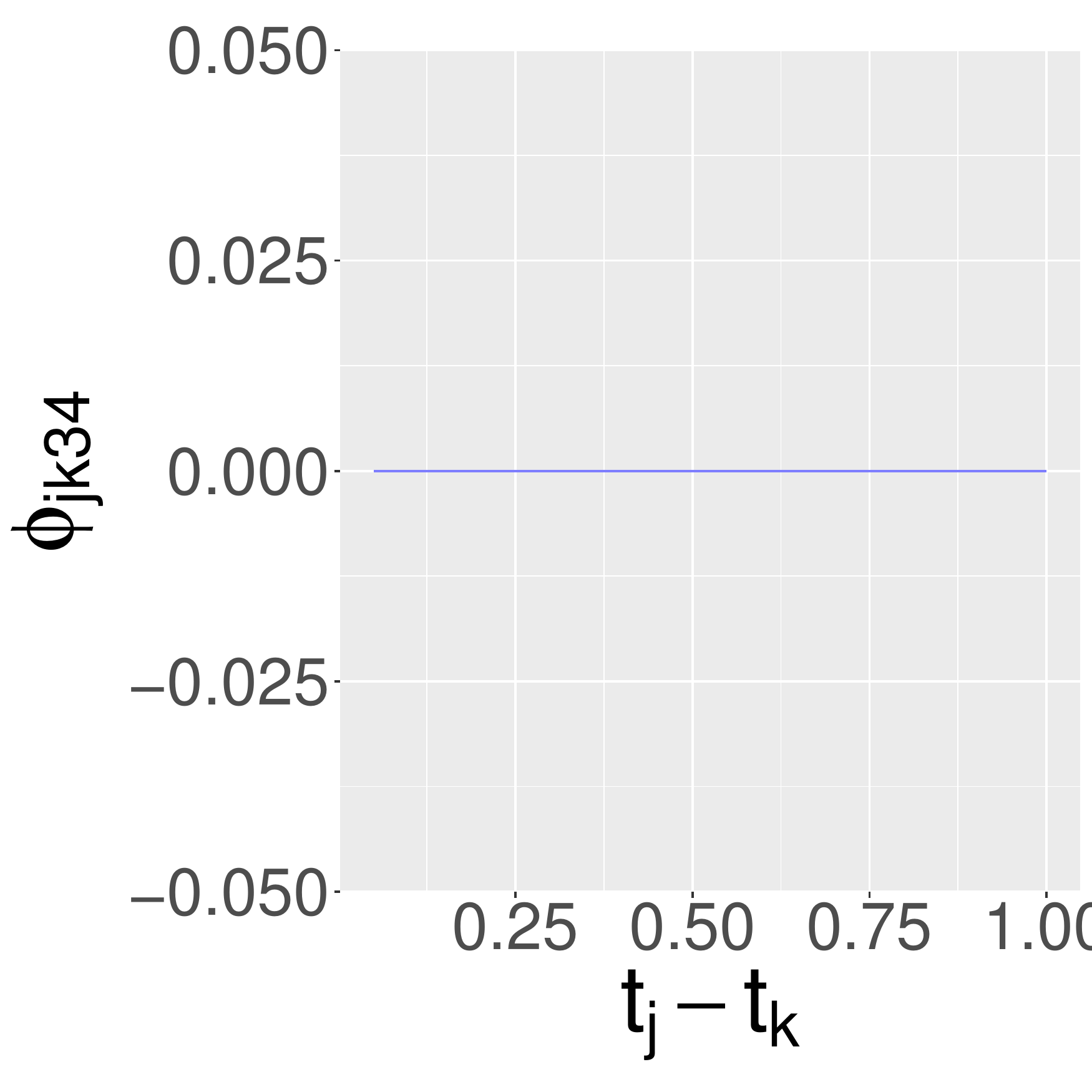} \\    
			\includegraphics[width=0.2\textwidth]{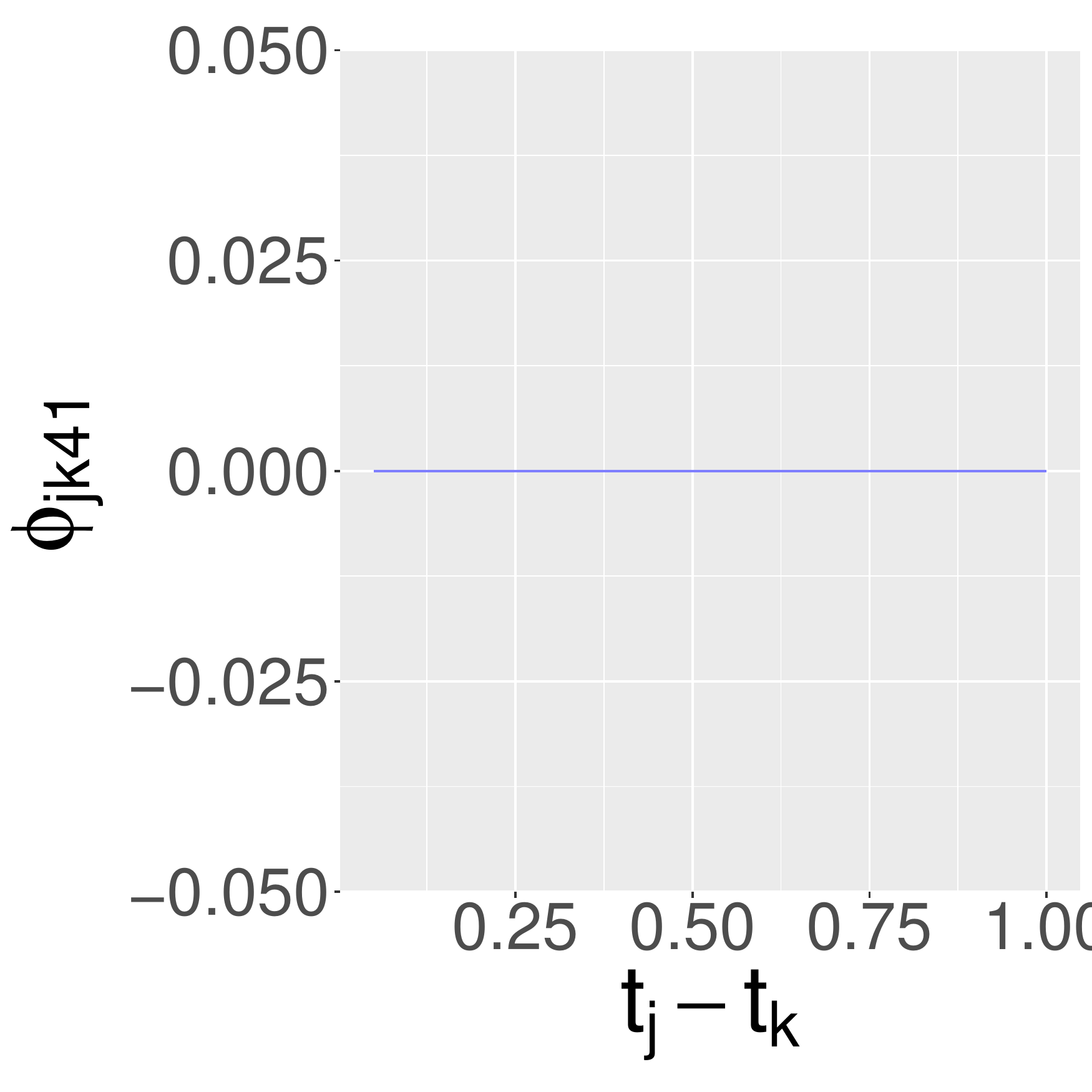} &  
			\includegraphics[width=0.2\textwidth]{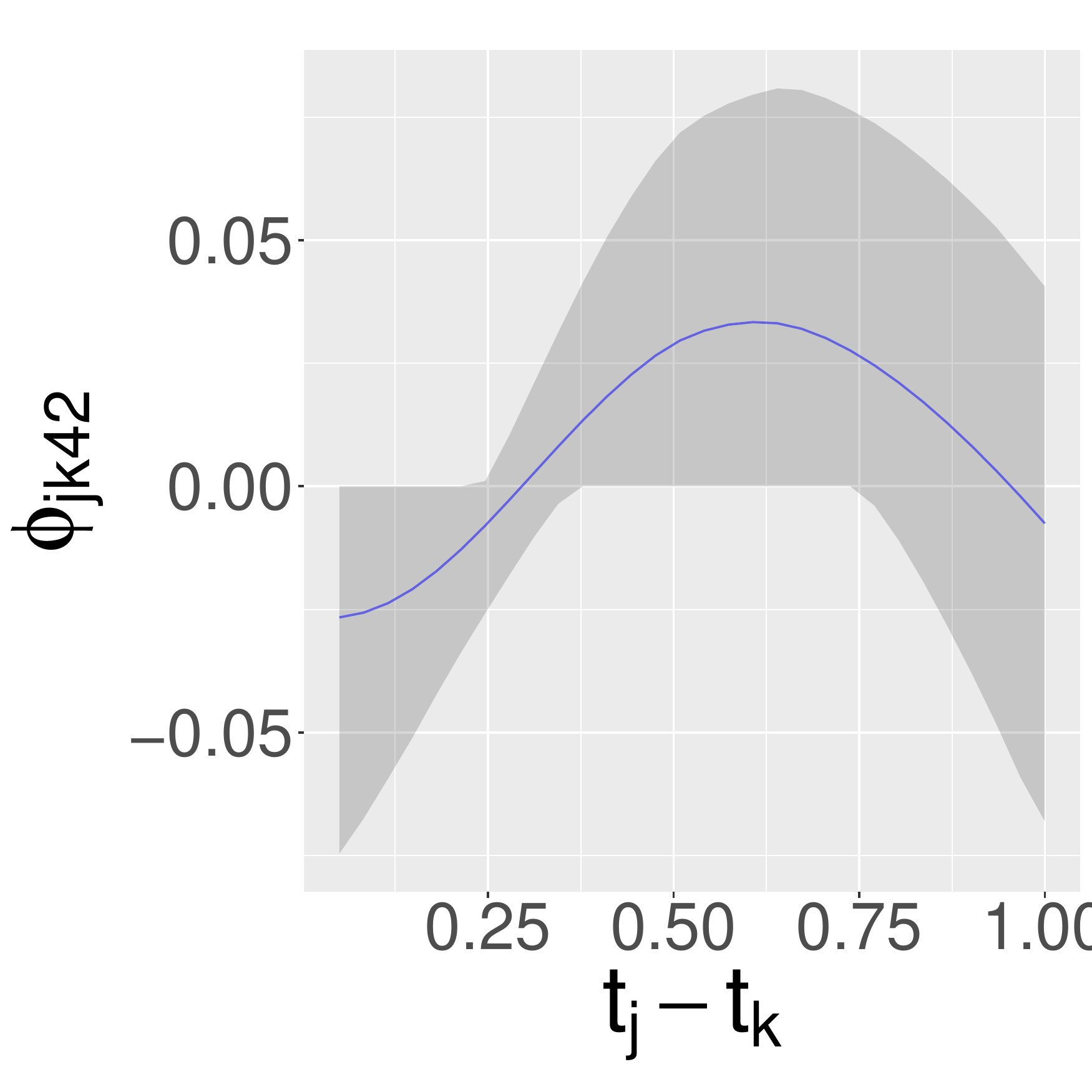} &  
			\includegraphics[width=0.2\textwidth]{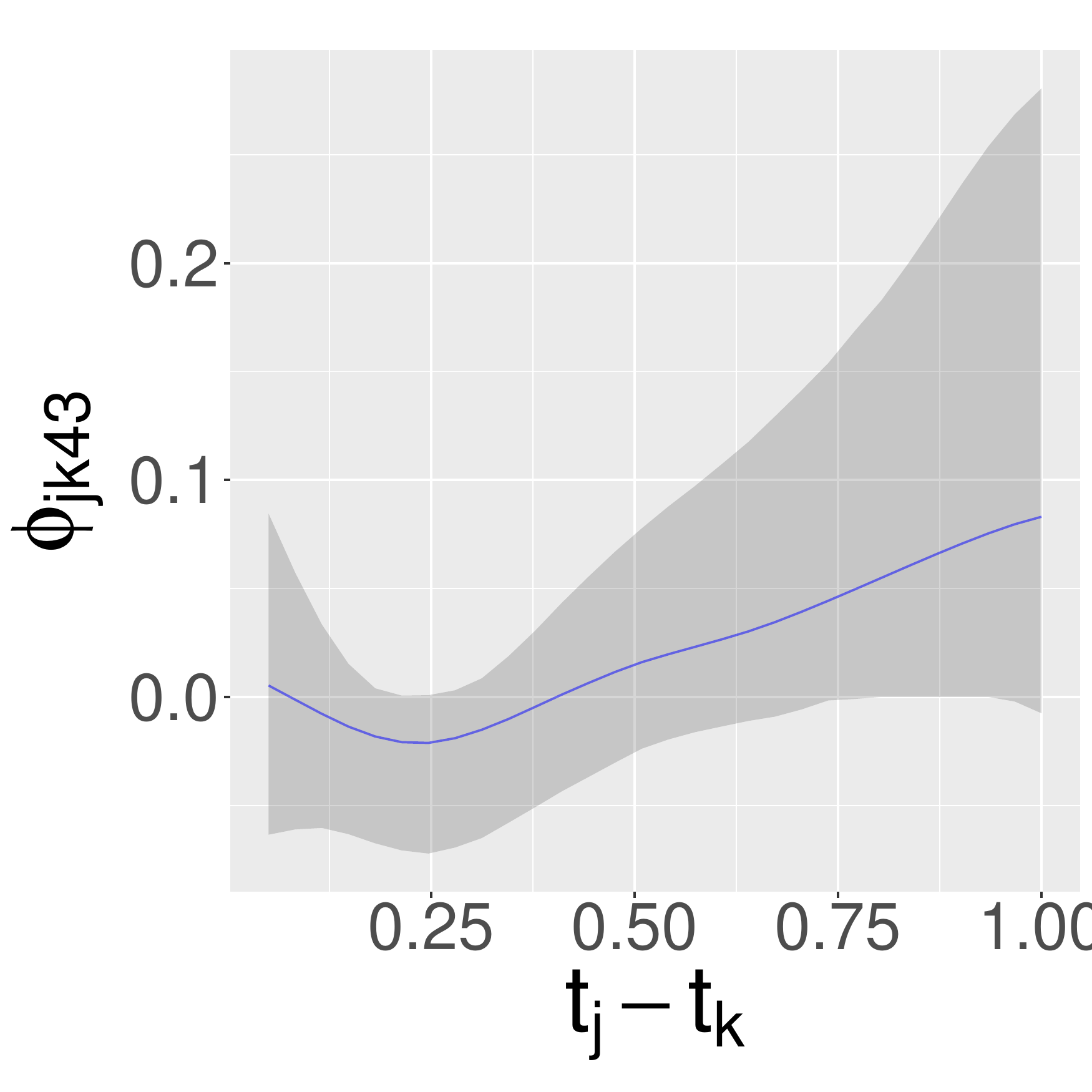} &
			\includegraphics[width=0.2\textwidth]{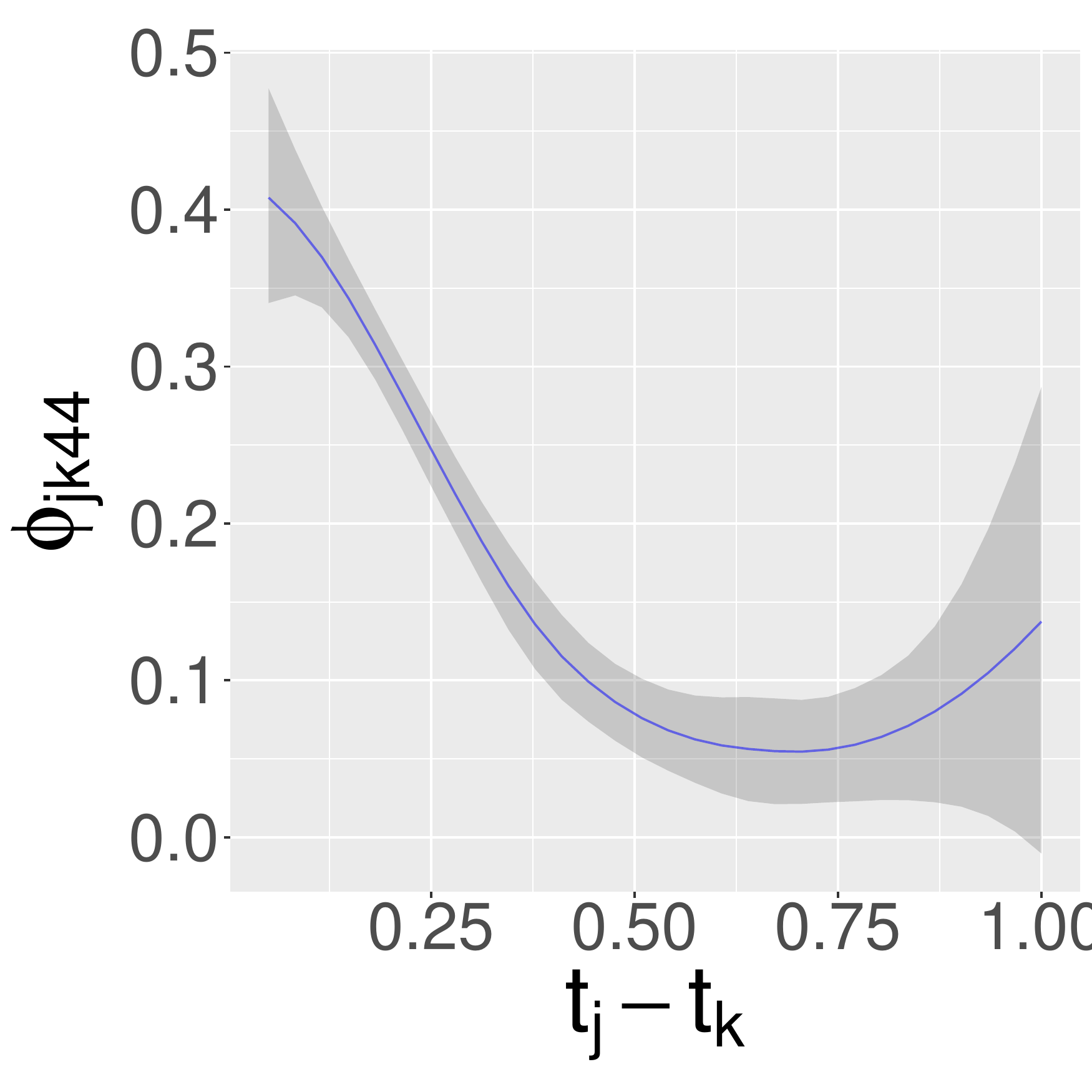} \\    
		\end{tabular}
	\end{center}
	\caption{Application results: autoregressive coefficient regression models. Posterior means and $80\%$ credible  intervals.}\label{Appauto}
\end{figure}

Results on the autoregressive coefficient regression models are shown in Figure \ref{Appauto}. 
Recalling the discussion around (\ref{pred}), the fitted models presented in the diagonal of Figure \ref{Appauto} correspond to 
the coefficients for predicting responses based on passed observations on the same response. These 4 curves have similar shapes: they start from around $0.4$ to $0.5$
and they decrease towards $0$. Based on univariate data analyses, \citet{PanMac06} and \citet{Papg12} 
reported similar curves.
The remaining plots include a variety of fitted curves: complex and clearly important (such as $\phi_{jk12}$ and $\phi_{jk13}$), 
complex but not very important (such as $\phi_{jk14}$ and $\phi_{jk23}$)
and flat at 0 (such as $\phi_{jk34}$ and $\phi_{jk41}$).
For fitting these curves, of the $192$ regression coefficients, on average, $62.6$ or $32.6\%$ were selected during MCMC sampling.

\begin{figure}
	\begin{center}
		\begin{tabular}{cccc}
			\includegraphics[width=0.2\textwidth]{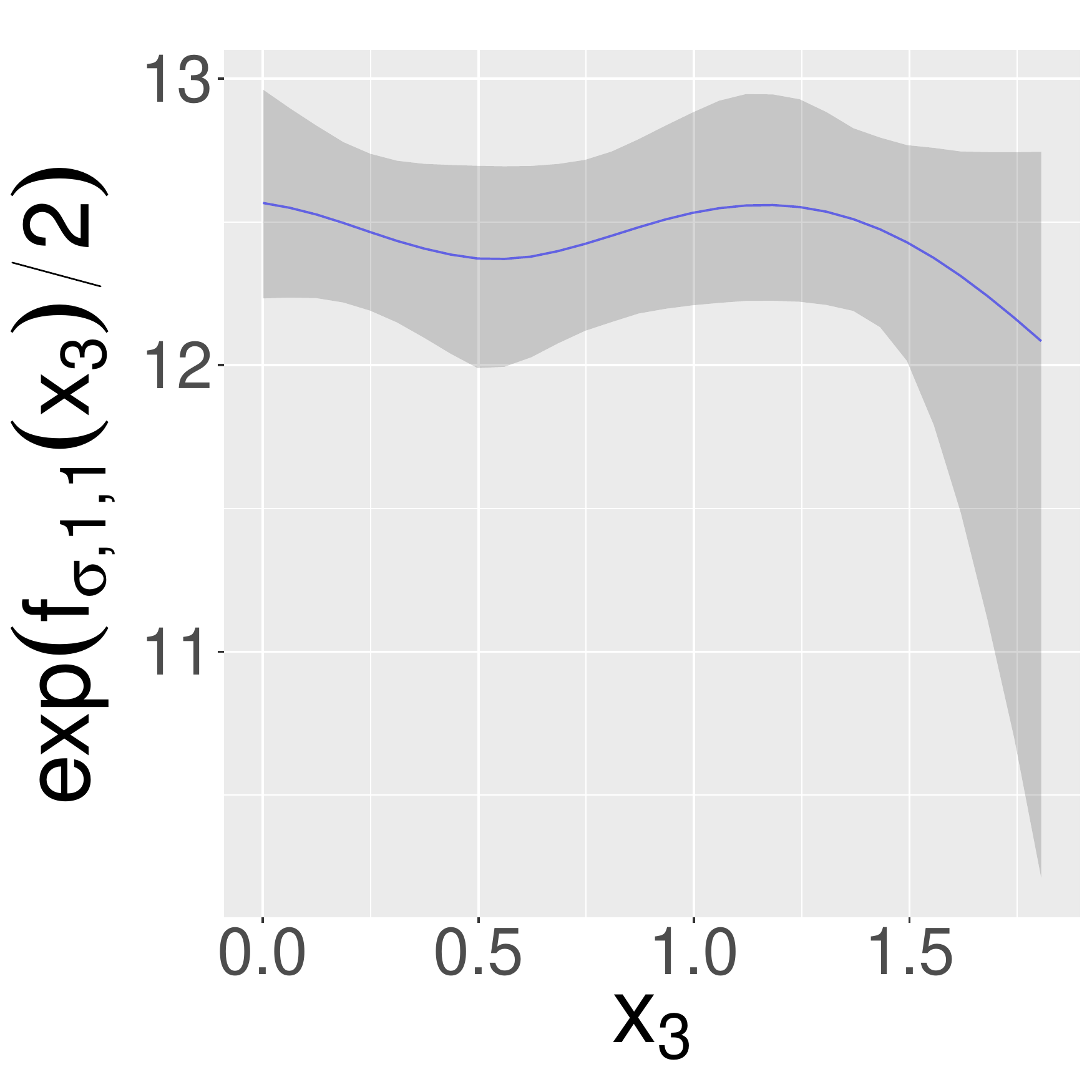} &  
			\includegraphics[width=0.2\textwidth]{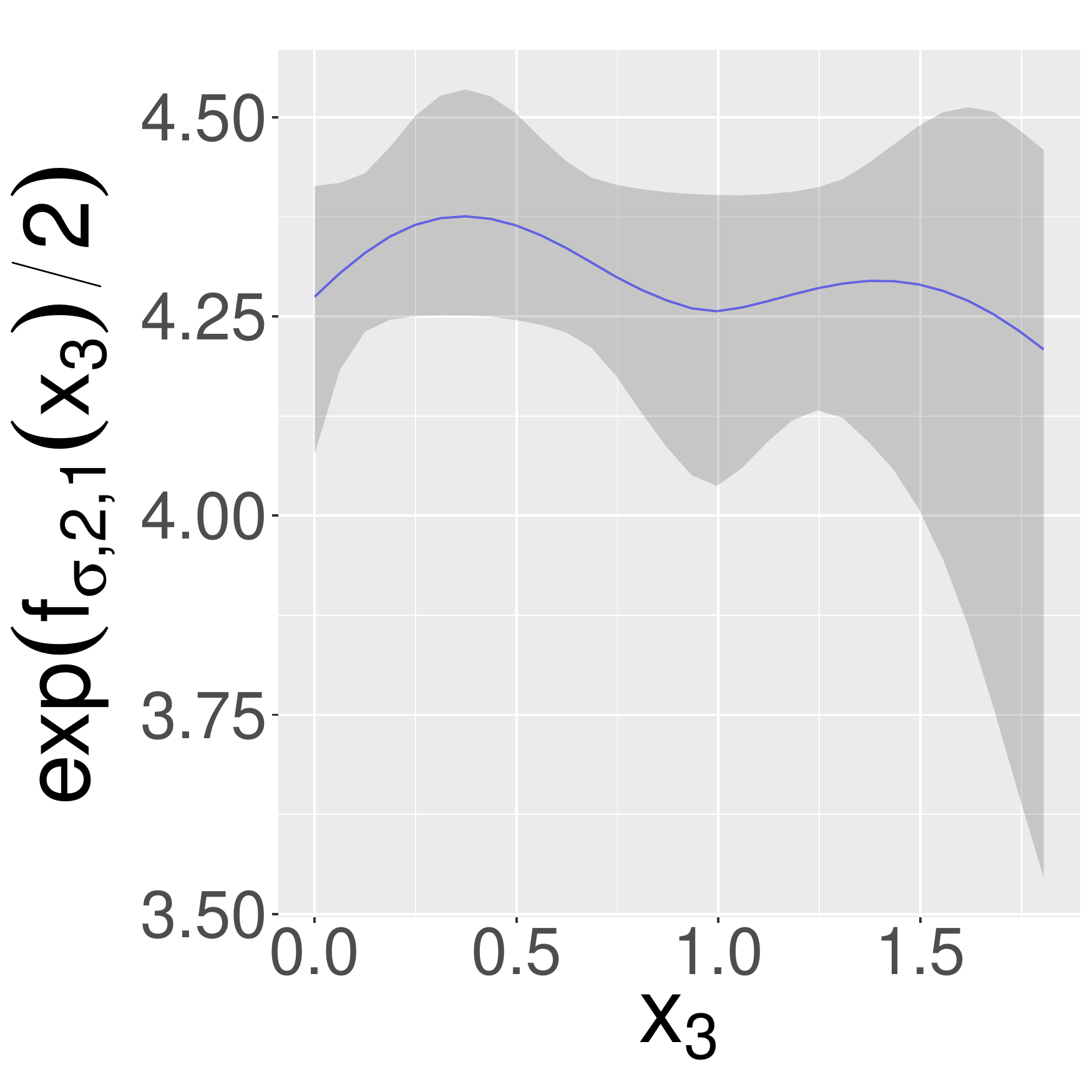} &
			\includegraphics[width=0.2\textwidth]{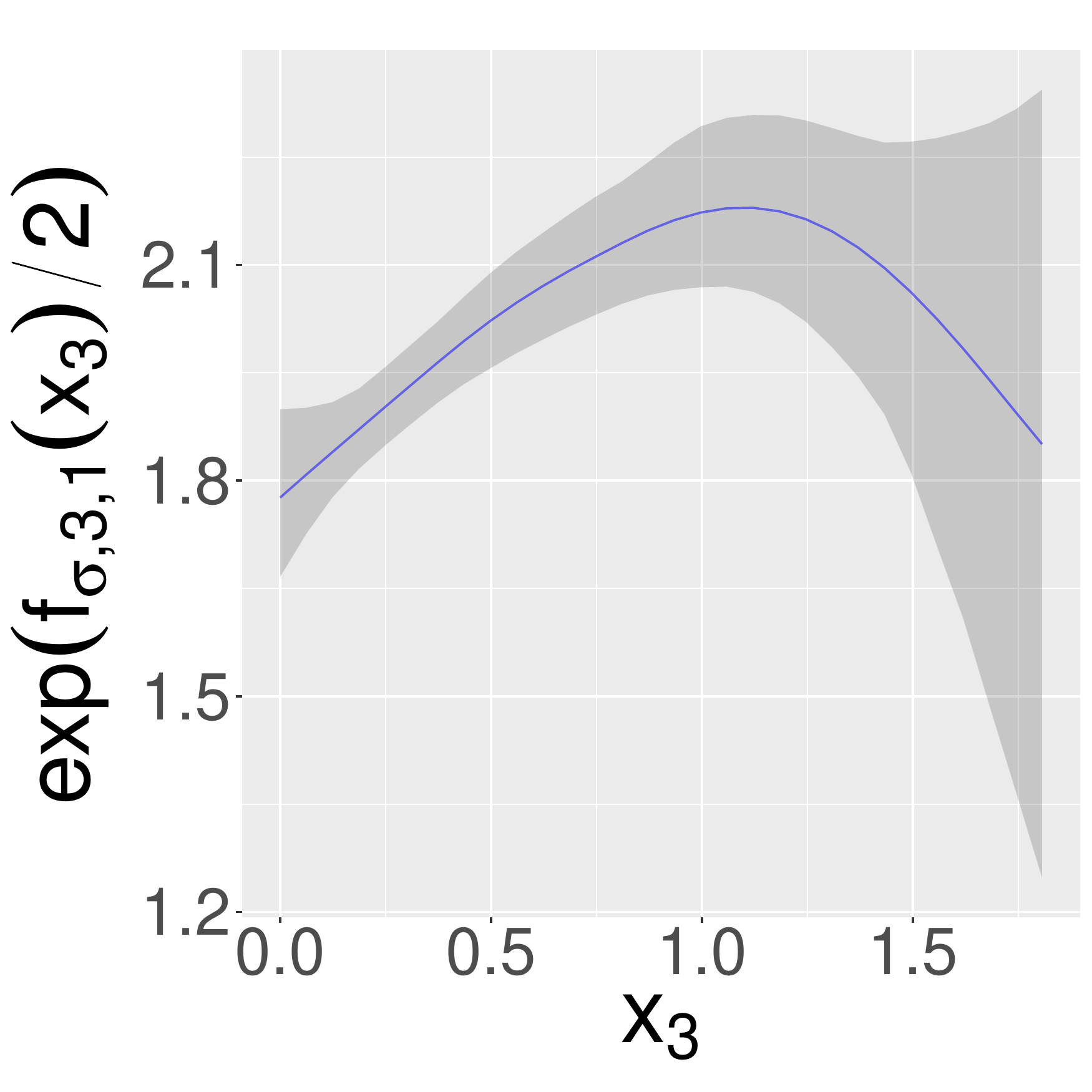} &  
			\includegraphics[width=0.2\textwidth]{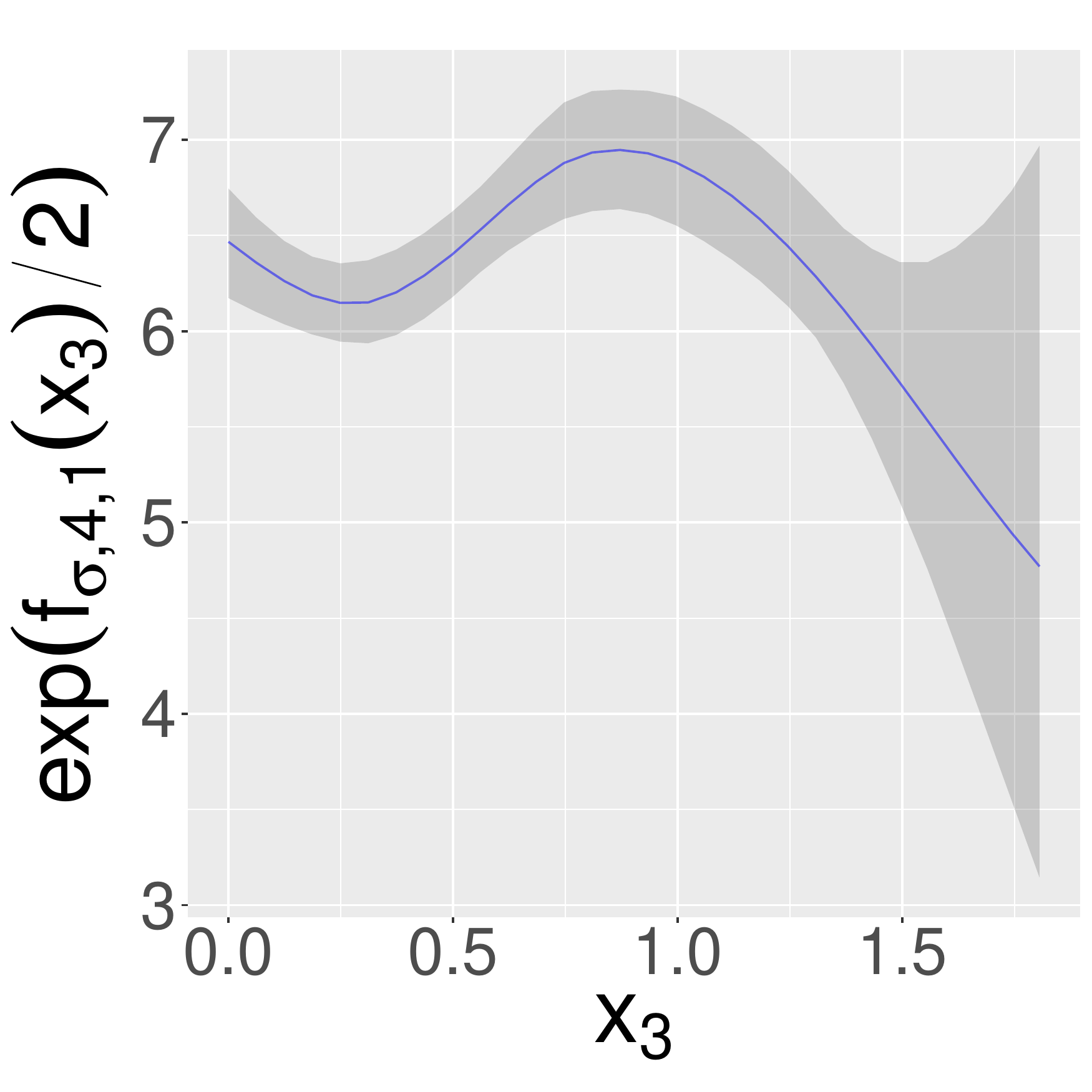} \\
			\includegraphics[width=0.2\textwidth]{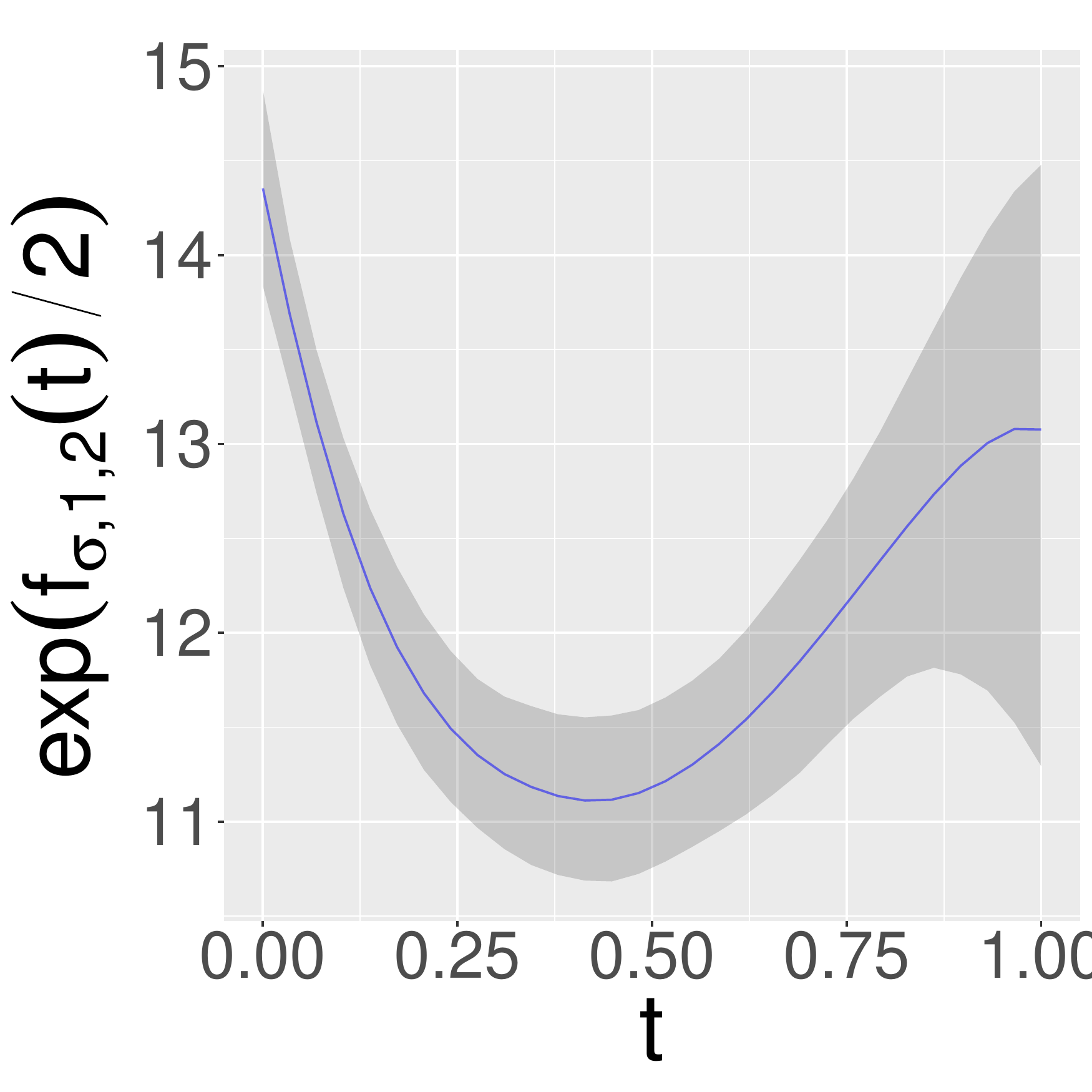} &  
			\includegraphics[width=0.2\textwidth]{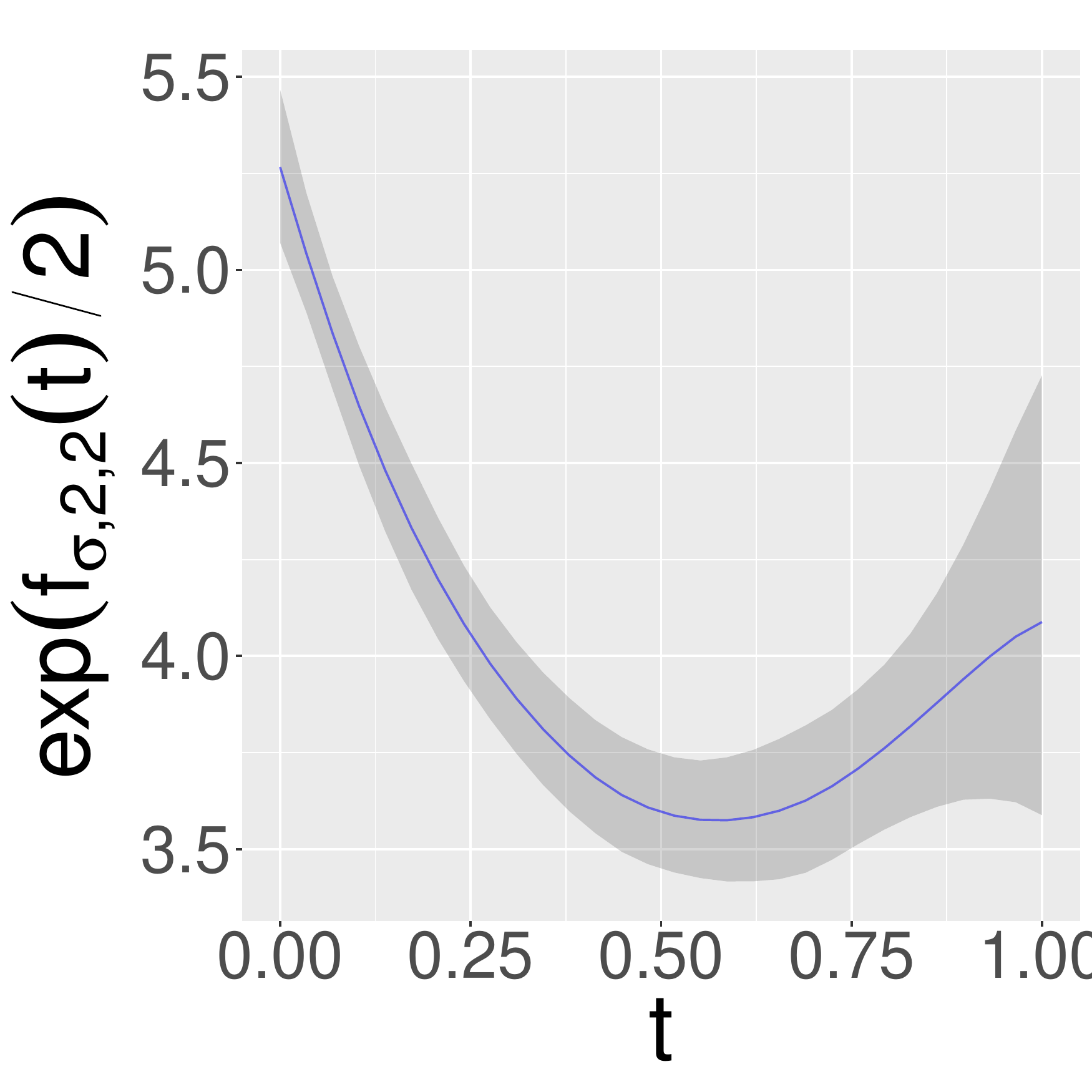} &
			\includegraphics[width=0.2\textwidth]{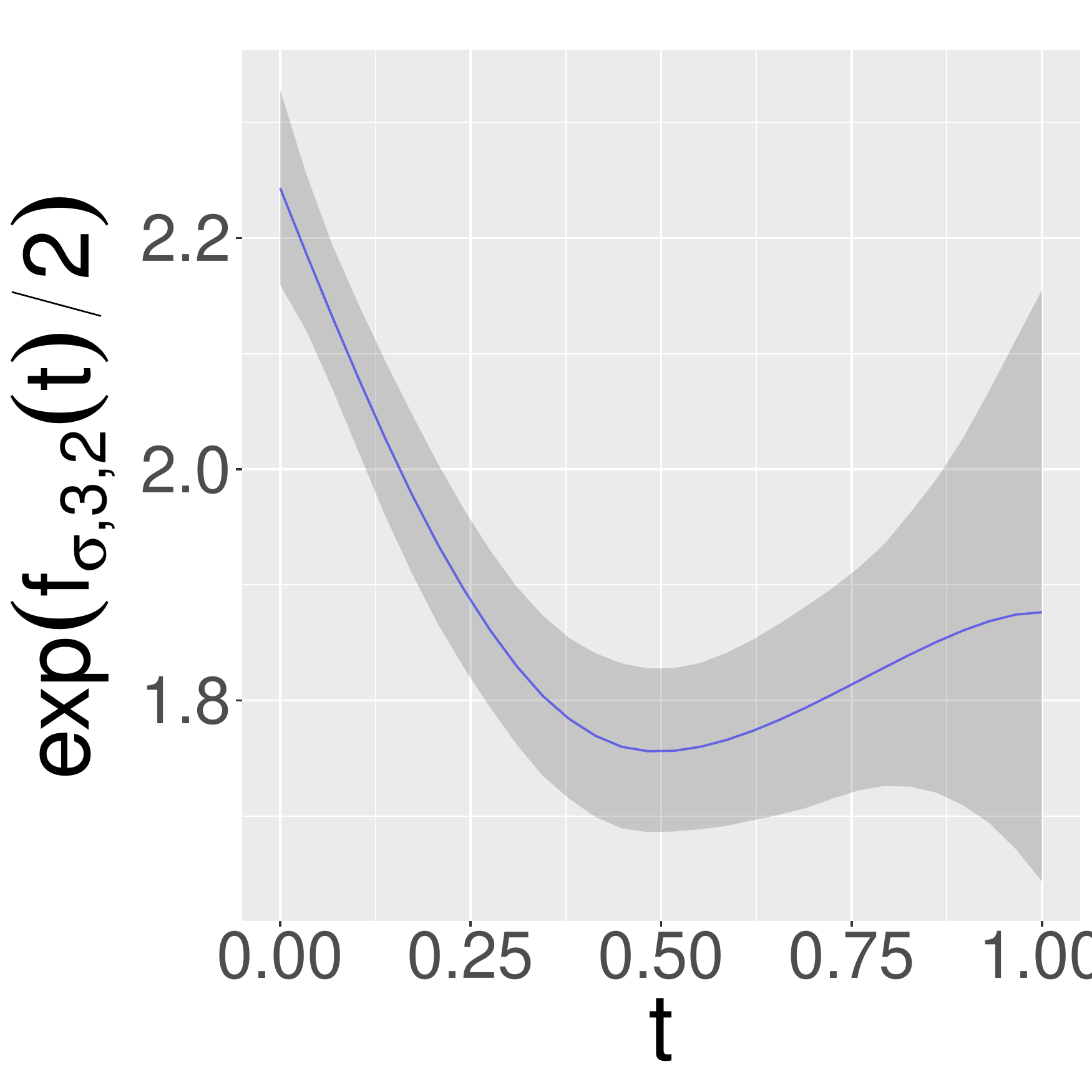} &  
			\includegraphics[width=0.2\textwidth]{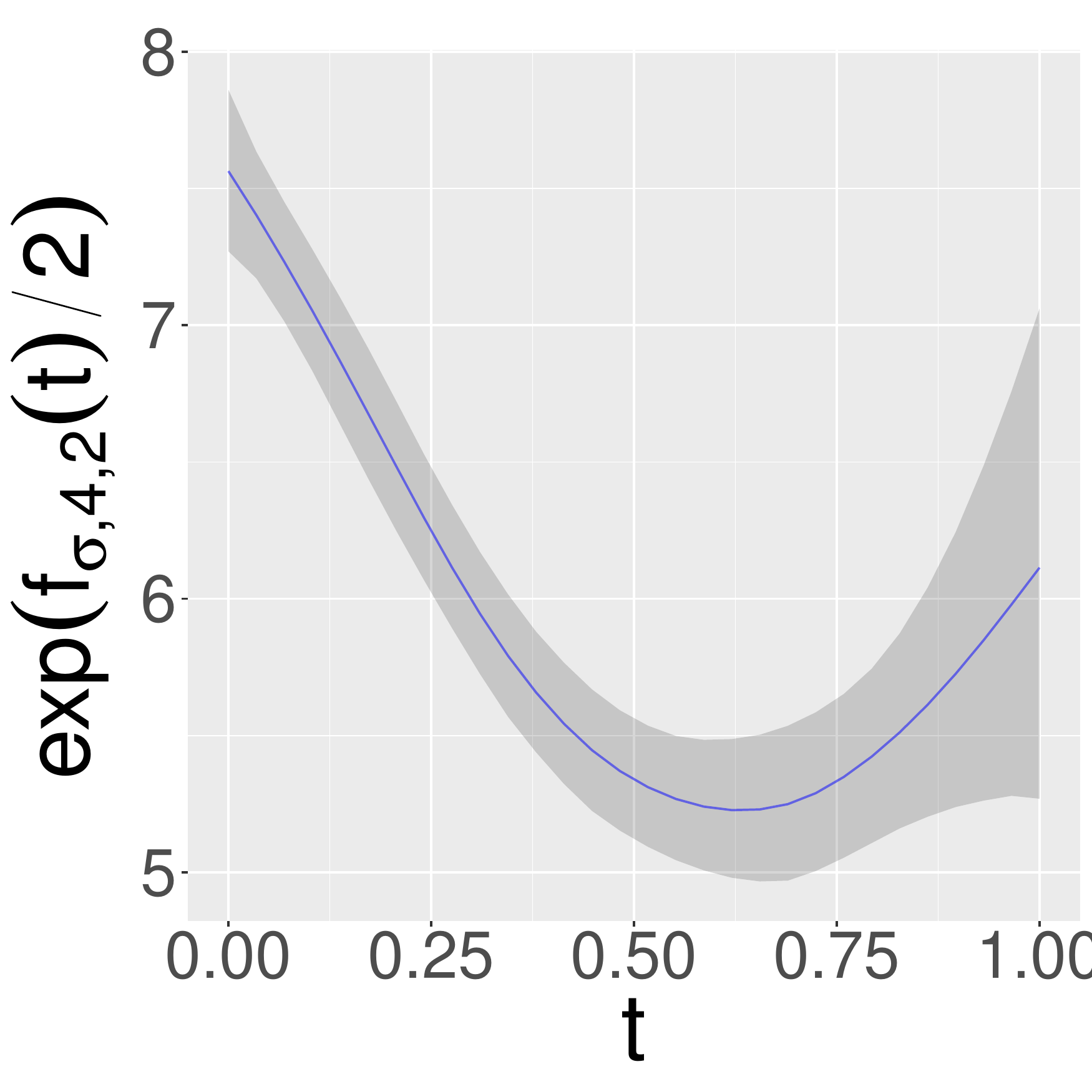} \\
		\end{tabular}
	\end{center}
	\caption{Application results: innovation standard deviation regression models. Posterior means and $80\%$ credible  intervals.
		Rows refer to the covariate effects and columns to the innovation standard deviations.}\label{Appsd}
\end{figure}

Results relating to innovation standard deviations are displayed in Figure \ref{Appsd}. The first row shows almost flat fitted curves 
$\exp(f_{\sigma,1,1}(x_{3})/2)$ and $\exp(f_{\sigma,2,1}(x_{3})/2)$, and more complex fitted curves
$\exp(f_{\sigma,3,1}(x_{3})/2)$ and $\exp(f_{\sigma,4,1}(x_{3})/2)$. The four fitted curves
$\exp(f_{\sigma,k,2}(t)/2)$, shown in the second row of Figure \ref{Appsd}, have similar shapes: they decrease as $t$ increases and they 
are characterized by increasing uncertainty as $t$ increases.          
For fitting these curves, of the $80$ regression coefficients, on average, $29.1$ or $36.4\%$ were selected during MCMC sampling.

\begin{figure}
	\begin{center}
		\begin{tabular}{ccc}
			\includegraphics[width=0.25\textwidth]{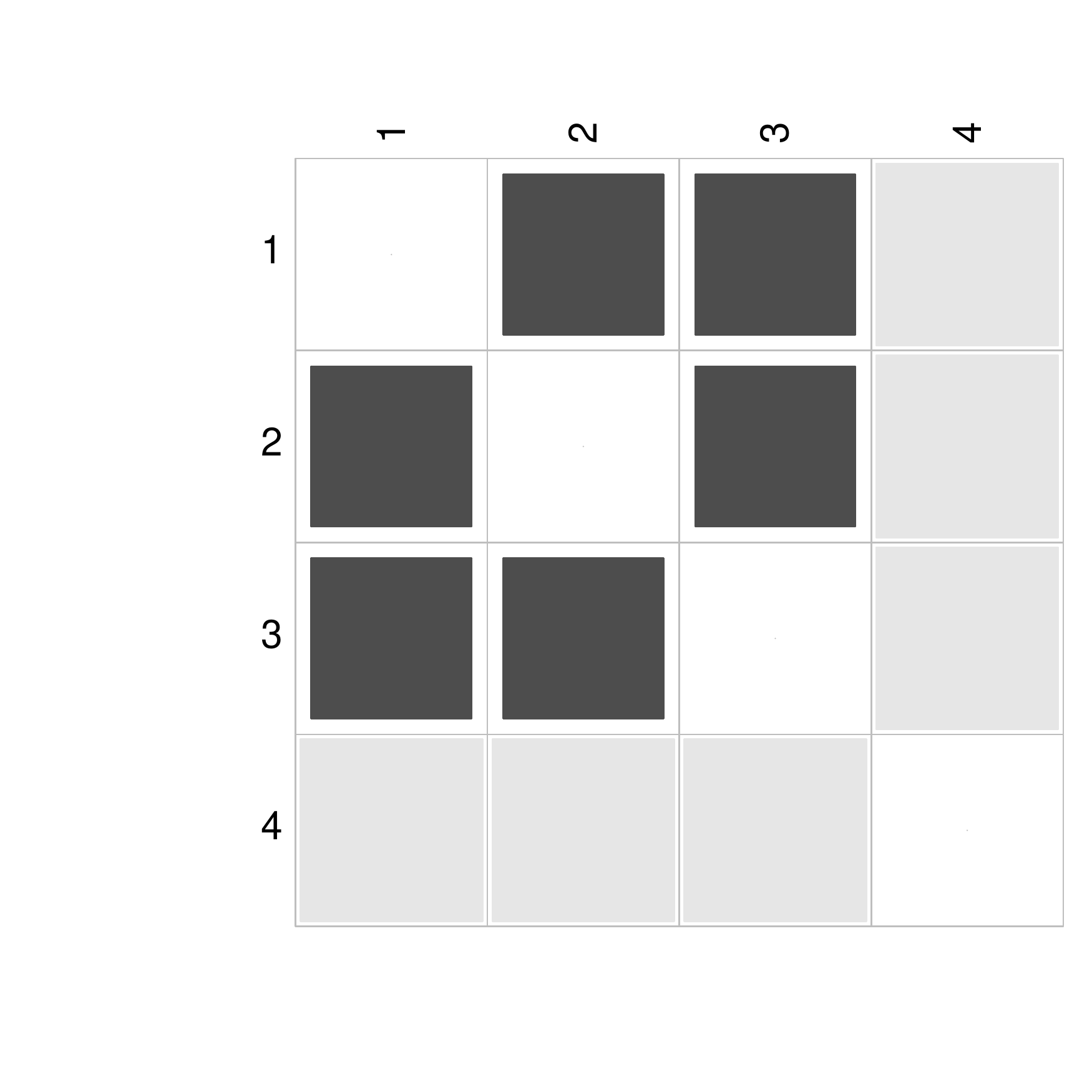} &  
			\includegraphics[width=0.25\textwidth]{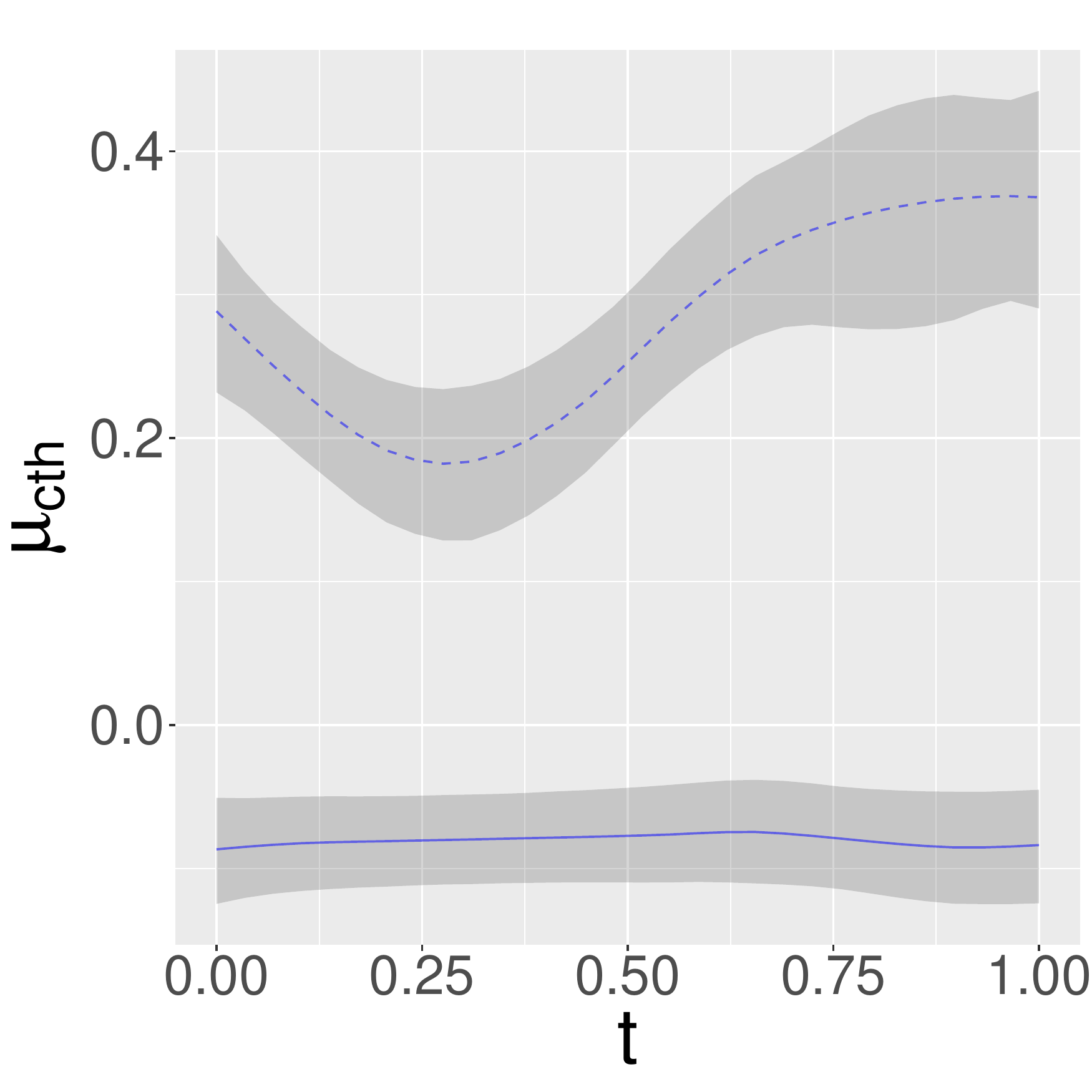} &
			\includegraphics[width=0.25\textwidth]{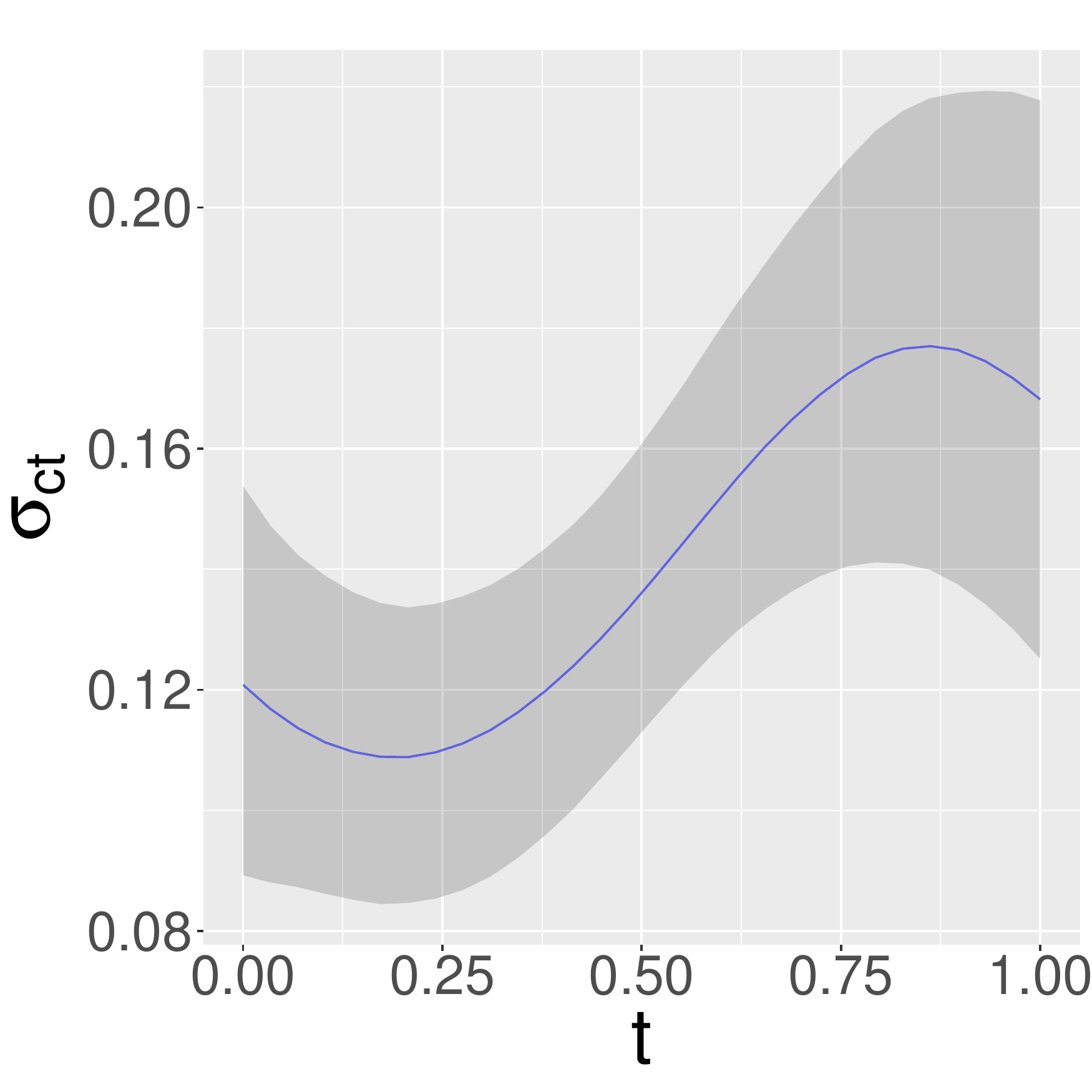} \\
		\end{tabular}
	\end{center}
	\caption{Application results: clustering structure of the elements of the $4 \times 4$ correlation matrices, posterior means and $80\%$ credible  intervals
		for $\mu_{cth}, h=1,2,$ and posterior mean and $80\%$ credible  interval for $\sigma_{ct}$.}\label{Appcor}
\end{figure}

Figure \ref{Appcor} shows the fitted models of the parameters of the innovation correlation matrices. 
The first plot shows the clustering structure that the MCMC sampler visited most often: 
for $94.5\%$ of the MCMC samples, two clusters were formed, the first being $\{(1,2), (1,3), (2,3) \}$ (denoted by the darker colour on the plot) 
while the second being $\{(1,4), (2,4), (3,4) \}$ (denoted by the lighter colour). 
While this clustering was obtained by a `grouped correlations' model, it is also compatible with the  
`grouped variables' model, the first cluster consisting of variables $1,2,3$ (the results on the cognitive tests) and the second consisting of  
variable $4$ (the test result on depressive symptomatology). The other 2 plots in the Figure show the fitted curves for $\mu_{cth}, h=1,2$,
and $\sigma_{ct}$. The fitted curve $\mu_{t1}$ starts at around $0.3$, decreases below $0.2$ and then increases to just below $0.4$. 
The fitted curve $\mu_{t2}$ is much simpler: it is constant just below $0$ for all time points. 
For fitting these curves, of the $66$ regression coefficients, on average, $23.7$ or $35.9\%$ were selected during MCMC sampling.
The fitted smooth curve for $\sigma_{ct}$ is shown in the third plot of Figure \ref{Appcor}. 
For fitting this curve, of the $11$ regression coefficients, on average, $5.7$ or $51.4\%$ were selected during MCMC sampling.

Hence, overall, for fitting the $3570$ unique elements of the $84 \times 84$ covariance matrix, the model selects, on average, $132.1$  
of the available $360$ parameters, per MCMC iteration. 

\begin{figure}
	\begin{center}
		\begin{tabular}{cccc}
			\includegraphics[width=0.20\textwidth]{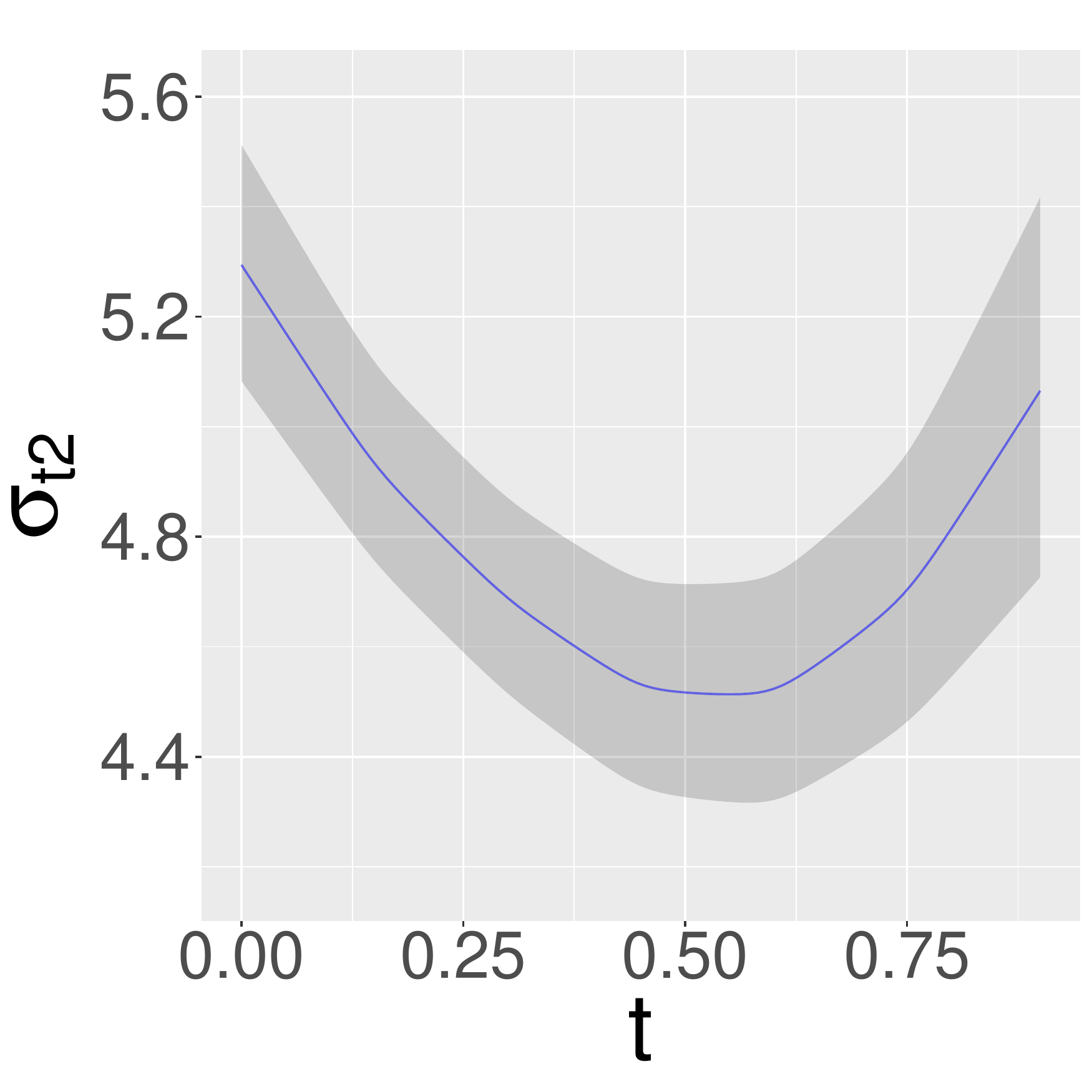} &  \includegraphics[width=0.20\textwidth]{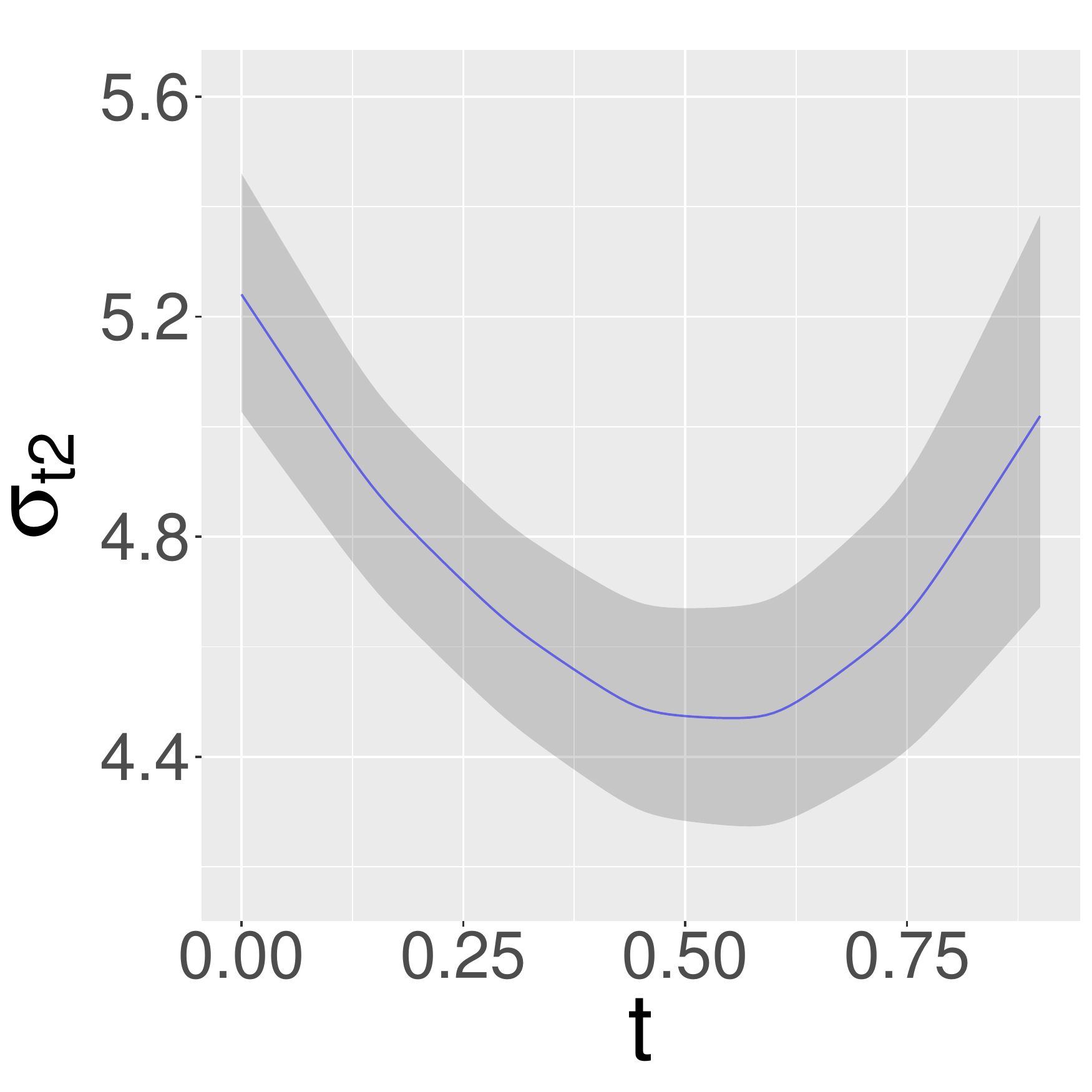} &
			\includegraphics[width=0.20\textwidth]{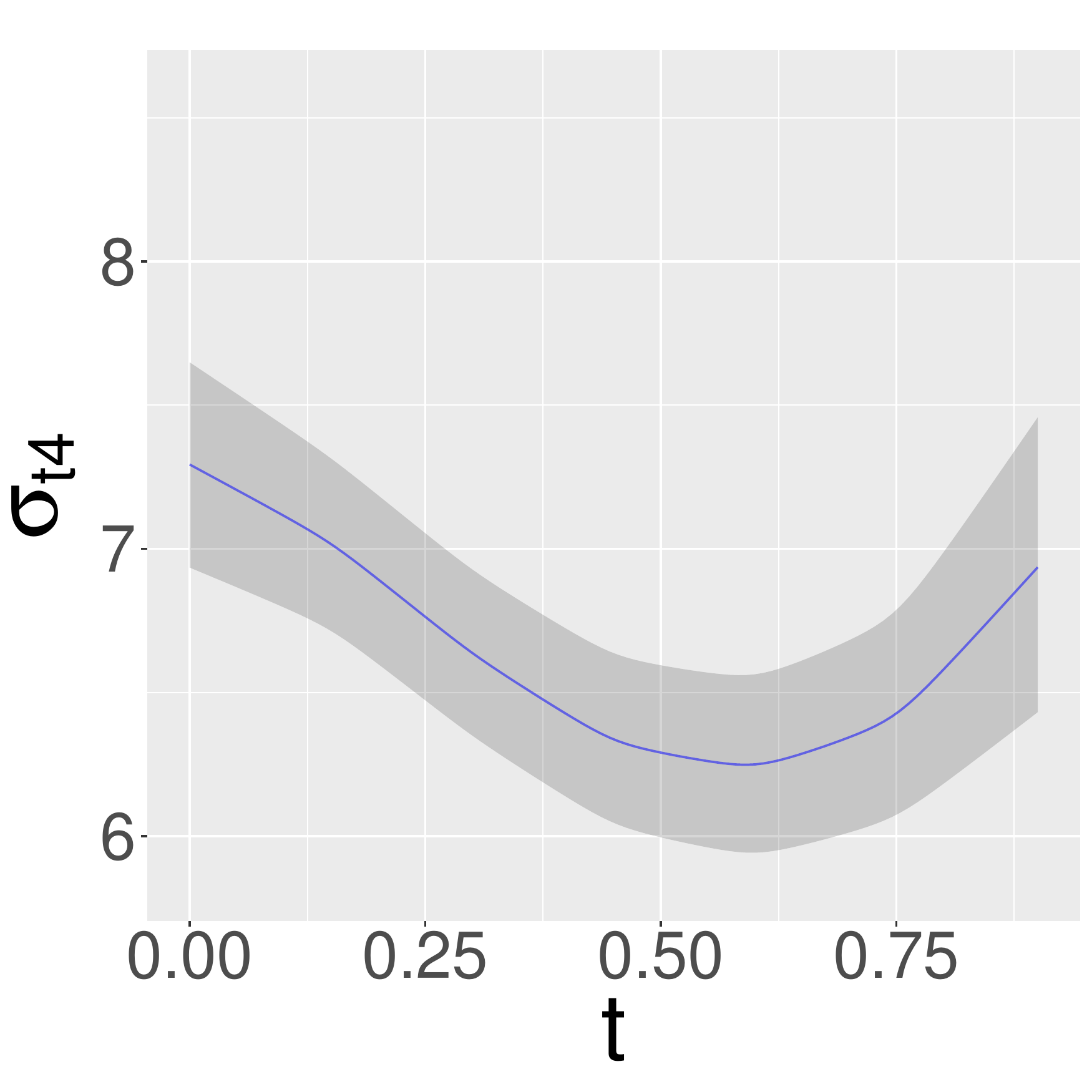} &  \includegraphics[width=0.20\textwidth]{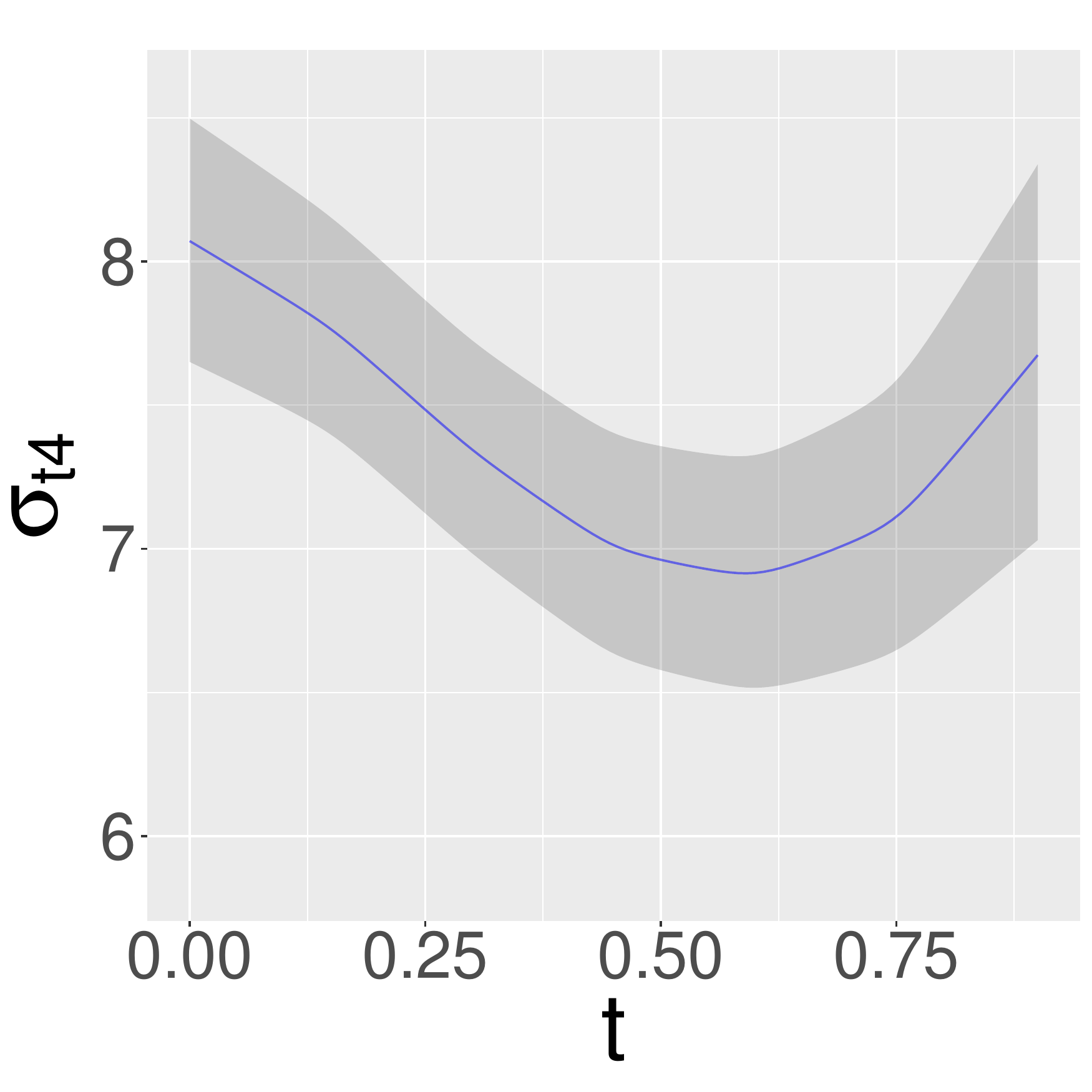} \\
		\end{tabular}
	\end{center}
	\caption{Application results: standard deviations of $Y_2$ and $Y_4$ over time, for ages $x_3=0.18$ and $x_3=0.73$, respectively. Posterior means and $80\%$ credible  intervals.}\label{AppSDfromSigma}
\end{figure}

\begin{figure}
	\begin{center}
		\begin{tabular}{ccc}
			\includegraphics[width=0.20\textwidth]{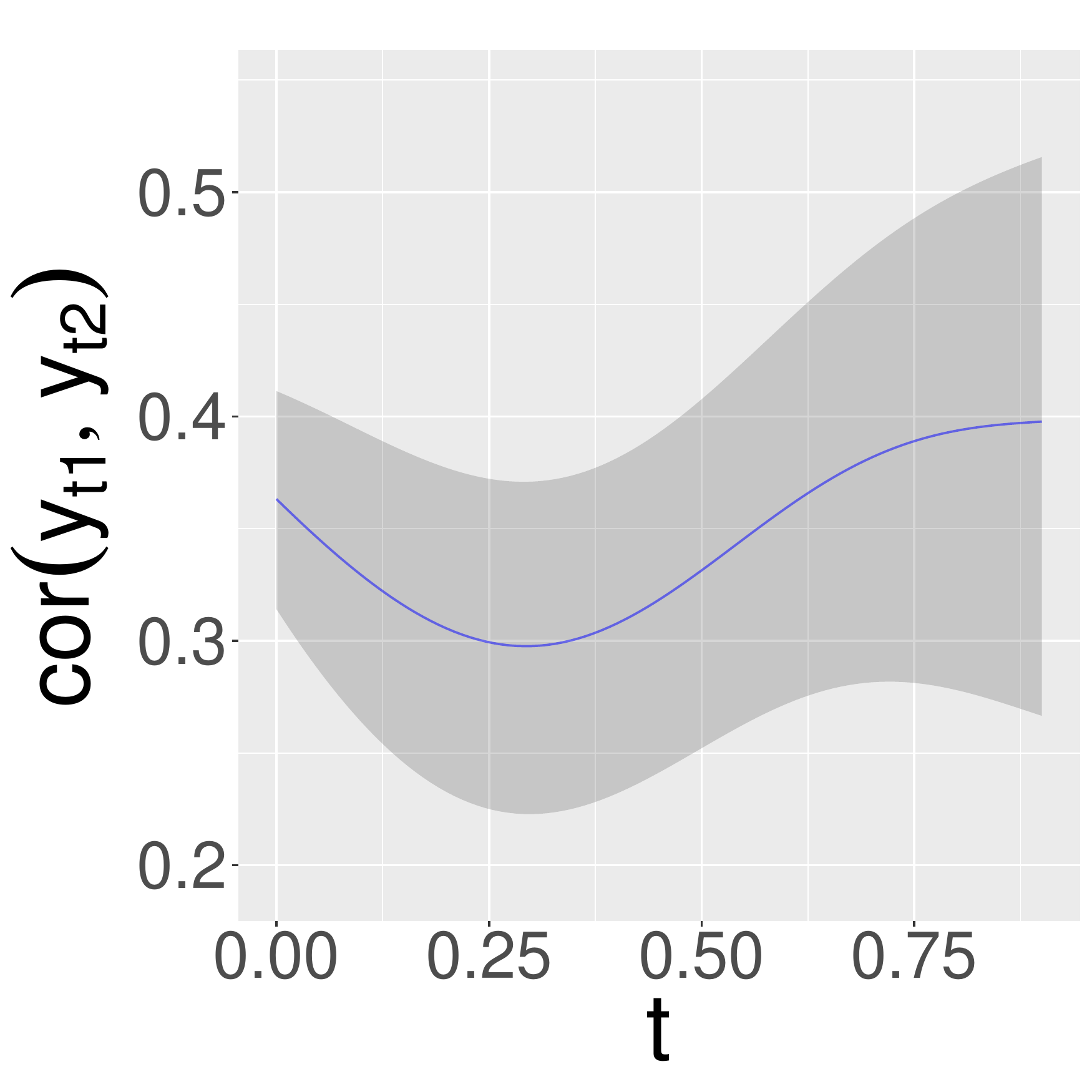} &  \includegraphics[width=0.20\textwidth]{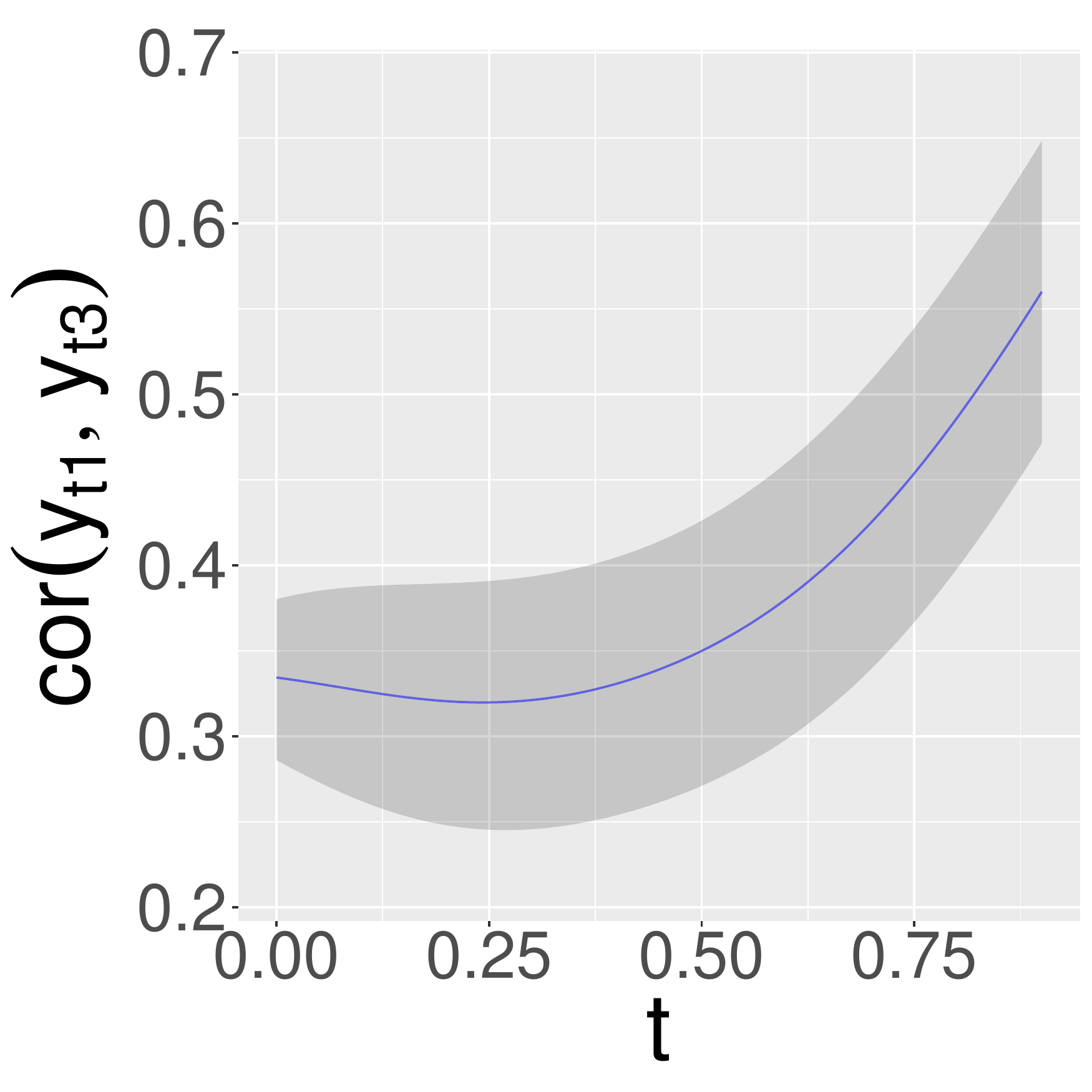} & \includegraphics[width=0.20\textwidth]{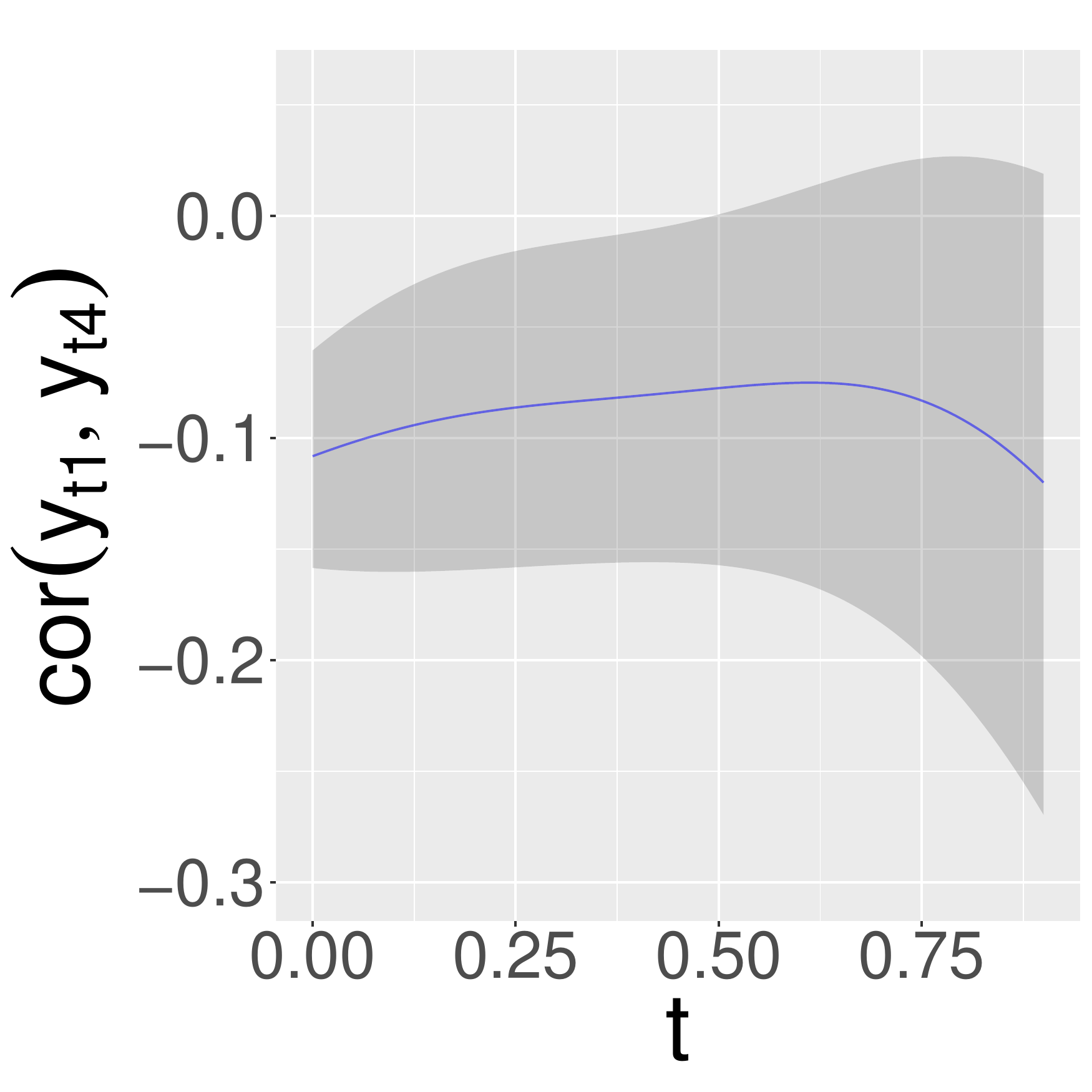} \\ 
			\includegraphics[width=0.20\textwidth]{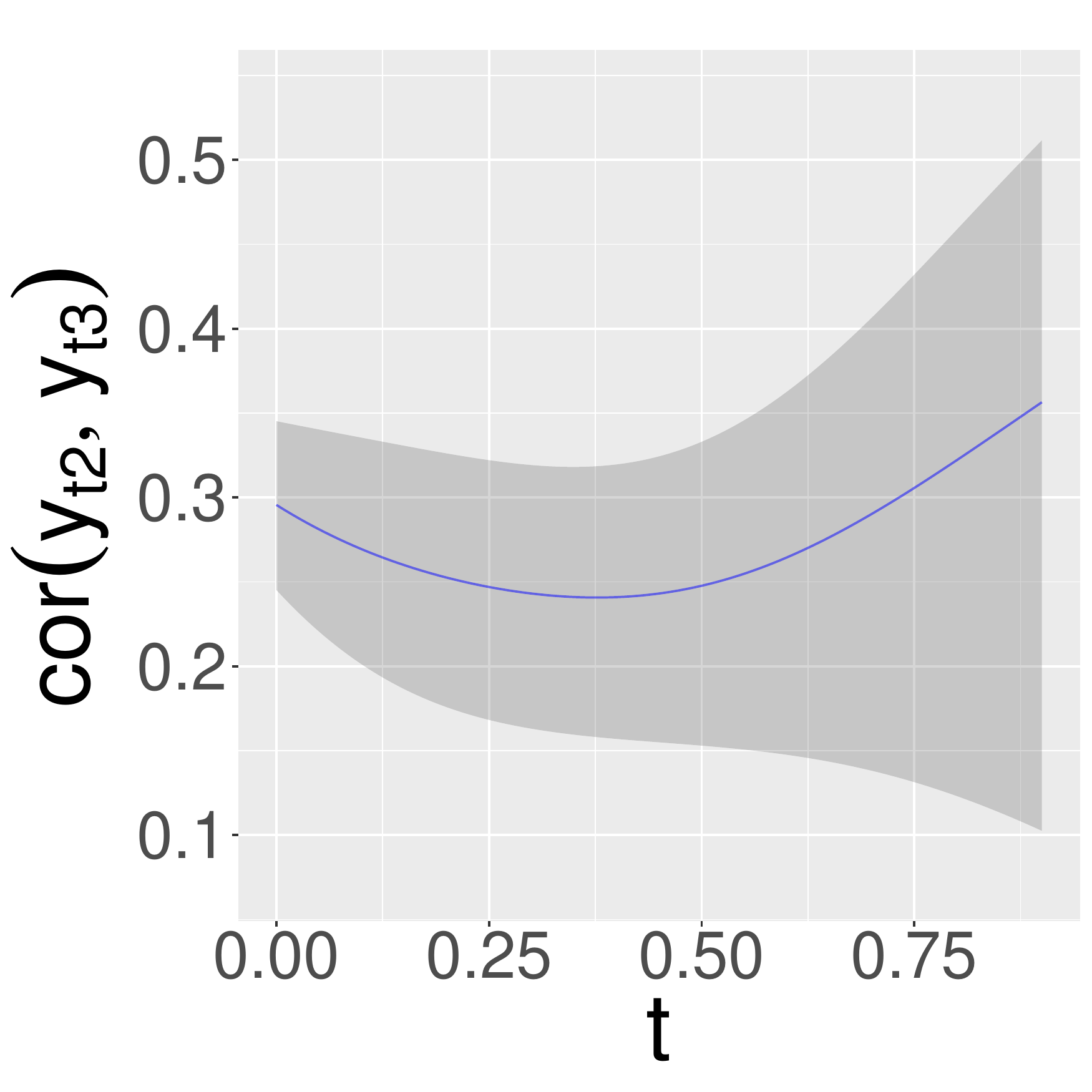} &  \includegraphics[width=0.20\textwidth]{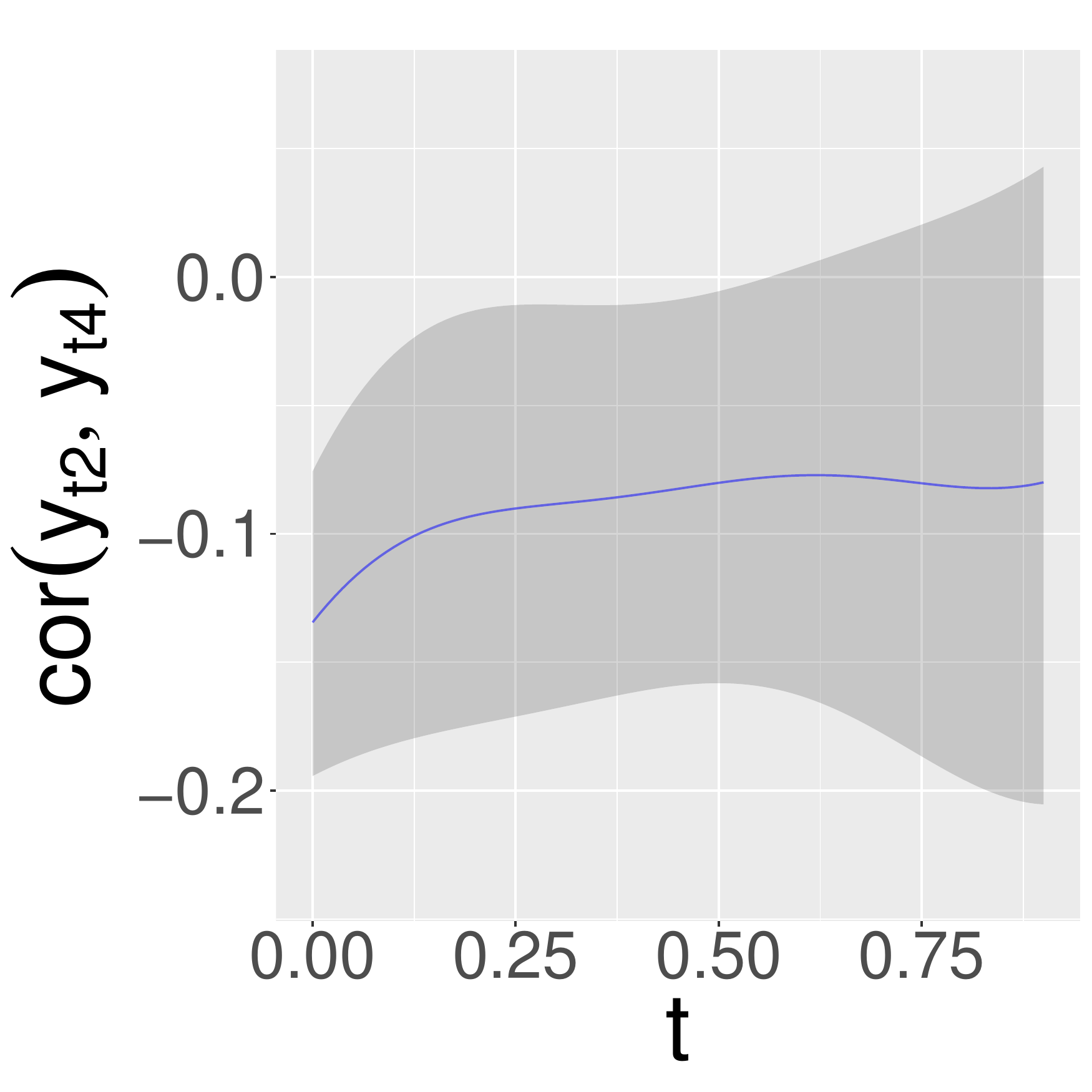} & \includegraphics[width=0.20\textwidth]{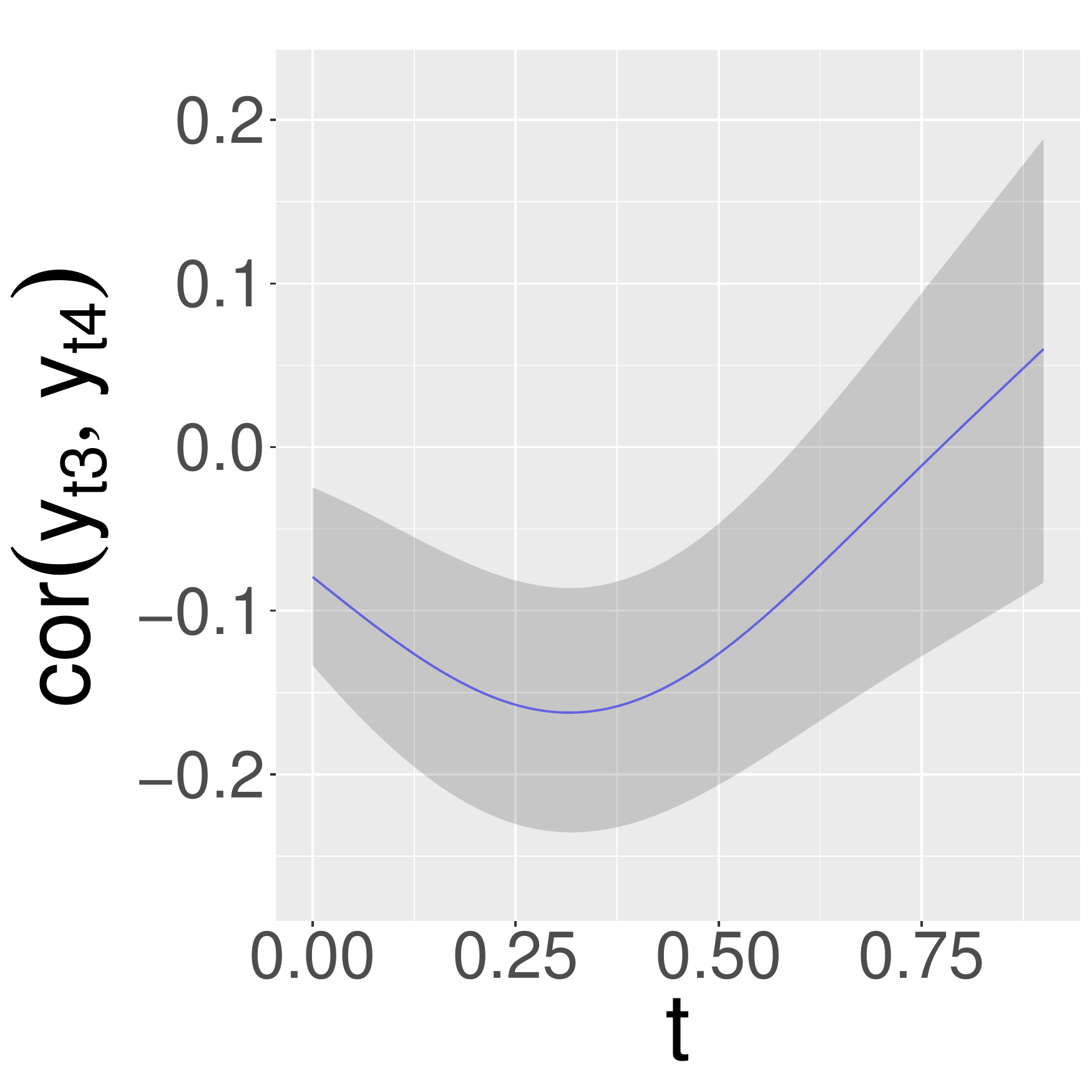} \\ 
		\end{tabular}
	\end{center}
	\caption{Application results: correlations over time. Posterior means and $80\%$ credible  intervals.}\label{AppCorfromSigma}
\end{figure}

\begin{figure}
	\begin{center}
		\begin{tabular}{cccc}
			\includegraphics[width=0.20\textwidth]{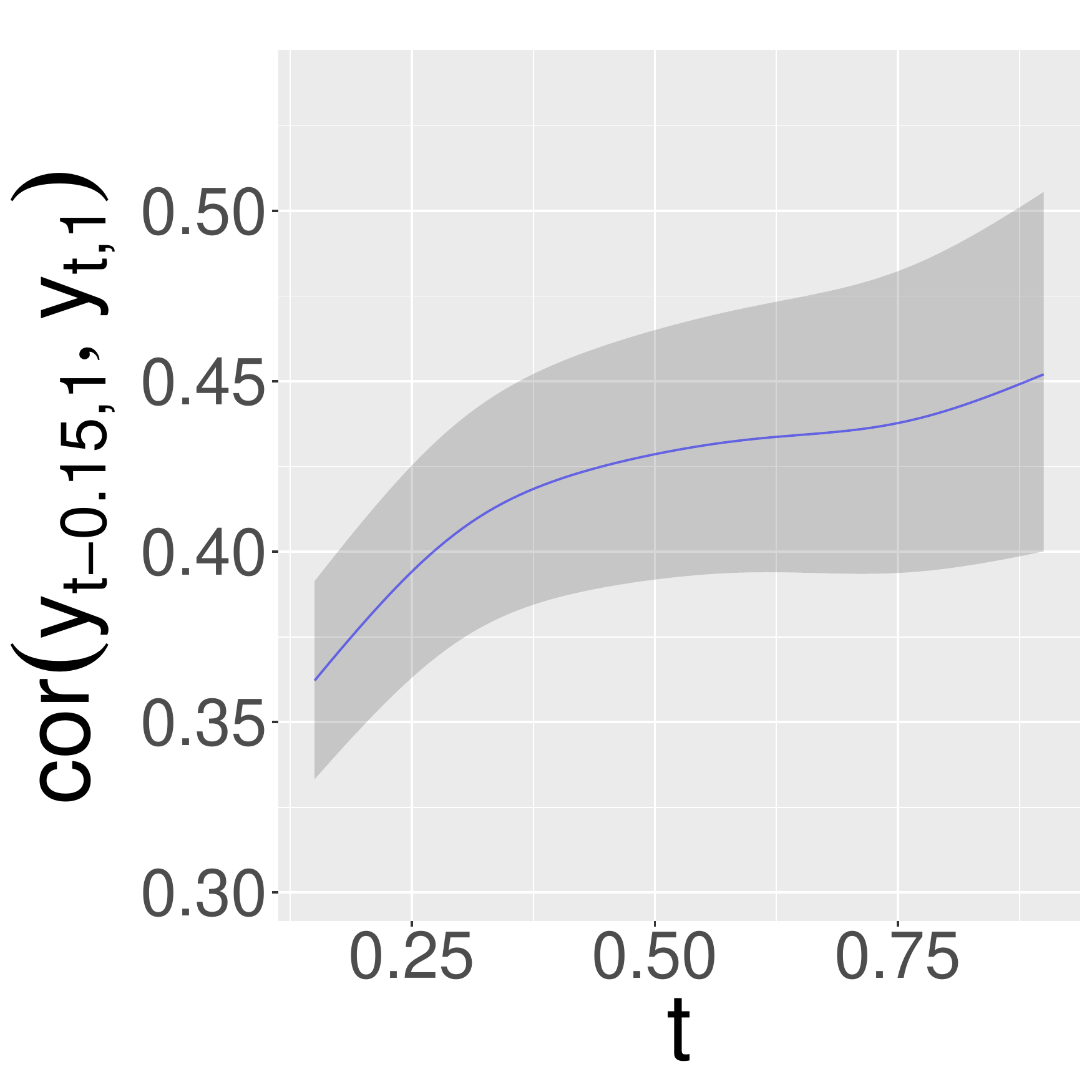} &  \includegraphics[width=0.20\textwidth]{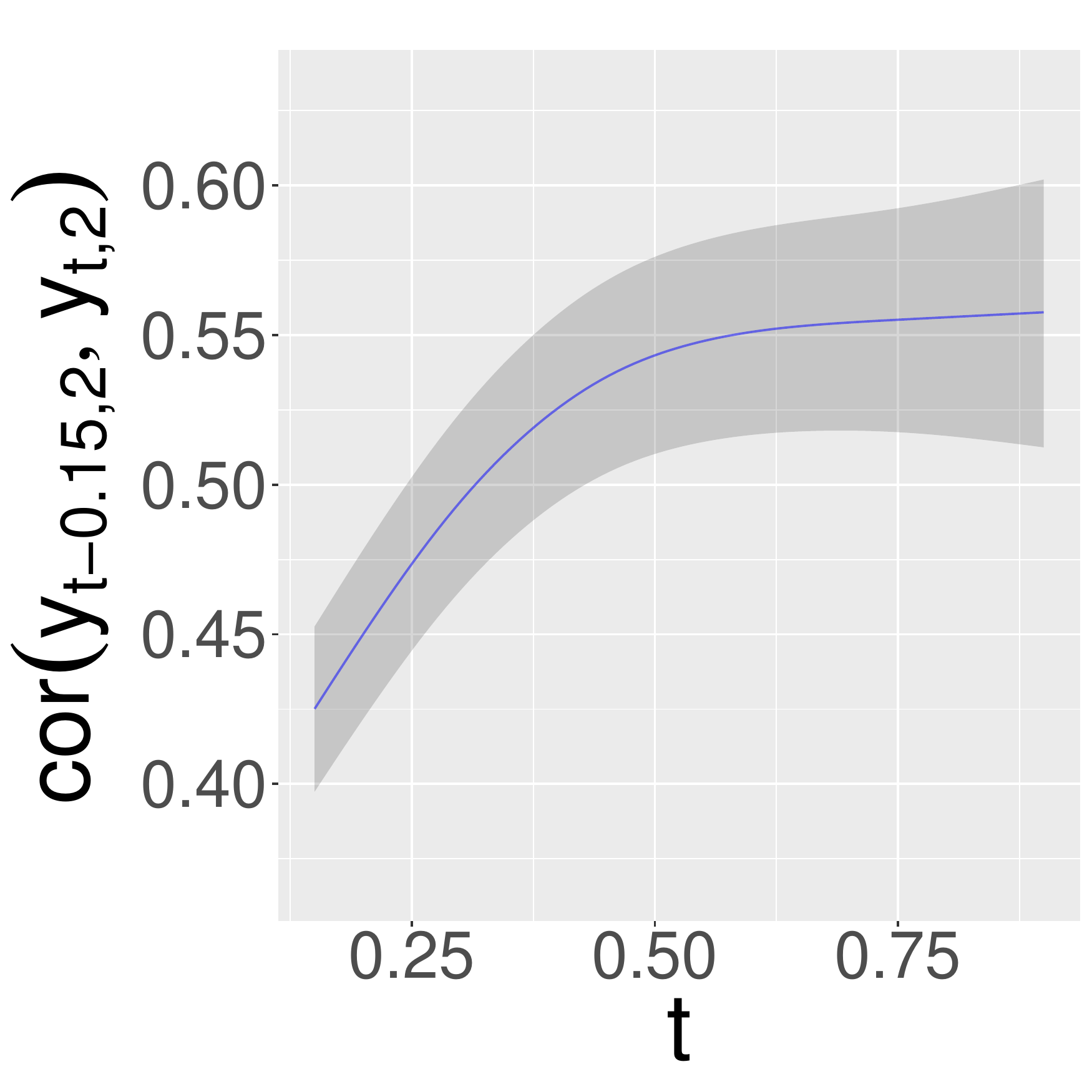} & 
			\includegraphics[width=0.20\textwidth]{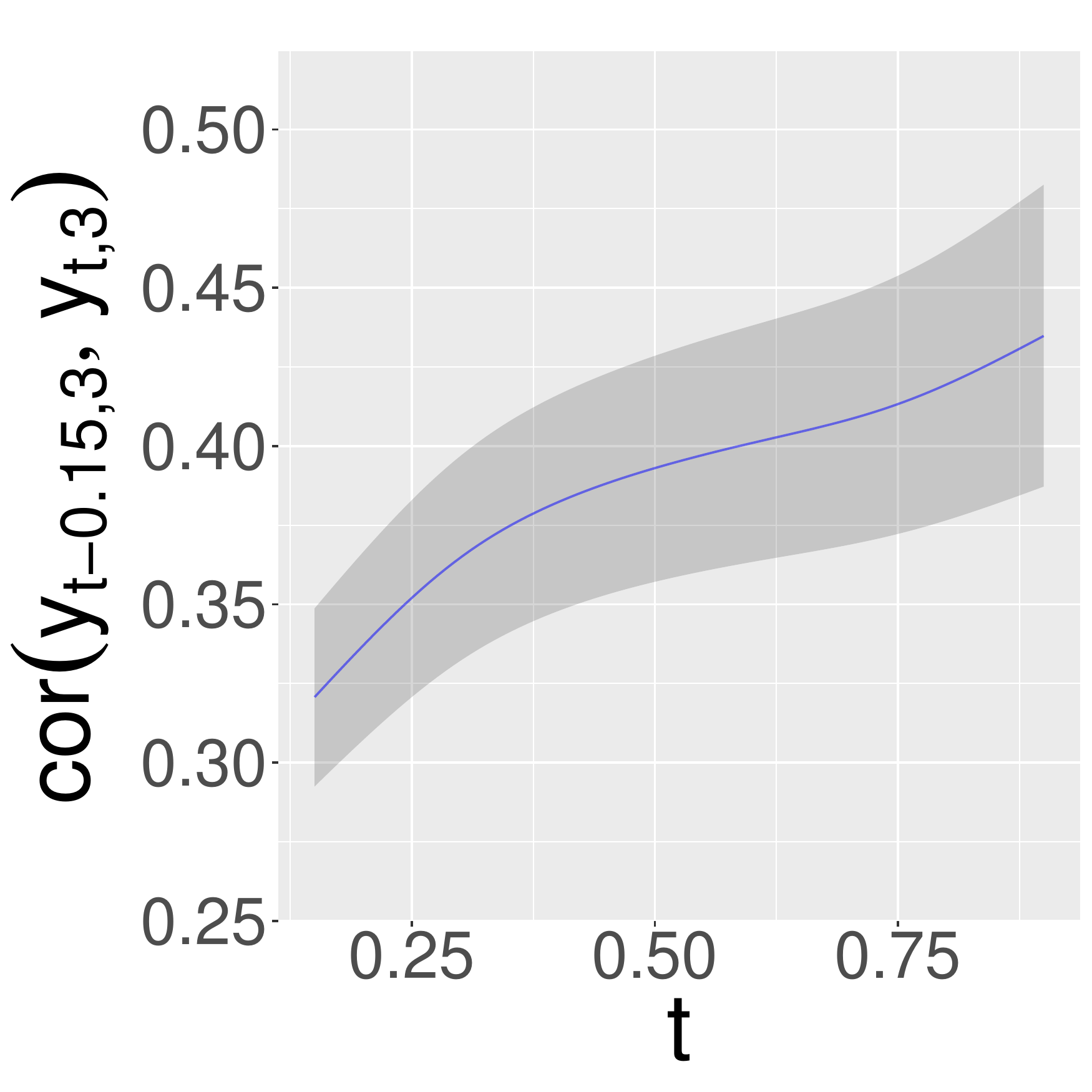} & \includegraphics[width=0.20\textwidth]{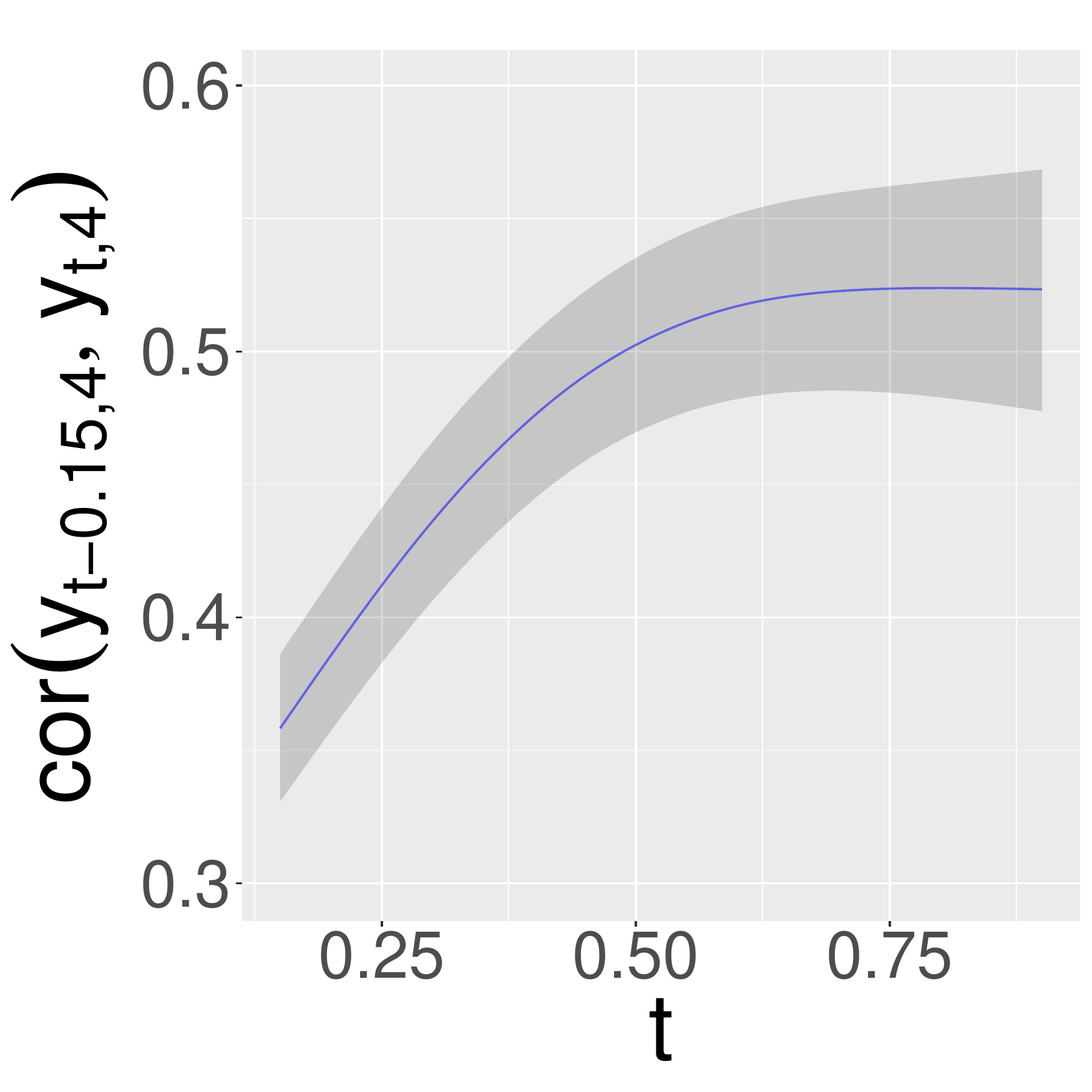}\\ 
		\end{tabular}
	\end{center}
	\caption{Application results: autocorrelations at lag $0.15$ over time. Posterior means and $80\%$ credible  intervals.}\label{AppAutoCorfromSigma}
\end{figure}

We conclude this section by providing an interpretation of the results in terms of the elements of the covariance matrix $\uSigma$ of the 4
responses. To do so, we construct two covariance matrices based on the sampled parameter values, 
both for time points $t=0,0.15,0.30,0.45,0.60,0.75,0.90$ and for $x_3$ equal to the lower $(x_3=0.18)$ and upper $(x_3=0.73)$ 
quartiles of age. With these choices we can examine the effects of time $t$ and age $x_3$ on the elements of $\uSigma$.  
For every iteration of the MCMC sampler (that is not discarded), we first obtain the innovation variances from (\ref{appinv}). 
Based on these and on the sampled correlation matrices $\uR$,
using (\ref{sep}), we reconstruct the innovation covariance matrices $\uD$. Further, we obtain the autoregressive coefficients from (\ref{appauto}),
and based on these we construct the generalized autoregressive matrices $\uPhi$, which, in turn, give us the matrix $\uL$ in (\ref{LD}). 
Lastly we rearrange (\ref{fd}) to obtain $\uSigma = \uL^{-1} \uD (\uL^{\top})^{-1}$.
This process results in $25 \times 10^3$ realizations of the $2$ covariance matrices that 
correspond to $x_3=0.18$ and $x_3=0.73$. Results are shown in Figures (\ref{AppSDfromSigma}), (\ref{AppCorfromSigma}) and (\ref{AppAutoCorfromSigma}). 

First, Figure (\ref{AppSDfromSigma}) shows the posterior means and $80\%$ credible intervals for the standard deviations of $Y_2$ and $Y_4$ over time. 
The standard deviations of $Y_1$ and $Y_3$  have similar shapes as the ones for  $Y_2$ and $Y_4$, and hence they are omitted. 
We can see that in all $4$ plots the standard deviations first decrease and then increase with time, $t$. The first two plots display the standard deviation of $Y_2$ for $x_3=0.18$ and $x_3=0.73$, respectively. We can see that age
has no effect on the standard deviation of $Y_2$. However, it does have an effect on the standard deviation of $Y_4$ as the curve in the fourth plot, when compared with that in the third plot, is shifted upwards.
Further, Figure (\ref{AppCorfromSigma}) shows how the correlations among responses evolve over time. We can see that some of the 
correlations exhibit an increasing trend
(cor$(y_{t1},y_{t3})$), others first decrease and then increase (cor$(y_{t1},y_{t2})$, cor$(y_{t2},y_{t3})$, and cor$(y_{t3},y_{t4})$), while others
are mostly flat (cor$(y_{t1},y_{t4})$ and cor$(y_{t2},y_{t4})$).
Lastly, Figure (\ref{AppAutoCorfromSigma}) displays autocorrelations at lag $0.15$ over time. These autocorrelations can be seen to be increasing over time.  

\section{Discussion}\label{discus}

The article describes a Bayesian framework for the analysis of multivariate longitudinal Gaussian responses, 
with nonparametric models for the means, the elements of the generalized autoregressive matrices, the elements of 
the innovation variance matrices, and the $2$ parameters of the innovation correlation matrices. The use of spike-slab priors 
is important here as it allows for automatic variable selection and function regularization, simultaneously in the $5$ submodels. Further, it obviates the need for model selection based on information criteria, see e.g. \citet{PanMack03}, 
and it allows for the uncertainty in the model choice to propagate in the model parameter estimates. In addition, the 
automatic variable selection allows each regression equation to have its own set of covariates. For example, looking back at
(\ref{appmean}), it may appear that the responses have the same mean model. 
However, the binary indicators for variable selection allow each response to have its own mean model. The same is true for the autoregressive coefficients and innovation variances.

The major assumption of the framework is that the errors have a multivariate Gaussian distribution.
It is worthwhile to relax this assumption, and to that end, in future work, we will be exploring the use of the multivariate skew-normal distribution \citep{Azz96}.

	\section{Supplementary material I: MCMC algorithm}\label{mcmc}
	
	Here we provide all details of the MCMC sampler of the three correlation models. We note that for all steps that involve tuning parameters, the values of these are chosen adaptively \citep{roberts_examples_2009}
	to achieve an acceptance probability of $20\% - 25\%$ \citep{Roberts2001c}.
	
	\subsection{MCMC algorithm for the common correlations model}\label{mcmc}
	
	Starting from the common correlations model, the algorithm proceeds as follows:
	
	\begin{enumerate}
		
		
		\item 
		
		The elements of $\ugamma_{kl},k=1,\dots,p,l=1,\dots,K,$ 
		are updated in random order and in blocks of random size \citep{Chan06}. Let $\ugamma_{Bkl}$ be a block
		of elements of $\ugamma_{kl}$.
		The proposed value for $\ugamma_{Bkl}$ is obtained from its prior with the remaining elements 
		of $\ugamma_{kl}$, denoted by $\ugamma_{Ckl}$, kept at their current value.   
		The proposal pmf is obtained from the binomial prior in (\ref{priorGamma}) with $\pi_{\mu kl}$ integrated out
		\begin{eqnarray}
			p(\ugamma_{Bkl}|\ugamma_{Ckl}) = \frac{p(\ugamma_{kl})}{p(\ugamma_{Ckl})}= 
			\frac{\text{Beta}(c_{\mu kl}+N(\ugamma_{kl}),d_{\mu kl}+q_{\mu l}-N(\ugamma_{kl}))}
			{\text{Beta}(c_{\mu kl}+N(\ugamma_{Ckl}),d_{\mu kl}+q_{\mu l}-L(\ugamma_{Bkl})-N(\ugamma_{Ckl}))},\nonumber
		\end{eqnarray}
		where $L(\ugamma_{Bkl})$ denotes the length of $\ugamma_{Bkl}$ i.e. the size of the block. 
		For this proposal pmf, the acceptance probability of the Metropolis-Hastings move reduces
		to the ratio of the likelihoods in (\ref{marginalY})
		\begin{eqnarray}
			\min\left\{1,(c_{\beta}+1)^{\{N(\gamma^C) - N(\gamma^P)\}/2} \exp\{(S^C-S^P)/2\}\right\},\nonumber 
		\end{eqnarray}
		where superscripts $P$ and $C$ denote proposed and currents values respectively. 
		
		
		\item 
		
		Parameter $c_{\beta}$ is updated from the marginal (\ref{marginalY}) and the IG$(a_{\beta},b_{\beta})$ prior 
		\begin{eqnarray}
			f(c_{\beta}|\dots) \propto (c_{\beta}+1)^{-\frac{N(\gamma)+p}{2}} \exp(-S/2)
			c_{\beta}^{-a_{\beta}-1} \exp(-b_{\beta}/c_{\beta}).\nonumber
		\end{eqnarray}
		To sample from the above, we utilize a normal approximation. Let
		$\ell(c_{\beta}) = \log\{f(c_{\beta}|\dots)\}$. We utilize a normal proposal density 
		$N(\hat{c}_{\beta},-g^2/\ell^{''}(\hat c_{\beta}))$, where $\hat{c}_{\beta}$ is the mode
		of $\ell(c_{\beta})$ found using a Newton-Raphson algorithm, 
		$\ell^{''}(\hat c_{\beta})$ is the second derivative of $\ell(c_{\beta})$ evaluated at the mode, and $g^2$ is a tuning parameter. 
		With superscripts $P$ and $C$ denoting proposed and currents values, the acceptance probability 
		is the minimum between one and
		\begin{equation}
			\frac{f(c_{\beta}^{P}|\dots)}{f(c_{\beta}^{C}|\dots)}
			\frac{N(c_{\beta}^{C};\hat{c}_{\beta},-g^2/\ell^{''}(\hat c_{\beta}))}
			{N(c_{\beta}^{P};\hat{c}_{\beta},-g^2/\ell^{''}(\hat c_{\beta}))}.\nonumber
		\end{equation}
		
		
		\item 
		
		Pairs $(\udelta_{kl}, \ualpha_{kl}), k=1,\dots,p, l=1,\dots,L,$ are updated simultaneously.
		Similarly to the updating of $\ugamma_{kl}$, the elements of $\udelta_{kl}$ are updated in random 
		order and in blocks of random size. Let $\udelta_{Bkl}$ denote a block. Blocks $\udelta_{Bkl}$ and the whole vector
		$\ualpha_{kl}$ are generated simultaneously. As was mentioned by \citet{Chan06}, generating the whole
		vector $\ualpha_{kl}$, instead of subvector $\ualpha_{Bkl}$, is necessary
		in order to make $\ualpha_{kl}$ consistent with the proposed value of $\udelta_{kl}$. 
		
		Generating the proposed value for $\udelta_{Bkl}$ is done in a similar way as was done for $\ugamma_{Bkl}$. 
		Let $\udelta^P_{kl}$ denote the proposed value of $\udelta_{kl}$. Next, we describe how  
		the proposed vale for $\ualpha_{\delta^P_{kl}}$ is obtained. 
		To avoid clutter, proposed values $\ualpha^P_{\delta^P_{kl}}$ will be denoted by the simpler $\ualpha_{kl}^P$.  
		The development that follows is in the spirit of \citet{Chan06} who built on the work of \citet{Gamerman1997}. 
		
		Let $\hat{\ubeta}_{\gamma}^C = \{c_{\beta}/(1+c_{\beta})\}(\utX_{\gamma}^{\top} \utX_{\gamma})^{-1}\utX_{\gamma}^{\top} \uty$
		denote the current value of the posterior mean of $\ubeta_{\gamma}^{\ast}$.
		Define the current squared residuals 
		\begin{equation}
			e_{ijk}^C = (y_{ijk} - (\ux_{ijk}^{\ast})^{\top} \hat{\ubeta}_{\gamma_k}^C)^2.\nonumber 
		\end{equation}
		These have an approximate $\sigma^2_{ijk} \chi^2_1$ distribution,
		where $\sigma^2_{ijk} = \sigma^2_{k} \exp(\uw_{ijk}^{\top} \ualpha_{\delta_k})$. 
		The latter defines a Gamma generalized linear model (GLM) for the squared 
		residuals with mean $\sigma^2_{ijk}$, which,
		utilizing a $\log$-link, can be thought of as Gamma GLM with an 
		offset term: $\log(\sigma^2_{ijk}) = \log(\sigma^2_{k}) + \uw_{ijk}^{\top} \ualpha_{\delta_k}$.
		Given $\udelta^P_{kl}$, the proposal density for $\ualpha^P_{kl}$ 
		is derived utilizing the one step iteratively reweighted least squares algorithm.
		This proceeds as follows. First define the transformed observations
		\begin{eqnarray}
			d_{ijk}^C(\ualpha^C_k) = \log(\sigma^2_{k}) + \uw_{ij}^{\top} \ualpha_{k}^C + 
			\frac{e_{ijk}^C-(\sigma^2_{ijk})^C}{(\sigma^2_{ijk})^C},\nonumber
		\end{eqnarray}
		where superscript $C$ denotes current values. Further, let $\ud_k^C$ denote the vector of $d_{ijk}^C$.
		
		Let $\usigma^2_k = (\sigma_{11k}^2,\dots,\sigma_{nn_nk}^2)^{\top}$. Model (\ref{lastvar}) for $\usigma^2_k$ can be expressed as 
		\begin{equation}
			\usigma^2_k = \sigma^2_{k} \exp(\uW_{\delta_k} \ualpha_{\delta_k}), \nonumber
		\end{equation}
		where $\uW_{\delta_k} = [\uw_{11k}, \dots,\uw_{nn_nk}]^{\top}$.  
		
		Next we define
		\begin{eqnarray}
			\uDelta(\udelta^P_{kl}) = (c_{\alpha k}^{-1}\uI + \uW_{\delta^P_{kl}}^{\top}\uW_{\delta^P_{kl}})^{-1}
			\text{\;and\;}
			\ahat(\udelta^P_{kl},\ualpha^C_k) =  \uDelta_{\delta^P_{kl}} \uW_{\delta^P_{kl}}^{\top} \ud_k^C, \nonumber
		\end{eqnarray}
		where $\uW_{\delta_{kl}}$ is a submatrix of $\uW_{\delta_k}$ that considers only
		the columns that pertain to the $l$th effect.
		The proposed value $\ualpha_{kl}^P$ is obtained from the multivariate normal
		distribution  
		$N(\ualpha_{kl}^P;\ahat(\udelta^P_{kl},\ualpha^C_{k}),h_{kl} \uDelta(\udelta^P_{kl}))$,
		where $h_{kl}$ is a tuning parameter.
		
		Let $N(\ualpha_{kl}^C;\ahat(\udelta^C_{kl},\ualpha^P_k),h_{kl}\uDelta(\udelta^C_{kl}))$ denote the proposal density 
		for taking a step in the reverse direction, from model $\udelta^P_{kl}$ to $\udelta^C_{kl}$. 
		Then the acceptance probability of the pair $(\udelta^P_{kl},\ualpha^P_{kl})$ is  
		\begin{eqnarray}
			\min\left\{1,
			\frac{|\uSigma^P|^{-\frac{1}{2}} \exp\{-S^P/2\}}
			{|\uSigma^C|^{-\frac{1}{2}} \exp\{-S^C/2\}}
			\frac{
				(2\pi c_{\alpha k})^{-\frac{N(\delta^P_{kl})}{2}}\exp\{-\frac{1}{2c_{\alpha k}} (\ualpha^P_{kl})^{\top} \ualpha^P_{kl}\}
			}{
				(2\pi c_{\alpha k})^{-\frac{N(\delta^C_{kl})}{2}}\exp\{-\frac{1}{2c_{\alpha k}} (\ualpha^C_{kl})^{\top} \ualpha^C_{kl}\}
			}
			\frac{
				N(\ualpha_{kl}^C;\ahat_{\delta^C_{kl}},h_{kl} \uDelta_{\delta^C_{kl}})
			}{
				N(\ualpha_{kl}^P;\ahat_{\delta^P_{kl}},h_{kl} \uDelta_{\delta^P_{kl}})
			}
			\right\},\nonumber
		\end{eqnarray}
		where the determinants, for centred variables, are equal to one, otherwise, the ratio of the determinants may be computed as
		$\prod_{i=1}^n \prod_{j=1}^{n_i} \{(\sigma^2_{ijk})^{C}/(\sigma^2_{ijk})^{P}\}^{1/2}$.
		
		
		\item 
		
		The full conditional of $\sigma^2_{k}, k=1,\dots,p,$ is given by
		\begin{eqnarray}
			f(\sigma^2_{k}|\dots) \propto |\uSigma|^{-\frac{1}{2}} \exp(-S/2) \xi(\sigma^2_{k}),\nonumber
		\end{eqnarray}
		where $\xi(\sigma^2_{k})$ denotes either the IG or half-normal prior.
		We follow a random walk algorithm obtaining proposed values $(\sigma^2_{k})^{(P)} \sim N((\sigma^2_{k})^{(C)},v_{k}^2)$, 
		where $v_{k}^2$ is a tuning parameter. 
		Proposed values are accepted with probability $f((\sigma^2_{k})^{(P)}|\dots) /f((\sigma^2_{k})^{(C)}|\dots)$, which reduces to
		\begin{eqnarray}
			\{(\sigma^2_{k})^{C}/(\sigma^2_{k})^{P}\}^{N/2} \exp\{(S^C-S^P)/2\}) \xi((\sigma^2_{k})^{(P)})/\xi((\sigma^2_{k})^{(C)}).\nonumber
		\end{eqnarray}
		
		
		\item 
		
		Concerning parameter $c_{\alpha k}, k=1,\dots,p$, the full conditional that corresponds to the IG$(a_{\alpha k},b_{\alpha k})$ prior is 
		another inverse Gamma density IG$(a_{\alpha k} + N(\delta_k)/2,b_{\alpha k}+\ualpha_{\delta_k}^{\top}\ualpha_{\delta_k}/2)$.
		
		The full conditional that corresponds to the half-normal prior $\sqrt{c_{\alpha k}} \sim N(0,\phi^2_{\alpha k}) I[\sqrt{c_{\alpha k}}>0]$ is 
		\begin{eqnarray}
			f(c_{\alpha k}|\dots) \propto c_{\alpha k}^{-N(\delta_k)/2} 
			\exp\{-\ualpha_{\delta_k}^{\top}\ualpha_{\delta_k}/2c_{\alpha k}\} \exp\{-c_{\alpha k}/2\phi^2_{\alpha k}\}I[\sqrt{c_{\alpha k}}>0].\nonumber
		\end{eqnarray}
		We obtain proposed values from $c_{\alpha k}^{(P)} \sim N(c_{\alpha k}^{(C)},v_{k}^2)$, where $c_{\alpha k}^{(C)}$ denotes the current value and $v_{k}^2$ denotes a tuning parameter. 
		Proposed values are accepted with probability $f(c_{\alpha k}^{(P)}|\dots)/f(c_{\alpha k}^{(C)}|\dots)$. 
		
		
		\item 
		
		Samples from the posterior of $\ubeta^{\ast}_{\gamma}$ are generated from
		\begin{eqnarray}
			\ubeta_{\gamma}^{\ast} | \dots \sim N(\frac{c_{\beta}}{1+c_{\beta}}(\utX_{\gamma}^{\top} \utX_{\gamma})^{-1}\utX_{\gamma}^{\top} \uty,
			\frac{c_{\beta}}{1+c_{\beta}}(\utX_{\gamma}^{\top} \utX_{\gamma})^{-1}),\nonumber
		\end{eqnarray}
		which is obtained using the first version of $Q$, just above (\ref{Q1}), and prior (\ref{gprior1}).
		
		
		\item 
		
		Pairs $(\uxi_{lmb}, \upsi_{lmb}), l,m=1,\dots,p, b=0,\dots,B,$ are updated simultaneously.
		Let $\uxi_{Blmb}$ denote a randomly chosen block of $\uxi_{lmb}$. The block $\uxi_{Blmb}$ and the whole 
		of $\upsi_{lmb}$ are updated simultaneously. Proposed values for $\uxi_{Blmb}$ are obtained from its 
		prior, in a similar way as was done for $\ugamma_{Bkl}$. 
		Now, to obtain a proposal for $\upsi_{lmb}$, note that the full conditional of $\upsi^{\ast}_{\xi}$, as obtained from its normal prior and the likelihood that corresponds to (\ref{Q4}), is 
		\begin{eqnarray}
			\upsi_{\xi}^{\ast} | \dots \sim N\left[\uA \left( \sum_{i=1}^n \sum_{j=1}^{n_i} \uV_{i j \xi \beta}^{\top} \uD_{ij}^{-1} \ur_{ij}\right),
			\uA \equiv \left(\uD_{c{\psi}}^{-1} + \sum_{i=1}^n \sum_{j=1}^{n_i} \uV_{i j \xi \beta}^{\top} \uD_{ij}^{-1} \uV_{i j \xi \beta}\right)^{-1} \right],\label{ppsi} \nonumber
		\end{eqnarray}
		where $\uD_{c{\psi}}$ is a diagonal matrix with diagonal elements equal to $c_{\psi lm}$,  each having multiplicity equal to the length of $\upsi_{\xi lm}^{\ast}$.
		We use as proposal the conditional of $\upsi_{lmb}$ given all other elements of $\upsi_{\xi}^{\ast}$, denoted by 
		$N(\upsi_{lmb}|\upsi_{Clmb})$. 
		Then, the acceptance probability of the pair $(\uxi^P_{lmb},\upsi^P_{lmb})$ is  
		\begin{eqnarray}
			\min\left\{1,
			\frac{\exp(-S^P/2)}
			{\exp(-S^C/2)}
			\frac{
				N(\upsi_{lmb}^P|\uzero,c_{\psi lm} \uI)
			}{
				N(\upsi_{lmb}^C|\uzero,c_{\psi lm} \uI)
			}
			\frac{
				N(\upsi_{lmb}^C|\upsi_{Clmb})
			}{
				N(\upsi_{lmb}^P|\upsi_{Clmb})
			}
			\right\}.\nonumber
		\end{eqnarray}
		
		
		\item 
		
		For parameter $c_{\psi}$ the full conditional that corresponds to the IG$(a_{\psi},b_{\psi})$ prior is 
		another inverse Gamma density, IG$(a_{\psi} + N(\xi)/2,b_{\psi}+(\upsi_{\xi}^{\ast})^{\top}\upsi_{\xi}^{\ast}/2)$.
		
		The full conditional that corresponds to the half-normal prior $\sqrt{c_{\psi}} \sim N(0,\phi^2_{\psi}) I[\sqrt{c_{\psi}}>0]$ is 
		\begin{eqnarray}
			f(c_{\psi}|\dots) \propto c_{\psi}^{-N(\xi)/2} 
			\exp[-(\upsi_{\xi}^{\ast})^{\top}\upsi_{\xi}^{\ast}/(2c_{\psi})] \exp(-c_{\psi}/2\phi^2_{\psi}) I(\sqrt{c_{\psi}}>0).\nonumber
		\end{eqnarray}
		We obtain proposed values from $c_{\psi}^{(P)} \sim N(c_{\psi}^{(C)},s_{\psi}^2)$, where $c_{\psi}^{(C)}$ is the current value and $s_{\psi}^2$ is a tuning parameter. Proposed values are accepted with probability  $f(c_{\psi}^{(P)}|\dots)/f(c_{\psi}^{(C)}|\dots)$. 
		
		
		\item 
		
		To sample from the full conditional of $\uR_{t}, t \in T,$ recall (\ref{priorR2})
		and (\ref{Qcom}). We have 
		\begin{eqnarray}\label{postRt}
			f(\uR_{t}|\dots) \propto |\uR_{t}|^{-\frac{n_{t}}{2}} 
			\text{etr}(-\uR_t^{-1} \uS_t/2) 
			\prod_{k<l} \exp[-(g(r_{tkl})-\theta_{tkl})^2/2\tau^2] J[g(r_{tkl}) \rightarrow r_{tkl}] I[\uR_{t} \in \mathcal{C}], \label{Rt}
		\end{eqnarray}
		where $n_t$ denotes the total number of observations at time point $t$, 
		$\text{etr}(.)=\exp(\text{tr}(.))$, and $\uS_t = \sum_{i \in O_t} \ucepsilon_{it}\ucepsilon_{it}^{\top}$. 
		
		To obtain a proposal density and sample from (\ref{Rt}) we utilize the method of \citet{xiao} and \citet{LD2006}.
		We start by considering symmetric pd and otherwise unconstrained matrices $\uE_t$ in place of $\uR_t$.  
		These are assumed to have an inverse Wishart prior $\uE_t \sim \text{IW}(\zeta_t,\uPsi_t), t \in T$, 
		with mean equal to the realization of $\uE_t$ from the previous iteration.  
		Given the inverse Wishart prior on $\uE_t$, we obtain the following easy to sample from inverse Wishart posterior
		\begin{eqnarray}
			g(\uE_{t}|\dots) \propto |\uPsi_t|^{\frac{\zeta_t}{2}} |\uE_{t}|^{-\frac{n_t+\zeta_t+p+1}{2}} \text{etr}\{-\uE_t^{-1} (\uS_t+\uPsi_t)/2\}.\label{Et}
		\end{eqnarray}
		
		We decompose $\uE_t=\uD_t^{1/2} \uR_t \uD_t^{1/2}$ into a diagonal matrix of variances  
		$\uD_t = \text{Diag}(d^2_{t1},\dots,d^2_{tp})$, 
		and a correlation matrix $\uR_t$. The Jacobian associated with this transformation is 
		$J(\uE_t \rightarrow \uD_t, \uR_t) = \prod_{k=1}^p (d_{tk})^{p-1} = |\uD_t|^{(p-1)/2}$. 
		It follows that the joint density for $(\uD_t, \uR_t)$ is  
		\begin{eqnarray}
			h(\uD_t, \uR_t|\dots) \propto |\uPsi_t|^{\frac{\zeta_t}{2}} |\uD_t|^{(p-1)/2} |\uE_{t}|^{-\frac{n_t+\zeta_t+p+1}{2}} \text{etr}\{-\uE_t^{-1} (\uS_t+\uPsi_t)/2\}.\label{DR}
		\end{eqnarray}
		
		Sampling from (\ref{DR}) at iteration $u+1$ proceeds by sampling $\uE_t^{(u+1)}$ from (\ref{Et}) and decomposing 
		$\uE_t^{(u+1)}$ into $(\uD_t^{(u+1)},\uR_t^{(u+1)})$. 
		Further, the pair $(\uD_t^{(u+1)},\uR_t^{(u+1)})$ is accepted as a sample from (\ref{Rt}) with probability
		\begin{eqnarray}
			\alpha = \min\left\{1,\frac{f(\uR_t^{(u+1)}|\dots) h(\uD_t^{(u)}, \uR_t^{(u)}|\dots)}
			{f(\uR_t^{(u)}|\dots) h(\uD_t^{(u+1)}, \uR_t^{(u+1)}|\dots)}\right\},\nonumber
		\end{eqnarray}
		where in $h(,|)$, $\uPsi_t = (\zeta_t-p-1) \uE_t^{(u)}$.
		We treat $\zeta_t$ as a tuning parameter.
		
		
		\item 
		
		To sample from the full conditional of $\utheta_t, t \in T$, first 
		write $f(\ur|\utheta,\tau^2) = \prod_{t \in T}\pi(\utheta_t,\tau^2) N(g(\ur);\utheta,\tau^2\uI)$
		for the likelihood of (\ref{priorR2}). Further, the prior for $\utheta$, as derived from (\ref{marginalTh}), is
		$\utheta \sim N(\uzero,\uSigma_{\theta})$ 
		where $\uSigma_{\theta}^{-1} = \sigma^{-2}\uD^{-1}(\uomega_{\varphi}) \left\{ 
		\uI - \frac{c_{\eta}}{1+c_{\eta}} \utZ_{\nu} (\utZ_{\nu}^{\top}\utZ_{\nu})^{-1}\utZ_{\nu}^{\top}
		\right\}\uD^{-1}(\uomega_{\varphi})$. Hence, it is easy to show that the posterior is
		\begin{eqnarray}\label{ptheta2}
			f(\utheta|\dots) = \left\{\prod_{t \in T} \pi(\utheta_t,\tau^2)\right\}
			N\left(\utheta; \tau^{-2} \uA g(\ur),\uA \equiv (\tau^{-2} \uI + \uSigma_{\theta}^{-1})^{-1}\right).
		\end{eqnarray}
		At iteration $u+1$ we sample $\utheta^{(u+1)}$ utilizing as proposal the normal distribution
		that appears on the right of (\ref{ptheta2}), ignoring the normalizing constants.  
		The proposed $\utheta^{(u+1)}$ is accepted with probability
		\begin{eqnarray}
			\min\left\{1,\frac{\prod_{t \in T} \pi(\utheta_{t}^{(u+1)},\tau^2)}{\prod_{t \in T} \pi(\utheta_{t}^{(u)},\tau^2)}\right\}.\nonumber
		\end{eqnarray}
		This, as was argued by \citet{Liechty} and \citet{Liechty2}, for a small value of $\tau^2$, can reasonably be assumed to be unity. 
		
		
		\item 
		
		The elements of $\unu$ are updated in random order and in blocks of random size. Let $\unu_B$ be such a block. The proposed value for $\unu_B$ is obtained from its prior with the remaining elements of $\unu$, denoted by $\unu_{B^C}$, kept at their current value, in a similar way as was done for $\ugamma_{Bkl}$.
		The acceptance probability reduces to the ratio of the likelihoods in (\ref{marginalTh})
		\begin{eqnarray}
			\min\left\{1,\frac{(c_{\eta}+1)^{-\frac{N(\nu^P)+1}{2}} \exp\{-(S^{\ast})^P/2\sigma^2\}}
			{(c_{\eta}+1)^{-\frac{N(\nu^C)+1}{2}} \exp\{-(S^{\ast})^C/2\sigma^2\}}\right\},\nonumber 
		\end{eqnarray}
		where superscripts $P$ and $C$ denote proposed and currents values respectively. 
		
		
		\item 
		
		Vectors $\uomega$ and $\uvarphi$ are updated following a similar approach as that for step 3. The elements of $\uvarphi$ are updated in random order and in blocks of random size. Let $\uvarphi_B$ denote a block. Blocks $\uvarphi_B$ and the whole vector $\uomega$ are generated simultaneously. 
		Generating the proposed value for $\uvarphi_B$ is done in a similar way as was done for $\ugamma_{Bkl}$. Let $\uvarphi^P$ denote the proposed value of $\uvarphi$. Next, we describe how the proposed vale for $\uomega_{\varphi^P}$ is obtained. 
		Let $\hat{\ueta}_{\nu}^C = \{c_{\eta}/(1+c_{\eta})\}(\utZ_{\nu}^{\top} \utZ_{\nu})^{-1}\utZ_{\nu}^{\top} \utTheta$
		denote the current value of the posterior mean of $\ueta_{\nu}$. Define the current squared residuals 
		\begin{equation}
			e_{tkl}^C = (\theta_{tkl} - (\uz^{\ast}_{t\nu})^{\top} \hat{\ueta}_{\nu}^C)^2,\nonumber 
		\end{equation}
		$t \in T,k=1,\dots,p-1,l>k$.
		These will have an approximate $\sigma^2_t \chi^2_1$ distribution, where $\sigma^2_t = \sigma^2 \exp(\uz_t^{\top} \uomega)$. The latter defines a Gamma generalized linear model (GLM) for the squared 
		residuals with mean $E(\sigma^2_t \chi^2_1) = \sigma^2_t = \sigma^2 \exp(\uz_t^{\top} \uomega)$, which, utilizing a $\log$-link, can be thought of as Gamma GLM with an offset term: $\log(\sigma^2_t) = \log(\sigma^2) + \uz_t^{\top} \uomega$. Given the proposed value of $\uvarphi$, denoted by $\uvarphi^P$, the proposal density for $\uomega_{\varphi^P}$ is derived utilizing the one step iteratively reweighted least squares algorithm. This proceeds as follows. First define the transformed observations
		\begin{eqnarray}
			d_{tkl}^C(\uomega^C) = \log(\sigma^2) + \uz_t^{\top} \uomega^C + \frac{e_{tkl}^C-(\sigma^2_t)^C}{(\sigma^2_t)^C},\nonumber
		\end{eqnarray}
		where superscript $C$ denotes current values. Further, let $\ud^C$ denote the vector of $d_{tkl}^C$.
		
		Next we define
		\begin{eqnarray}
			\uDelta(\uvarphi^P) = (c_{\omega}^{-1}\uI + \uZ_{\varphi^P}^{\top}\uZ_{\varphi^P})^{-1}
			\text{\;and\;}
			\ohat(\uvarphi^P,\uomega^C) =  \uDelta_{\varphi^P} \uZ_{\varphi^P}^{\top} \ud^C,\nonumber 
		\end{eqnarray}
		where $\uZ$ is the design matrix $\uZ^{\ast}$ without the intercept column. The proposed value  $\uomega_{\varphi^P}^P$ is obtained from a multivariate normal distribution with mean $\ohat(\uvarphi^P,\uomega^C)$ and covariance $h \uDelta(\uomega^P)$, denoted as 
		$N(\uomega_{\varphi^P}^P;\ohat(\uvarphi^P,\uomega^C),h \uDelta(\uvarphi^P))$, where $h$ is a tuning parameter.
		
		Let  $N(\uomega_{\varphi^C}^C;\ohat(\uvarphi^C,\uomega^P),h\uDelta(\uvarphi^C))$ denote the proposal density 
		for taking a step in the reverse direction, from model $\uvarphi^P$ to $\uvarphi^C$. Then the acceptance probability of the pair $(\uvarphi^P,\uomega^P_{\varphi^P})$ is  
		\begin{eqnarray}
			\min\left\{1,
			\frac{|D^2(\uomega^P_{\varphi^P})|^{-\frac{1}{2}} \exp\{-S^P/2\sigma^2\}}
			{|D^2(\uomega^C_{\varphi^C})|^{-\frac{1}{2}} \exp\{-S^C/2\sigma^2\}}
			\frac{
				(2\pi c_{\omega})^{-\frac{N(\varphi^P)}{2}}\exp\{-\frac{1}{2c_{\omega}} (\uomega^P_{\varphi^P})^{\top} \uomega^P_{\varphi^P}\}
			}{
				(2\pi c_{\omega})^{-\frac{N(\varphi^C)}{2}}\exp\{-\frac{1}{2c_{\omega}} (\uomega^C_{\varphi^C})^{\top} \uomega^C_{\varphi^C}\}
			}
			\frac{
				N(\uomega_{\varphi^C}^C;\ahat_{\varphi^C},h \uDelta_{\varphi^C})
			}{
				N(\uomega_{\varphi^P}^P;\ahat_{\varphi^P},h \uDelta_{\varphi^P})
			}
			\right\}.\nonumber
		\end{eqnarray}
		
		
		\item 
		
		We update $\sigma^2$ utilizing the marginal (\ref{marginalTh}) and the either of the two prior specifications. 
		The full conditional that corresponds to the IG$(a_{\sigma},b_{\sigma})$ prior is
		\begin{eqnarray}
			f(\sigma^2|\dots) \propto (\sigma^2)^{-\frac{Md}{2}-a_{\sigma}-1} \exp\{-(S^{\ast}/2+b_{\sigma})/\sigma^2\},\nonumber
		\end{eqnarray}
		where $M$ denotes the number of unique observational times and $d = p (p-1)/2$. The above is recognized as IG$(Md/2+a_{\sigma},S^{\ast}/2+b_{\sigma})$.
		
		The full conditional that corresponds to the half-normal prior  $\sigma \sim N(\sigma;0,\phi^2_{\sigma}) I[\sigma>0]$ is
		\begin{eqnarray}
			f(\sigma^2|\dots) \propto (\sigma^2)^{-\frac{Md}{2}} \exp\{-S^{\ast}/(2\sigma^2)\} \exp\{-\sigma^2/(2\phi^2_{\sigma})\} I[\sigma>0].\nonumber
		\end{eqnarray}
		Proposed values are obtained from $\sigma^2_p \sim N(\sigma^2_c,f_1^2)$, where $\sigma^2_c$ denotes the current value and $f_1^2$ is a tuning parameter. Proposed values are accepted with probability $f(\sigma^2_p|\dots)/f(\sigma^2_c|\dots)$. 
		
		
		\item 
		
		Parameter $c_{\eta}$ is updated from the marginal (\ref{marginalTh}) and the IG$(a_{\eta},b_{\eta})$ prior 
		\begin{eqnarray}
			f(c_{\eta}|\dots) \propto (c_{\eta}+1)^{-\frac{N(\nu)+1}{2}} \exp\{-S^{\ast}/2\sigma^2\}
			(c_{\eta})^{-a_{\eta}-1} \exp\{-b_{\eta}/c_{\eta}\}.\nonumber
		\end{eqnarray}
		To sample from the above, we utilize a normal approximation to it. Let
		$\ell(c_{\eta}) = \log\{f(c_{\eta}|\dots)\}$. We utilize a normal proposal density 
		$N(\hat{c}_{\eta},-g^2/\ell^{''}(\hat c_{\eta}))$ where $\hat{c}_{\eta}$ is the mode
		of $\ell(c_{\eta})$, found using a Newton-Raphson algorithm, 
		$\ell^{''}(\hat c_{\eta})$ is the second derivative of $\ell(c_{\eta})$ evaluated at the mode, 
		and $g^2$ is a tuning parameter. 
		At iteration $u+1$ the acceptance probability is the minimum between one and
		\begin{equation}
			\frac{f(c_{\eta}^{(u+1)}|\dots)}{f(c_{\eta}^{(u)}|\dots)}
			\frac{N(c_{\eta}^{(u)};\hat{c}_{\eta},-g^2/\ell^{''}(\hat c_{\eta}))}
			{N(c_{\eta}^{(u+1)};\hat{c}_{\eta},-g^2/\ell^{''}(\hat c_{\eta}))}.\nonumber
		\end{equation}
		
		
		\item 
		
		Concerning parameter $c_{\omega}$, the full conditional that corresponds to the IG$(a_{\omega},b_{\omega})$ prior is 
		another inverse Gamma density IG$(a_{\omega}+N(\varphi)/2,b_{\omega}+\uomega_{\varphi}^{\top}\uomega_{\varphi}/2)$.
		
		The full conditional that corresponds to the half-normal prior $\sqrt{c_{\omega}} \sim N(\sqrt{c_{\omega}};0,\phi^2_{c_{\omega}}) I[\sqrt{c_{\omega}}>0]$ is 
		\begin{eqnarray}
			f(c_{\omega}|\dots) \propto c_{\omega}^{-N(\varphi)/2} 
			\exp\{-\uomega_{\varphi}^{\top}\uomega_{\varphi}/(2c_{\omega})\} \exp\{-c_{\omega}/(2\phi^2_{c_{\omega}})\}I[\sqrt{c_{\omega}}>0].\nonumber
		\end{eqnarray}
		We obtain proposed values $c_{\omega}^{P} \sim N(c_{\omega}^{C},f_2^2)$, where $c_{\omega}^{C}$ denotes the current value and $f_2^2$ is a tuning parameter. Proposed values are accepted with probability $f(c_{\omega}^{P}|\dots)/f(c_{\omega}^{C}|\dots)$.
		
		
		\item 
		
		The sampler utilizes the marginal in (\ref{marginalTh}) to improve mixing. 
		However, if samples are required from the posterior of $\ueta$, they can be generated
		from
		\begin{eqnarray}
			\ueta_{\nu} | \dots \sim N(\frac{c_{\eta}}{1+c_{\eta}}(\utZ_{\nu}^{\top} \utZ_{\nu})^{-1}\utZ_{\nu}^{\top} \utTheta,
			\frac{\sigma^2 c_{\eta}}{1+c_{\eta}}(\utZ_{\nu}^{\top} \utZ_{\nu})^{-1}),\nonumber
		\end{eqnarray}
		where $\ueta_{\nu}$ is the non-zero part of $\ueta$. 
	\end{enumerate}

	\subsection{MCMC algorithm for the grouped correlations model}
	
	The algorithm for the `grouped correlations' model requires $3$ additional steps (the first $3$ below) and a modification to a step from the algorithm of the `common correlations' model (the last one below). We describe these next. 
	
	\begin{enumerate}
		
		\item 
		
		Update $v_h \sim\text{Beta}(d_h+1,d-\sum_{l=1}^hd_h+\alpha^{\ast}), h=1,\dots,H-1,$
		where $d=p(p-1)/2$ is the number of correlations, and $d_h$ is the number of correlations allocated in the $h$th cluster. 
		Given $v_h$, update the weights $w_h, h=1,\dots,H$. 
		
		\item 
		
		To calculate the cluster assignment probabilities first 
		define $\utheta_{kl} = (\theta_{1kl},\dots,\theta_{Mkl})^{\top}$ to be the profile of the $(k,l)$ correlation. Based on (\ref{modtheta2}), we have the following posterior probability 
		\begin{eqnarray}
			P(\lambda_{kl}=h|\dots) \propto w_{h} N(\utheta_{kl};\uZ^{*}_{kl \nu_h} \ueta^{*}_{\nu_h},\sigma^2 \uD_{kl}^2(\uomega_{\varphi})).\nonumber 
		\end{eqnarray} 
		Here matrix $\uD_{kl}^2$ is diagonal of dimension $M$ with elements 
		given by $\exp\{\uz_{\varphi t}^{\top} \uomega_{\varphi}\},t \in T$, and matrix $\uZ_{kl\nu_h}^{\ast}$ has $M$ rows 
		given by $(\uz_{th}^{\ast})^{\top},t \in T,$ i.e. it is the subset of $\uZ^{\ast}_{\nu_h}$ that corresponds to 
		$\utheta_{kl}$. Vector $\ueta_{\nu_h}^{*}$ is imputed from its posterior, or if the cluster is empty
		from prior (\ref{m2etaB}).
		
		\item 
		
		We update concentration parameter $\alpha^{\ast}$ using the method described by \citet{EscobarWest}.
		With the $\alpha^{*} \sim \text{Gamma}(a_{\alpha*},b_{\alpha*})$ prior,
		the posterior can be expressed as a mixture of two Gamma distributions 
		\begin{equation}\label{alphadist2}
			\alpha^{\ast}|\eta,k \sim \pi_{\eta} \text{Gamma}(a_{\alpha*}+k,b_{\alpha*}-\log(\eta)) + (1-\pi_{\eta}) 
			\text{Gamma}(a_{\alpha*}+k-1,b_{\alpha*}-\log(\eta)),
		\end{equation}
		where $k$ is the number of non-empty clusters, 
		$\pi_{\eta} = (a_{\alpha*}+k-1)/\{a_{\alpha*}+k-1+n(b_{\alpha*}-\log(\eta))\}$ and 
		\begin{equation}\label{etadist}
			\eta|\alpha^{\ast},k \sim \text{Beta}(\alpha^{\ast}+1,d). 
		\end{equation}
		Hence, the algorithm proceeds as follows: with $\alpha^{\ast}$ and $k$ fixed at their current values, we sample $\eta$ from
		(\ref{etadist}). Then, based on the same $k$ and the newly sampled value of $\eta$, we sample a new $\alpha^{*}$  from (\ref{alphadist2}). 
		
		\item 
		
		A modification to step 12. of the previous algorithm is needed. Here, we calculate the squared residuals based on 
		$\hat{\ueta}_{\nu_h}^C = \{c_{\eta}/(1+c_{\eta})\}(\utZ_{\nu_h}^{\top} \utZ_{\nu_h})^{-1}\utZ_{\nu_h}^{\top} \utTheta_h$.
		Furthermore, function $S^{*}$ is needed for this step.  
		
	\end{enumerate}
	
	\subsection{MCMC algorithm for the grouped variables model}
	
	The algorithm for the `grouped variables' model differs from that of the `grouped correlations' model in the calculation of the cluster assignment probabilities.
	Here, they are computed as follows.
	Recall that $\utheta_{kl} = (\theta_{1kl},\dots,\theta_{Mkl})^{\top}$ and let
	$w_{h}$ be the prior probability that a variable is assigned to cluster $h$.
	We have the following posterior probability
	\begin{eqnarray}
		P(\lambda_{k}=h|\dots) \propto 
		w_{h} \prod_{l \neq k} N(\utheta_{kl};\uZ^{*}_{kl \nu_{h \lambda_l}} \ueta^{\ast}_{\nu_{h\lambda_l}},\sigma^2 \uD_{kl}^2(\uomega_{\varphi})).\nonumber 
	\end{eqnarray}
	
	\section{Supplementary material II: Application}
	
	Here we provide the Figure that displays the effects of the $4$ covariates on the means of the $4$ responses. 
	
	\begin{figure}[!h]
		\begin{center}
			\begin{tabular}{cccc}
				\includegraphics[width=0.2\textwidth]{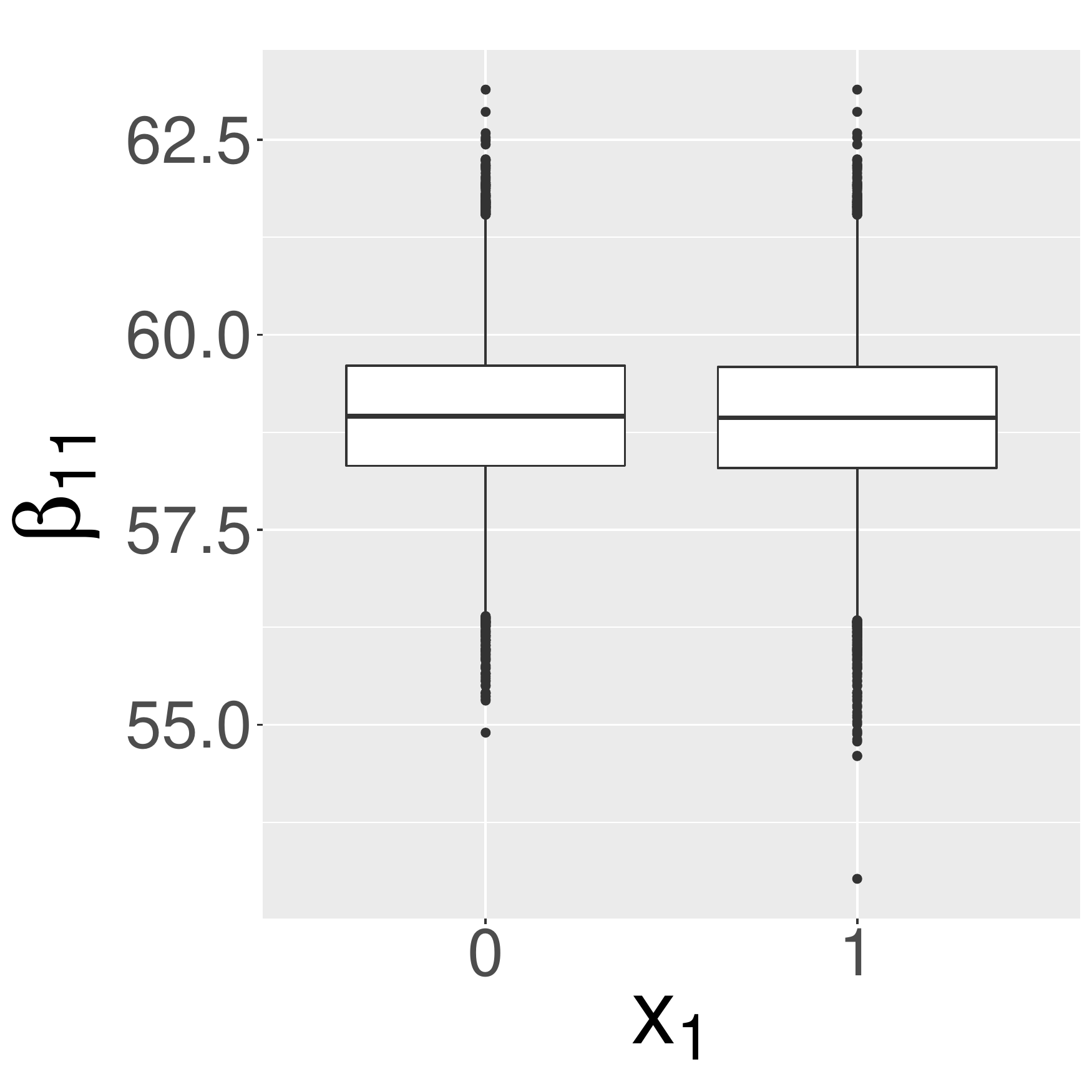} &  
				\includegraphics[width=0.2\textwidth]{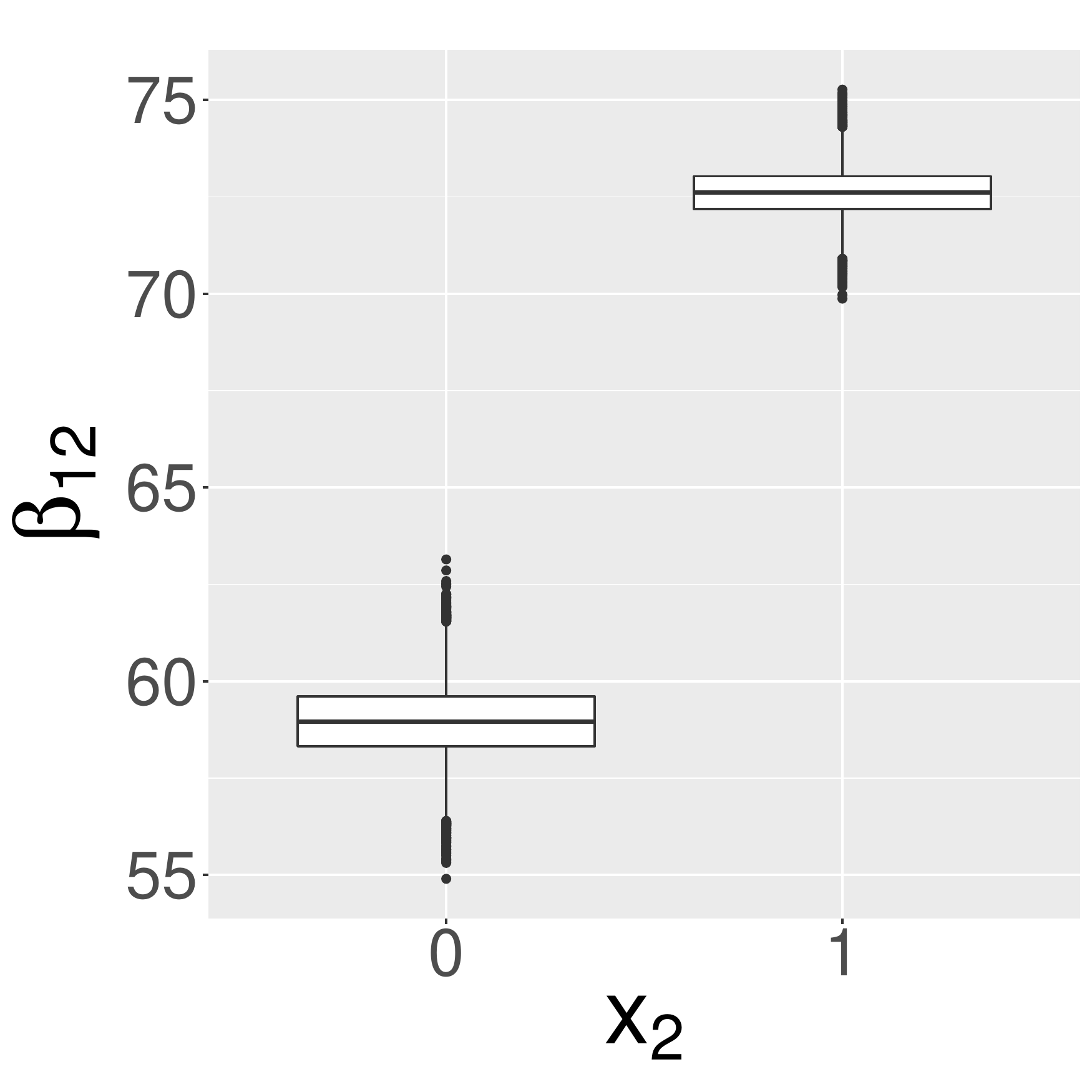} &  
				\includegraphics[width=0.2\textwidth]{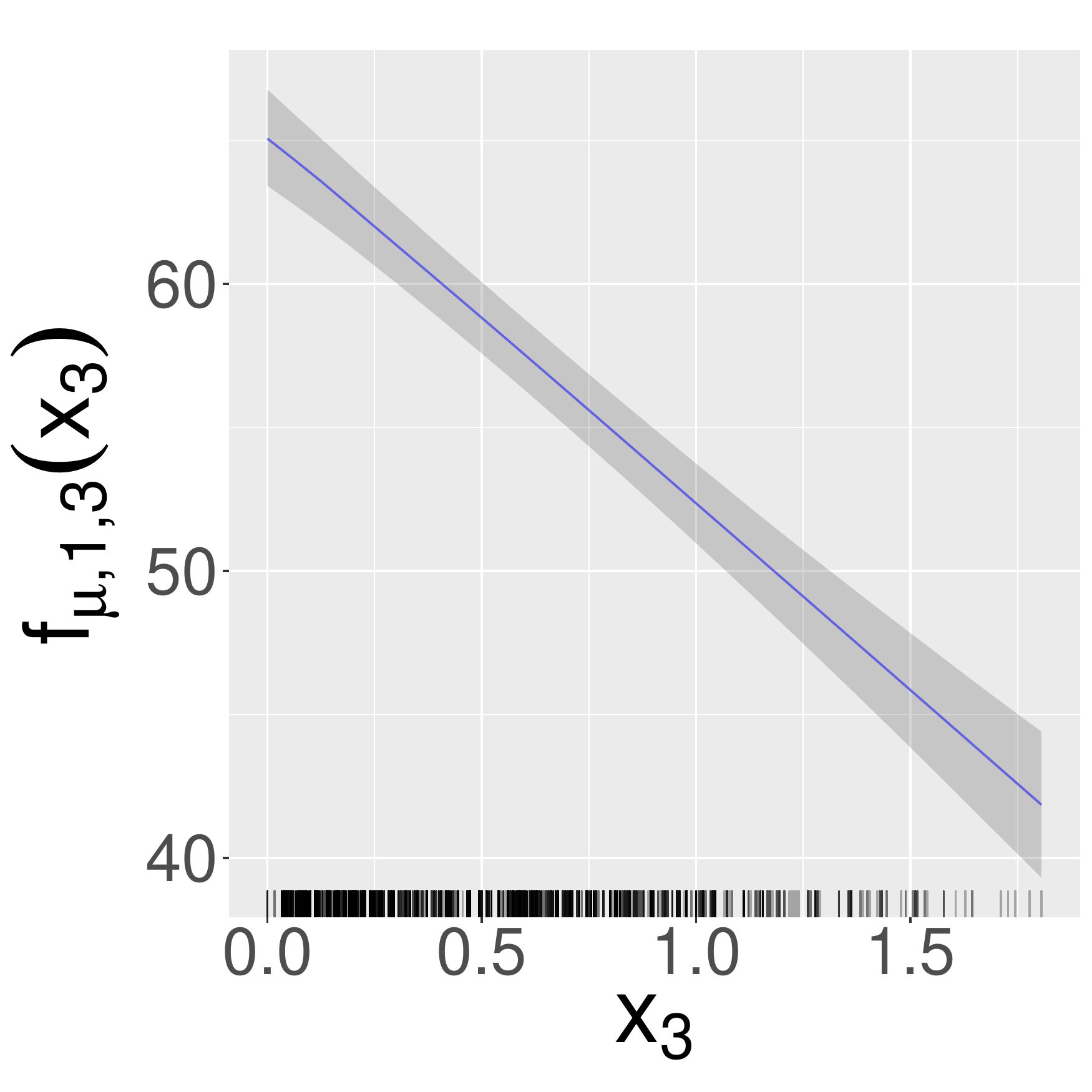} &
				\includegraphics[width=0.2\textwidth]{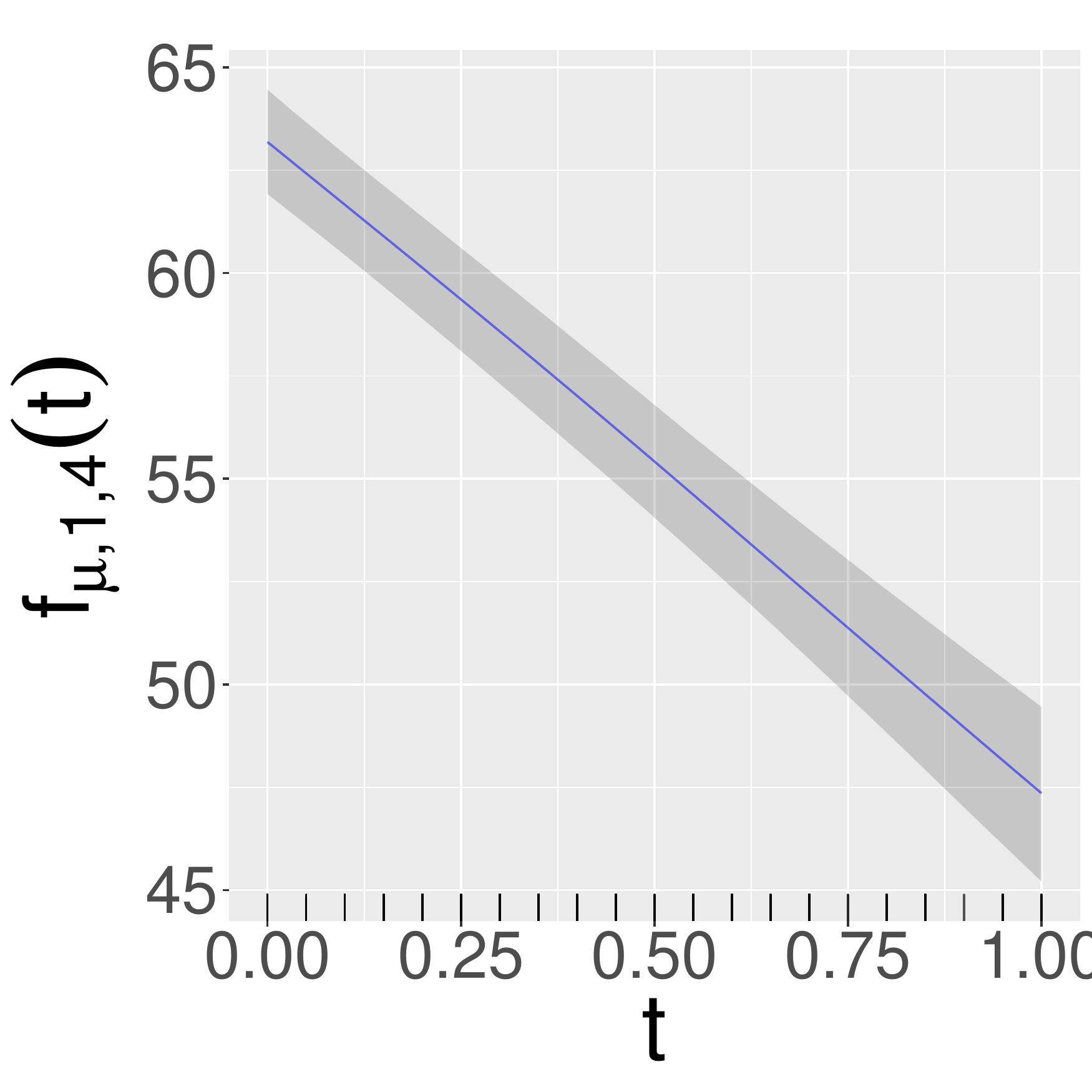} \\    
				\includegraphics[width=0.2\textwidth]{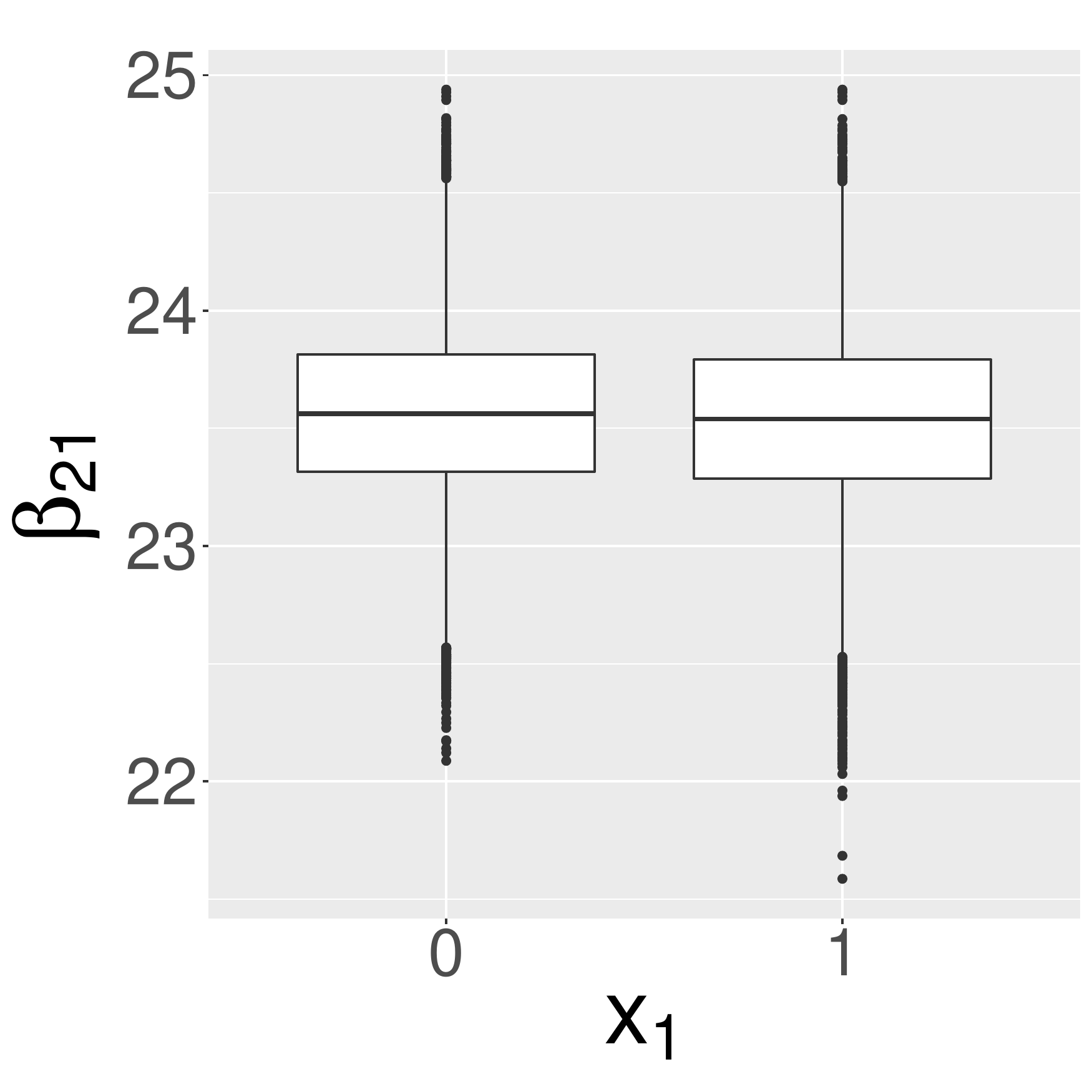} &  
				\includegraphics[width=0.2\textwidth]{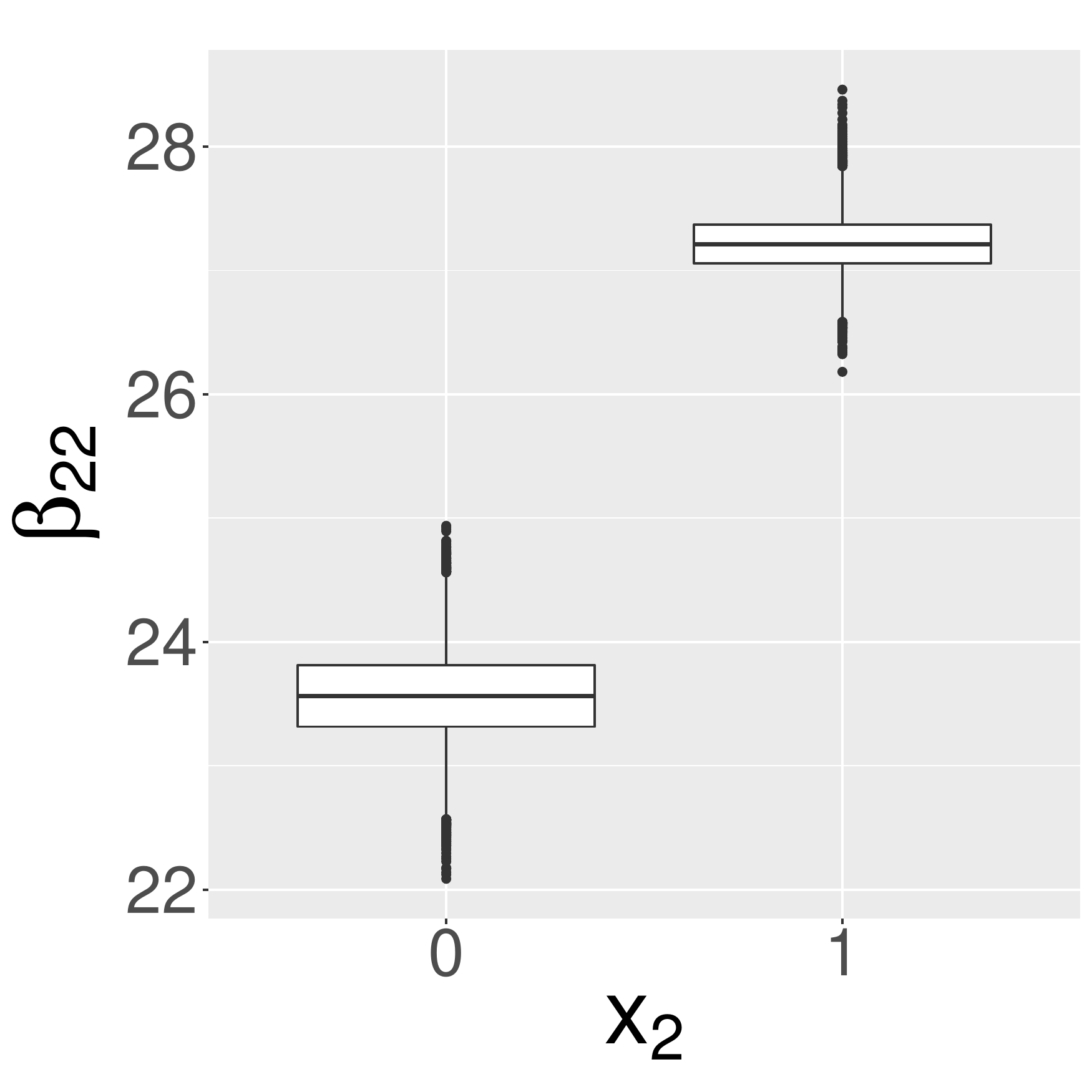} &  
				\includegraphics[width=0.2\textwidth]{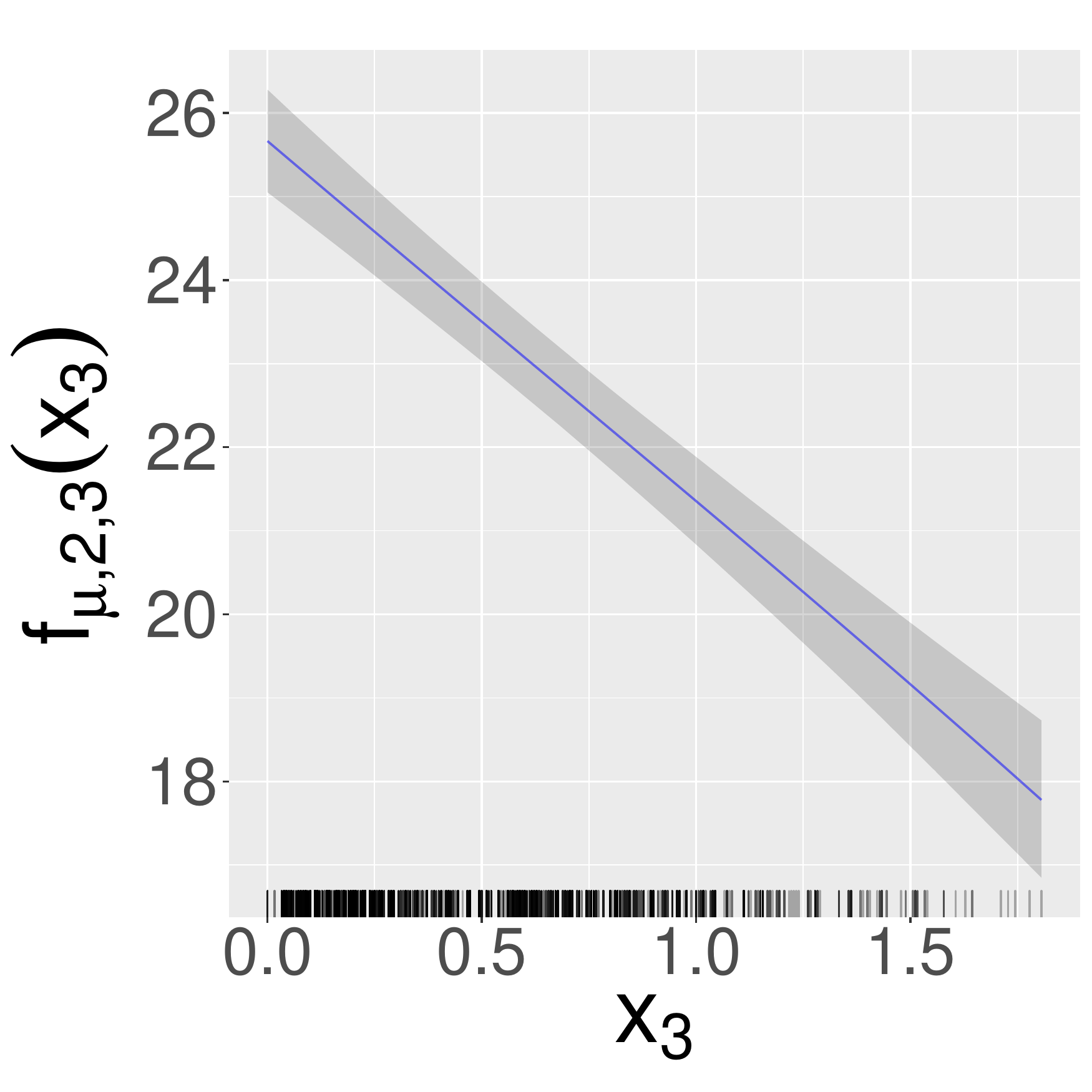} &
				\includegraphics[width=0.2\textwidth]{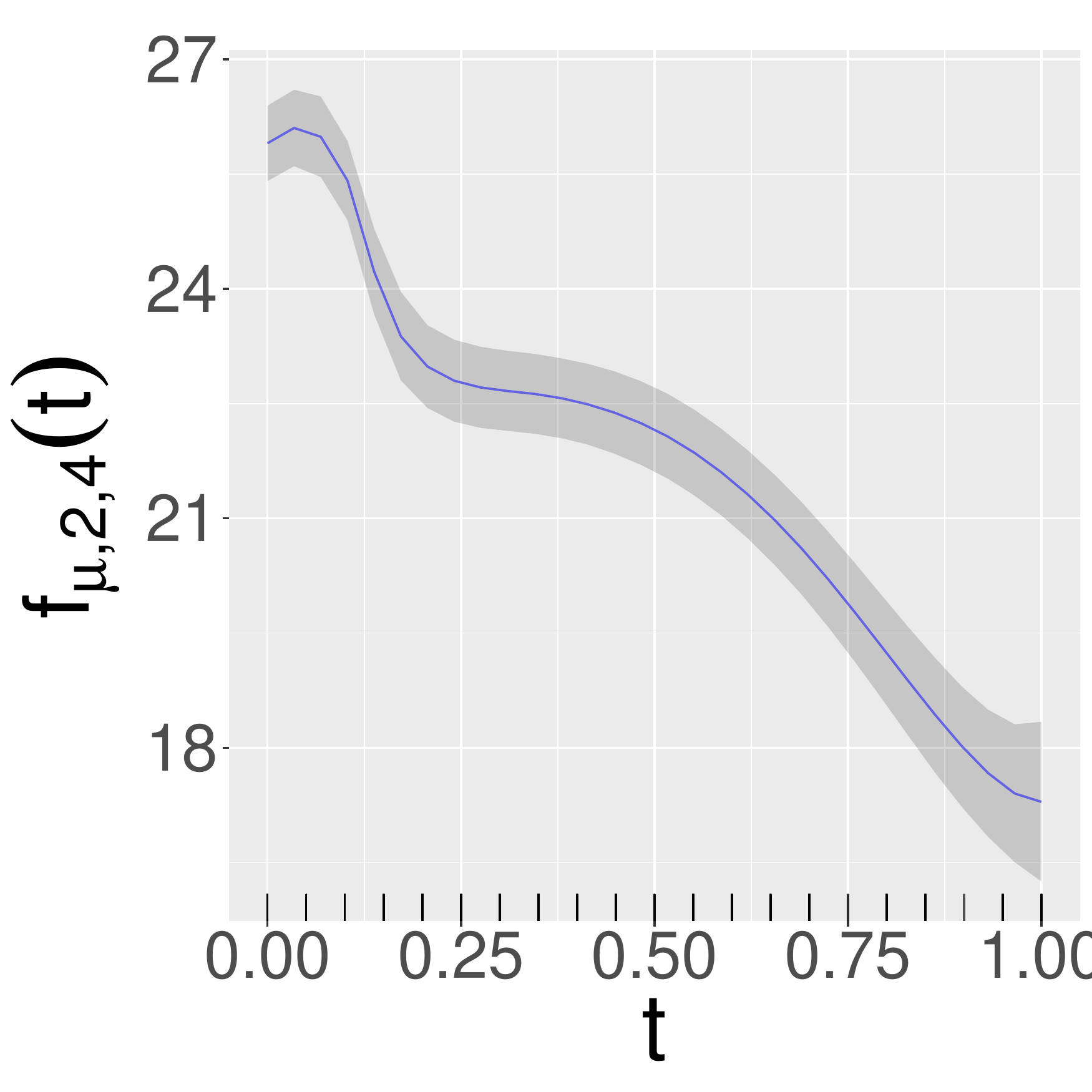} \\    
				\includegraphics[width=0.2\textwidth]{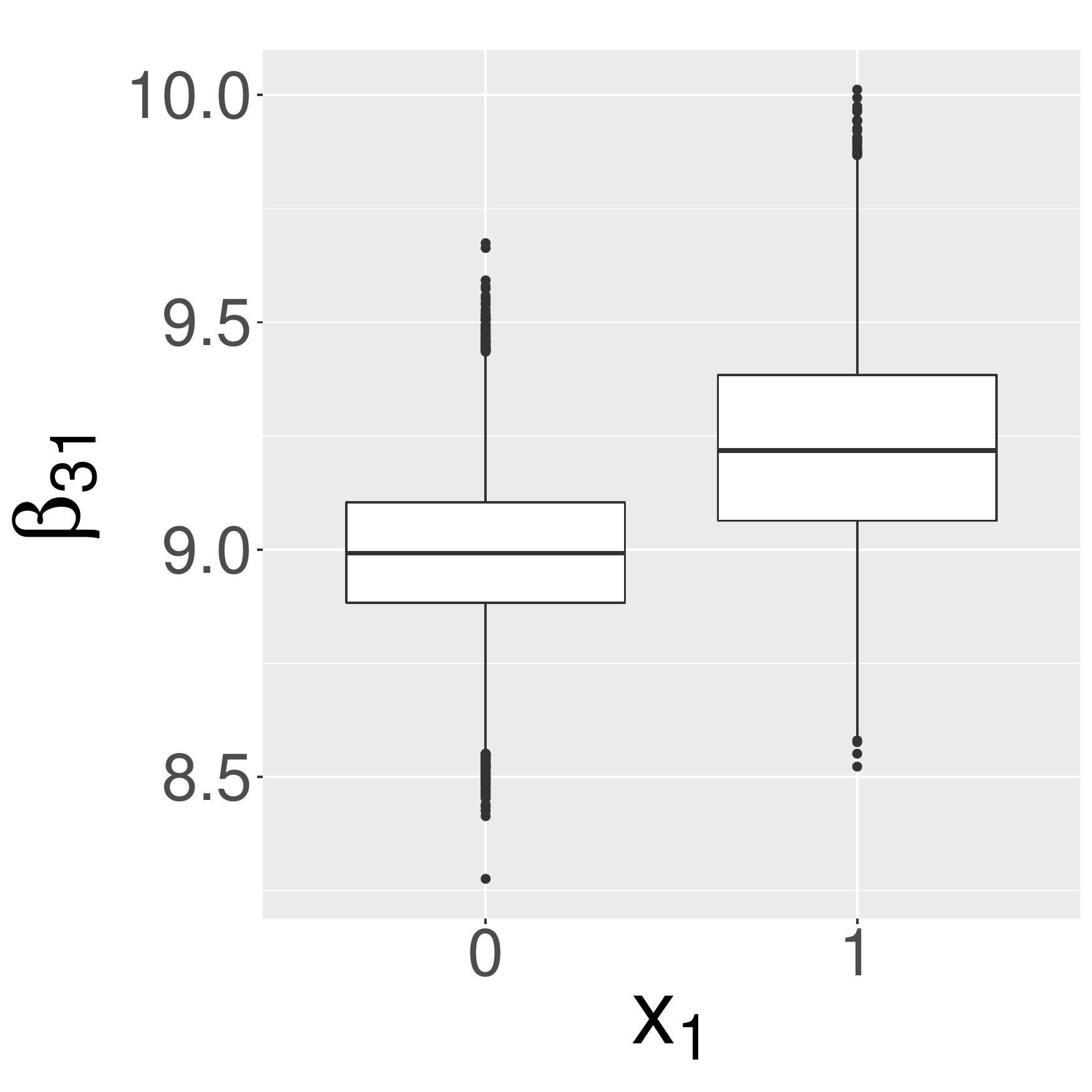} &  
				\includegraphics[width=0.2\textwidth]{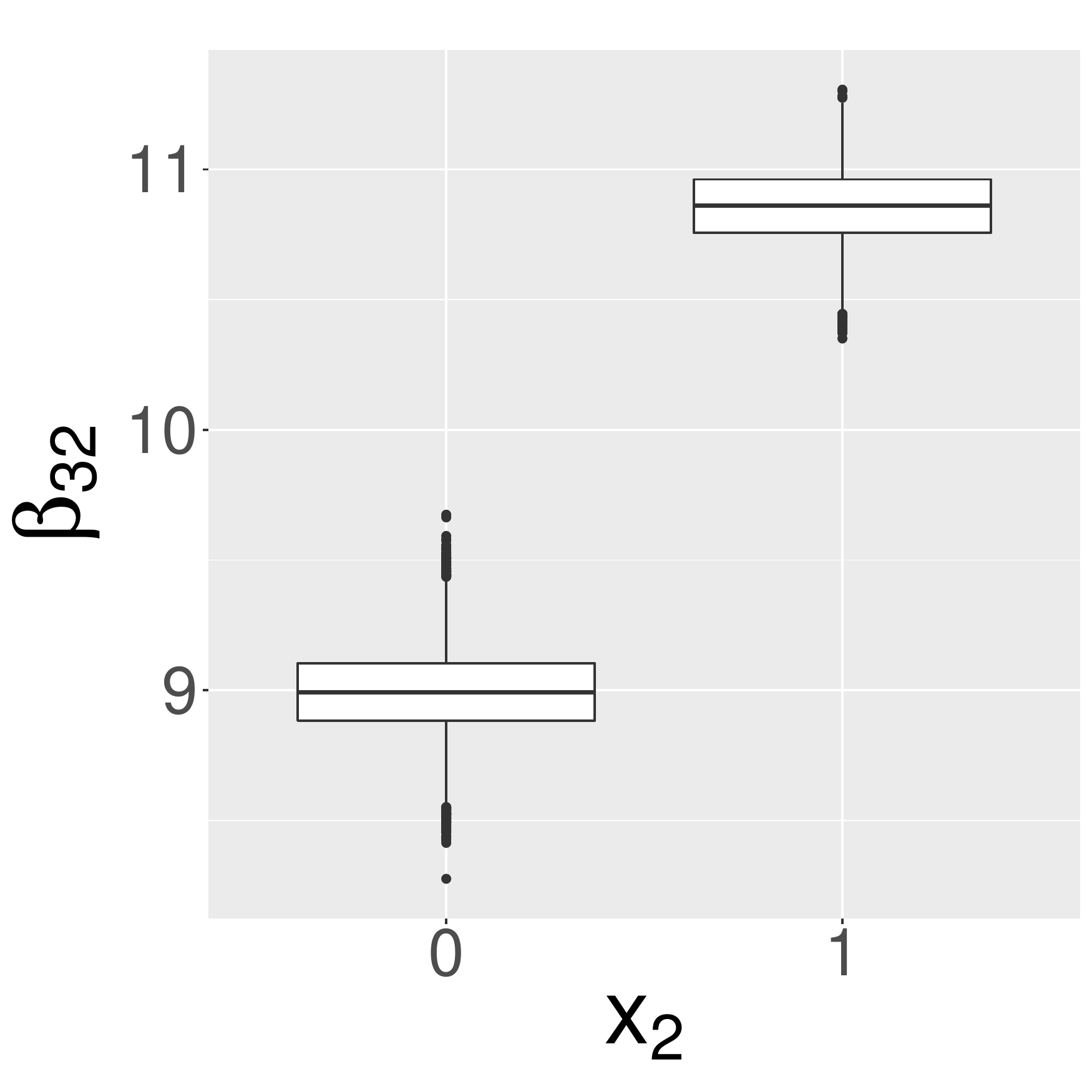} &  
				\includegraphics[width=0.2\textwidth]{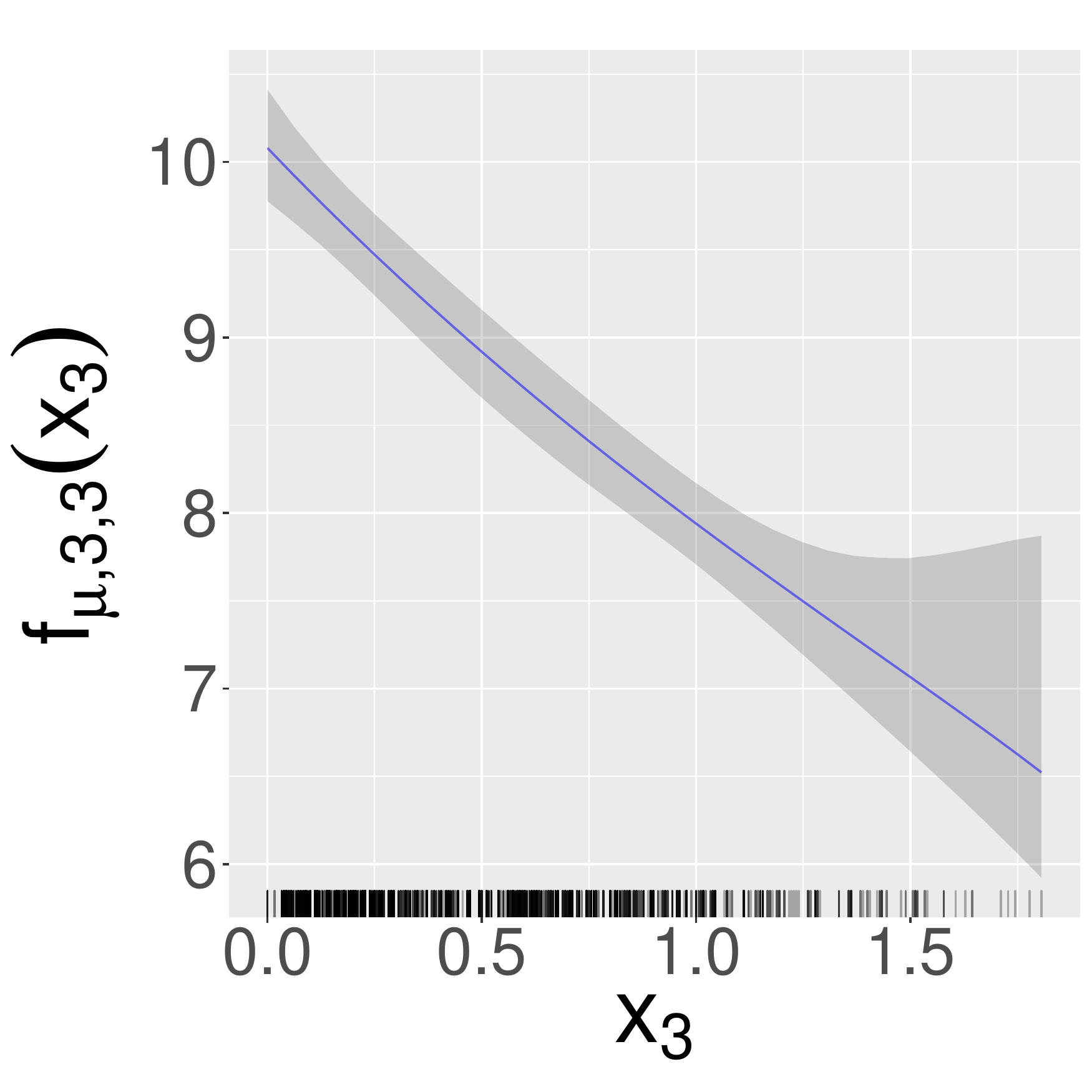} &
				\includegraphics[width=0.2\textwidth]{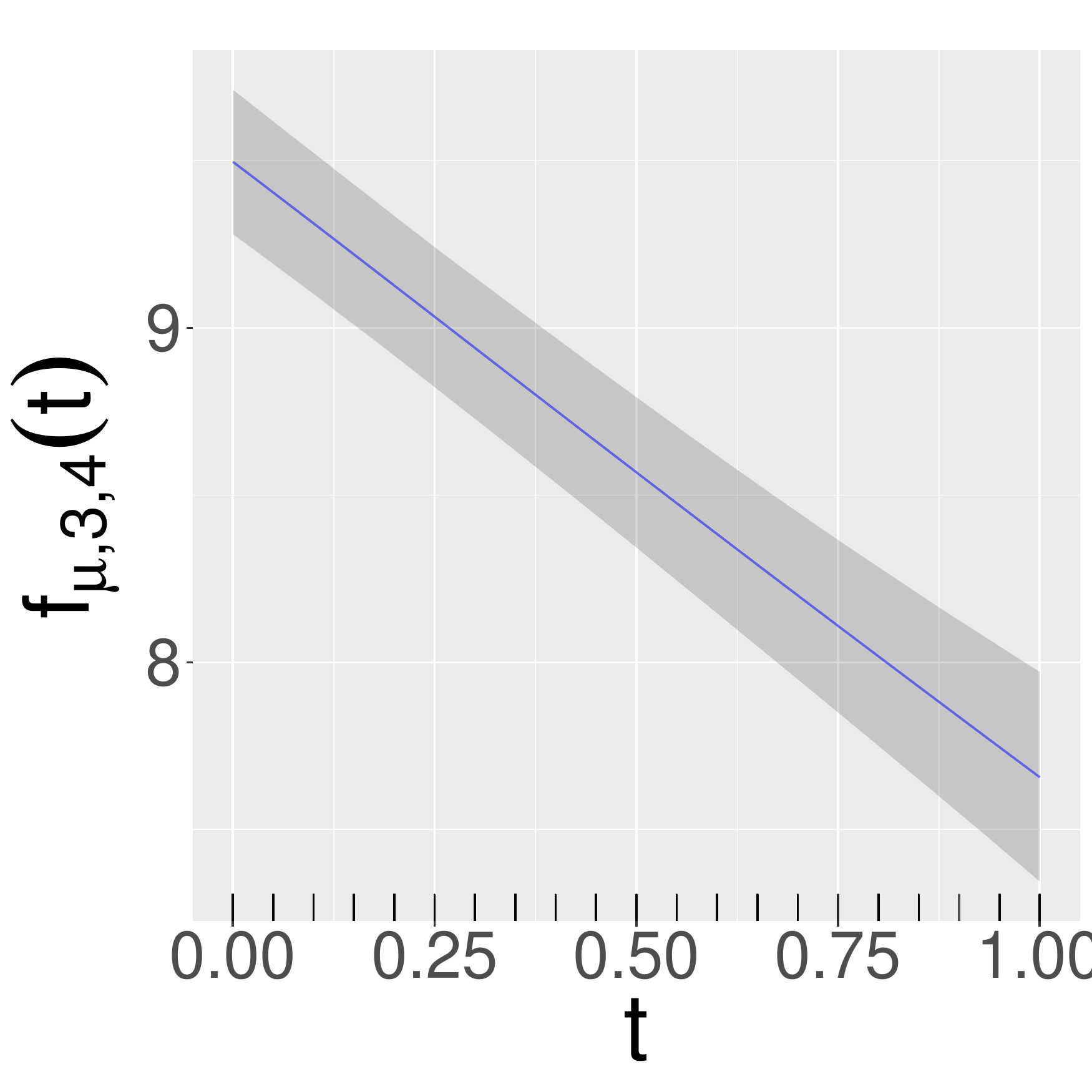} \\    
				\includegraphics[width=0.2\textwidth]{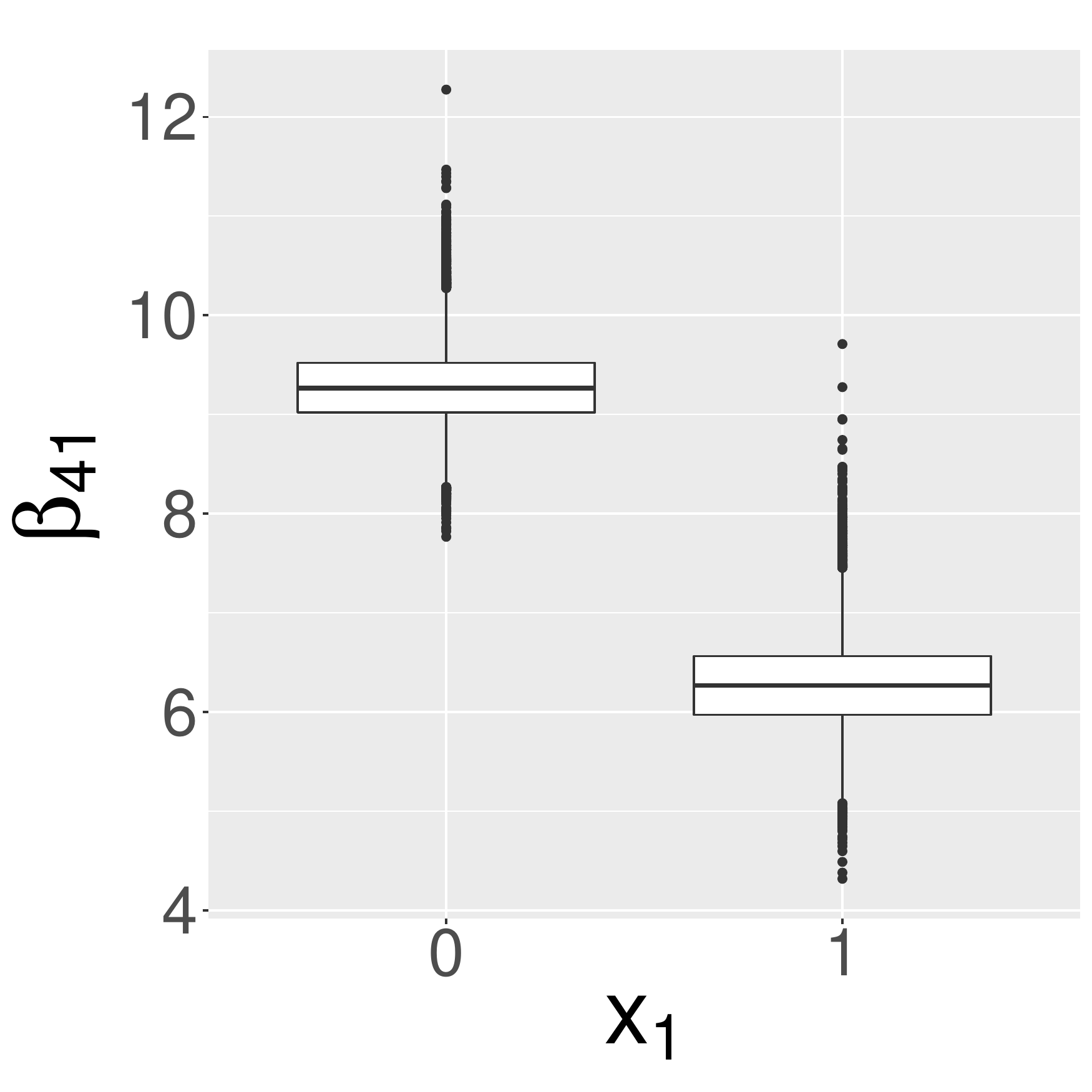} &  
				\includegraphics[width=0.2\textwidth]{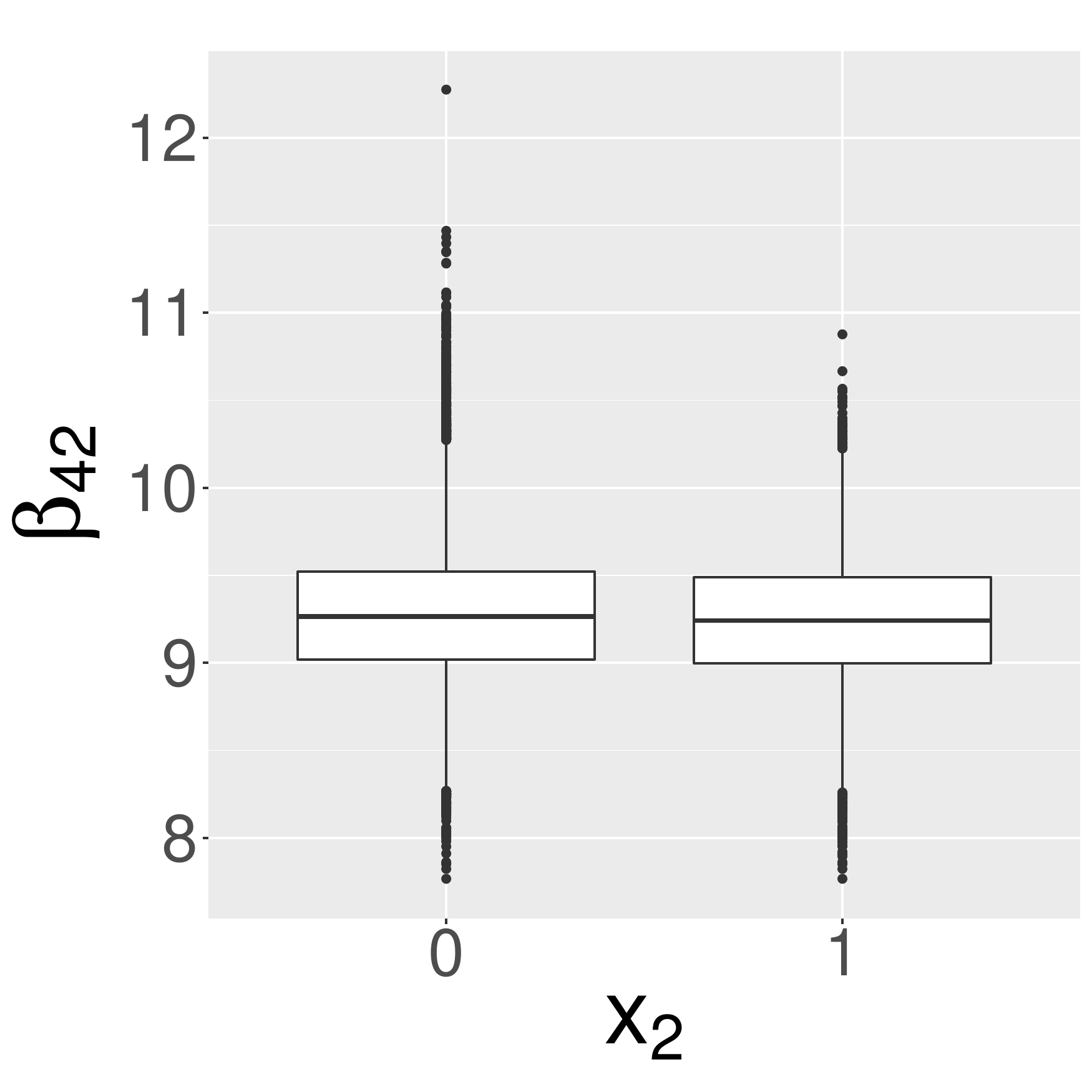} &  
				\includegraphics[width=0.2\textwidth]{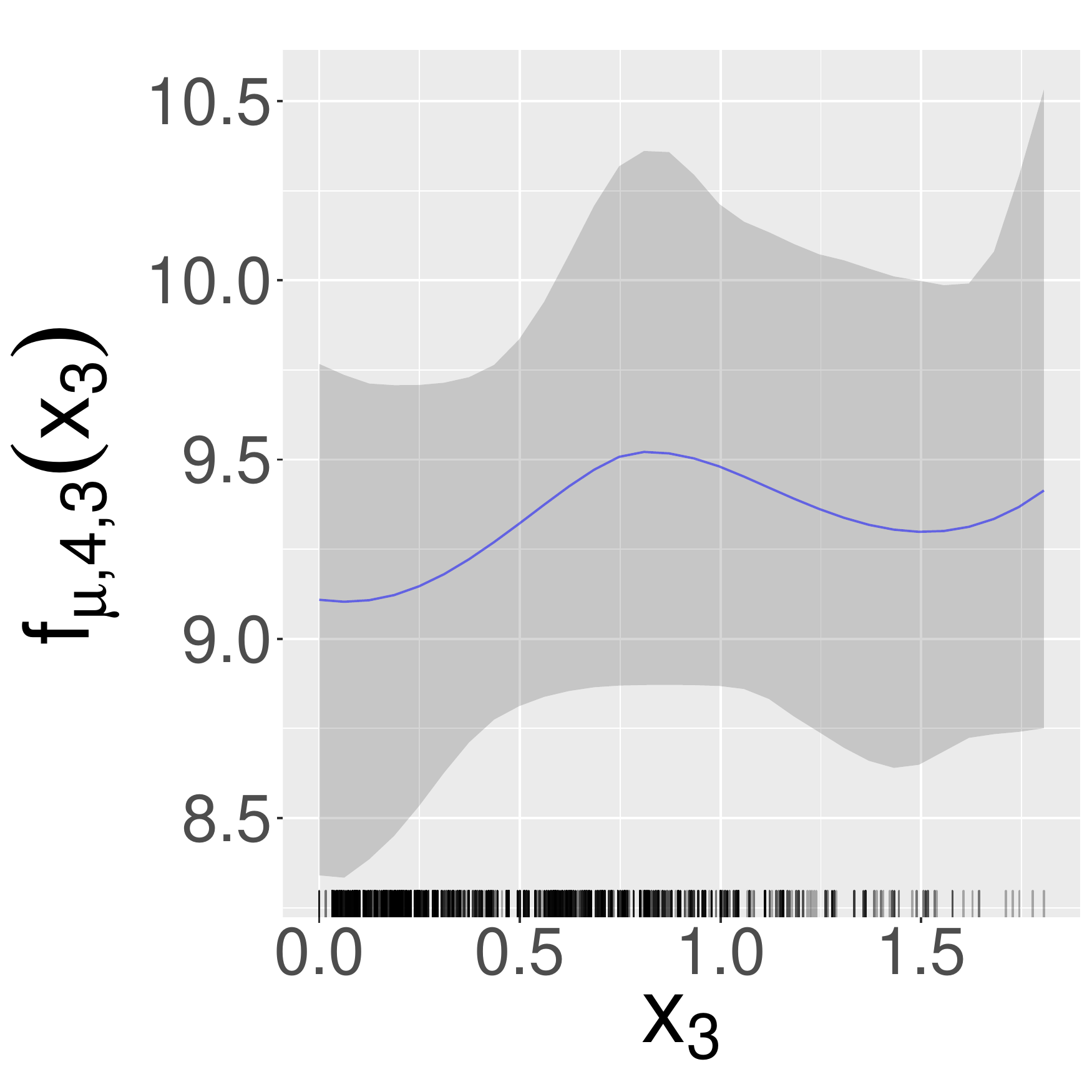} &
				\includegraphics[width=0.2\textwidth]{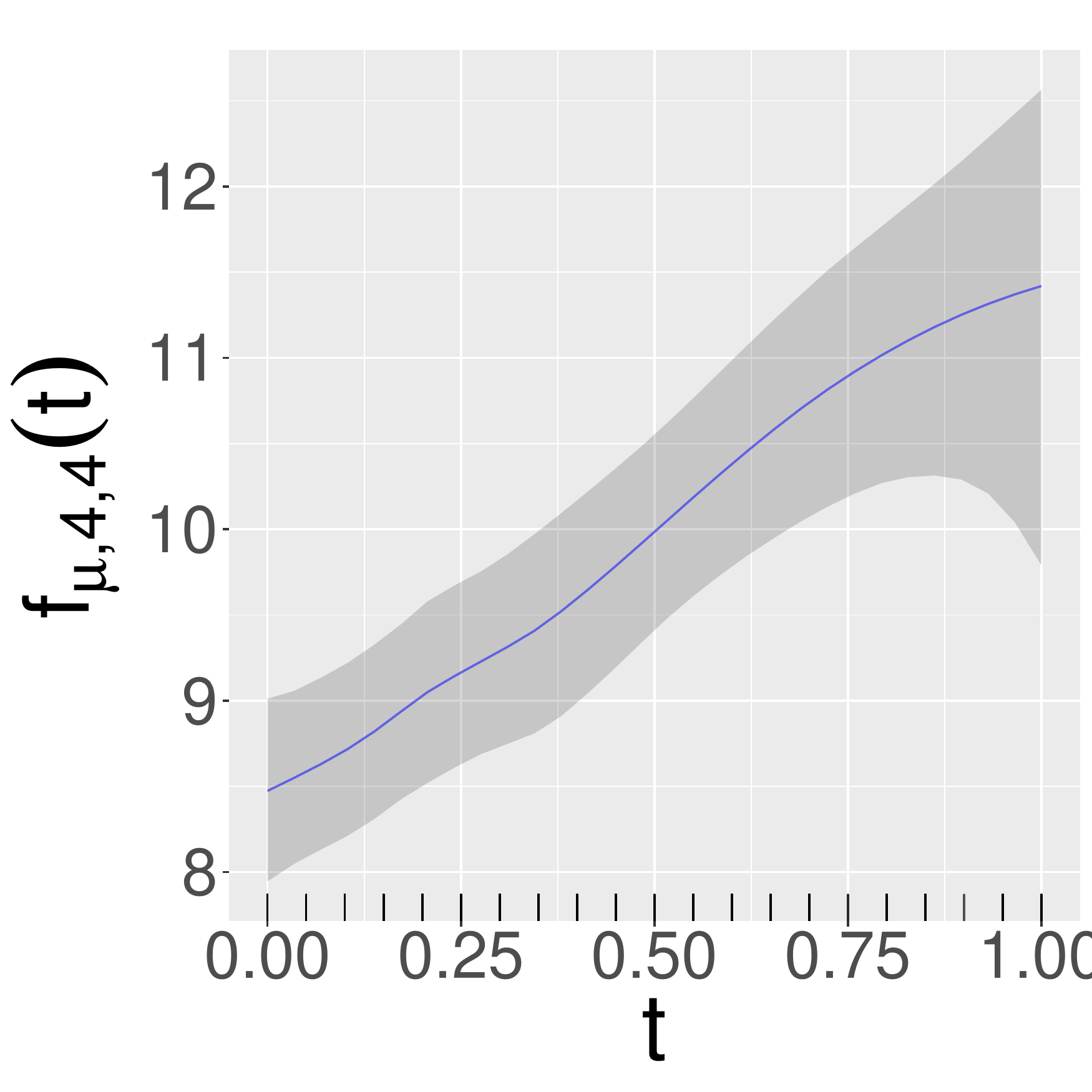} \\    
			\end{tabular}
		\end{center}
		\caption{Application results: mean regression models. Posterior means and $80\%$ credible  intervals.
			Rows refer to the response means and columns to the covariate effects.}\label{Appmeans}
	\end{figure}
\newpage	
	\bibliographystyle{apalike2}
	\bibliography{all.bib}

\end{document}